%% file: 15705.tex
\documentclass{aa}
\usepackage{graphicx,color}
\usepackage{natbib,longtable,amssymb}
\bibpunct{(}{)}{;}{a}{}{,}

\newcommand{\dataset}{data set}
\newcommand{\Dataset}{Data set}

\newcommand{\si}{{Si~{\sc{ii}}~$\lambda$4000}}
\newcommand{\sifive}{{Si~{\sc{ii}}~$\lambda$5800}}
\newcommand{\sisix}{{Si~{\sc{ii}}~$\lambda$6150}}

\def\lsim{\raise0.3ex\hbox{$<$}\kern-0.75em{\lower0.65ex\hbox{$\sim$}}}
\def\gsim{\raise0.3ex\hbox{$>$}\kern-0.75em{\lower0.65ex\hbox{$\sim$}}}

\newcommand{\comment}[1]{}

\newcommand{\xclude}[1]{} 
\newcommand{\fixme}[1]{}
\newcommand{\new}[1]{#1}

\newcommand{\halfcol}{1}
\newcommand{\mycol}{5.5cm}
\newcommand{\modifierat}[1]{#1}
\newcommand{\refcom}[1]{#1}
\newcommand{\newnew}[1]{#1}

\newcommand{\lang}[1]{#1}

\newcommand{\nosniaspectra}{169}
\newcommand{\nosneia}{141}
\newcommand{\nosniaspectralc}{127}

\newcommand{\nobefcont}{116}

\newcommand{\nosniaspecmeas}{89}

\newcommand{\sdssminepoch}{-9}
\newcommand{\sdssmaxepoch}{20}

\newcommand{\nospeccfa}{162}
\newcommand{\noobjcfa}{19}
\newcommand{\nospecscp}{79}
\newcommand{\noobjscp}{16}
\newcommand{\nospecsuspect}{421}
\newcommand{\noobjsuspect}{40}

\newcommand{\distksallstretch}{5\%}

\newcommand{\distmedianrefredshifts}{0.01}

\newcommand{\distmedianrefstretch}{0.94}
\newcommand{\distmedianrefcolour}{0.10}
\newcommand{\distmediansdssredshifts}{0.17}

\newcommand{\distmediansdssstretch}{0.96}
\newcommand{\distmediansdsscolour}{0.05}

\begin{document}

\title{Spectral properties of Type Ia supernovae up to $z\sim0.3$\thanks{\newnew{Based on observations collected at the European Organisation for Astronomical Research in the Southern Hemisphere, Chile, in the ESO programmes 077.A-0437, 078.A-0325, 079.A-0715 and 080.A-0024. Also based on observations with the Nordic Optical Telescope acquired in the programmes with proposal numbers 34-004, 35-023 and 36-010.}}}

\author{J. Nordin\inst{1,2}, L. \"{O}stman\inst{1,2,3}, A. Goobar\inst{1,2},
R. Amanullah\inst{1,2}, R.~C. Nichol\inst{4}, M. Smith\inst{4,5}, J. Sollerman\inst{2,6,7},
B.~A. Bassett\inst{5,8,9}, J. Frieman\inst{10,11}, P.~M. Garnavich\inst{12,13},
 G. Leloudas\inst{6}, M. Sako\inst{14}, D.~P. Schneider\inst{15}
}

\institute{
Department of Physics, Stockholm University, 106 91 Stockholm, Sweden. 
\and
Oskar Klein Centre for Cosmo Particle Physics, AlbaNova, 106 91 Stockholm, Sweden. 
\and
Institut de F\'{i}sica d'Altes Energies, 08193 Bellaterra, Barcelona, Spain. 
\and
Institute of Cosmology and Gravitation, Portsmouth PO13FX, United Kingdom. 
\and
Department of Mathematics and Applied Mathematics, University of Cape Town, Cape Town, South Africa. 
\and
Dark Cosmology Centre, Niels Bohr Institute, University of Copenhagen, DK-2100, Denmark.  
\and
Astronomy Department, Stockholm University, AlbaNova University Center, 106 91 Stockholm, Sweden. 
\and
South African Astronomical Observatory, Cape Town, South Africa. 
\and
African Institute for Mathematical Sciences, Muizenberg, Cape Town, South Africa
. 
\and
Center for Particle Astrophysics, Fermi National Accelerator Laboratory, Batavia, Illinois 60510, USA. 
\and 
Kavli Institute for Cosmological Physics, University of Chicago, Chicago, Illinois 60637, USA. 
\and
Harvard-Smithsonian Center for Astrophysics, 60 Garden Street, Cambridge, MA 02138, USA. 
\and
Department of Physics, 225 Nieuwland Science Hall, University of Notre Dame, Notre Dame, IN 46556, USA. 
\and
Department of Physics and Astronomy, University of Pennsylvania, Philadelphia, PA 19104, USA. 
\and
Department of Astronomy and Astrophysics, Pennsylvania State University, University Park, PA 16802 USA. 
}

 \date{Received Sep 07, 2010; accepted Nov 18, 2010}

\abstract
{}
%
{Spectroscopic observations of Type Ia supernovae \modifierat{obtained} at the New Technology Telescope (NTT) and the Nordic Optical Telescope (NOT), in conjunction with the SDSS-II Supernova Survey, are analysed. We use spectral indicators measured up to a month after the lightcurve peak luminosity to characterise the supernova properties, and examine these
for potential correlations with host galaxy type, lightcurve shape,
colour excess\lang{,} and redshift. }
%
{Our analysis is based on {\nosniaspecmeas} Type 
Ia supernovae at a redshift interval $z=0.05-0.3$, for which
\lang{multiband} SDSS photometry is available. A lower-$z$ 
spectroscopy reference sample was used for comparisons over cosmic time.
We present measurements of time series of 
pseudo equivalent widths and line velocities of the main 
spectral features in Type Ia supernovae. }
%
{\new{Supernovae with shallower features are found predominantly among the intrinsically brighter slow declining supernovae}. We detect the strongest correlation
between lightcurve stretch and the {\si} absorption feature, which also
correlates with the estimated mass and star formation rate of the host galaxy.
We also report a tentative correlation between colour excess and spectral properties. If confirmed, this would suggest that \new{moderate} reddening of    
Type Ia supernovae is dominated by effects in the explosion
or its immediate environment, as opposed to extinction by interstellar
dust. 
}
{}

\authorrunning{Nordin, \"{O}stman et al.}

\maketitle


\section{Introduction}
Cosmological distance measurements based on  Type Ia supernovae (SNe Ia) 
\lang{led} to the discovery of the accelerated expansion of the Universe 
about a decade ago 
\citep{1998AJ....116.1009R,1999ApJ...517..565P}, \lang{which requires that ``dark energy'' exist.} \lang{This mysterious, hypothetical energy is one of the} biggest puzzles in contemporary cosmology and fundamental physics.

With the ever increasing statistical precision on the density and \lang{equation-of-state} parameter of dark energy, we are now reaching a point where
systematic uncertainties are the limiting factors 
\citep{2006A&A...447...31A,2007ApJ...666..694W,2008ApJ...686..749K,2009ApJ...700.1097H,2010ApJ...716..712A}. 
This is emphasised in the \lang{first-year} SDSS-II cosmology results
\citep{2009ApJS..185...32K,2009ApJ...703.1374S,2010MNRAS.401.2331L}.
Two of the major (known) sources of systematic uncertainties when
using SNe Ia to measure cosmological distances are the corrections for
the colour-brightness relation and a possible drift with redshift of
the SN brightness (e.g. \citealt{2008JCAP...02..008N}). These
shortcomings are related to our lack of \lang{any} detailed understanding of the
underlying physics preceding and during the explosion, including the
progenitor system \lang{and both} the circumstellar and interstellar
environments.

Optical spectroscopy provides an excellent testbed for
understanding SNe Ia: differences in explosion properties will likely
modify spectral features. Although detailed 3D modelling may be
needed to extract physical information (or parameters) from observations,
empirical techniques could provide important hints for future modelling. For
example, comparisons
between spectral properties for SNe with different host-galaxy type,
lightcurve parameters and redshifts may be used to infer trends that affect
their use as distance indicators.

There are indications of a population drift with redshift \citep{2006ApJ...648..868S}, which
could be detectable as spectral evolution if the average (composite) spectra in
different redshift bins are compared. Redshift dependencies in composite spectra have also been suggested by \citet{2008ApJ...684...68F} and \citet{2009ApJ...693L..76S}.
\refcom{While the metal content of the Universe increases with time, it is unclear to what degree increased metallicity actually propagates through the progenitor and explosion mechanisms to the outer ejecta of the supernova. If the element distribution of the ejecta is affected, this will likely change the observed spectrum \citep{2000ApJ...530..966L,2008MNRAS.391.1605S}.}

An important question is whether the available low-$z$ SN data
sets correctly sample the demographics of SNe Ia at cosmological
distances, or if there are subtypes of SNe not (yet) observed
in the smaller local samples, but present at higher redshifts. Either
case could yield an evolution of the average spectrum. Our data set at
intermediate redshifts provides a useful sample to close the ``gap''
in studies currently available in the literature. We also study individual spectra
and are thus potentially able to disentangle shifting population
demographics from
new subtypes.

It is also important to explore if there is any relation between SN Ia
spectral features and broad-band colours. The colour-brightness
relation of SNe Ia does not seem to match the Milky-Way dust extinction law
\citep{riess96,tripp98,krisciunas00,altavilla04,reindl05,2006A&A...447...31A,guy07,nobili07}. The
explanation may be connected to e.g., interactions in the
circumstellar environment \citep{2008ApJ...686L.103G} or intrinsic SN
properties, in which case a correlation of properties of spectral
features and broad-band colours could be expected. Thus, comparisons
of spectra of SNe with different colours may help in the understanding of the intrinsic colour \modifierat{dispersion} in SNe Ia and allow us to disentangle the
various components entering the colour-brightness relation and its
possible evolution with redshift, critical for precision cosmology.

%
Relations between lightcurve parameters and host galaxy
properties have been presented recently in
\citet{2010ApJ...715..743K,2010arXiv1003.5119S} and \citet{2010arXiv1005.4687L}. These empirical findings call for further scrutiny; one way to do so is through comparisons with properties of SN Ia spectra.

%
 Although a number of spectral feature comparisons of SN
spectra have been performed in recent years
\citep[e.g.][]{2005AJ....130.2788H,2005ApJ...623.1011B,2006PASP..118..560B,2006AJ....131.1648B,2007AA...470..411G,2008ApJ...684...68F,2008A&A...477..717B,2008ApJ...674...51E,2009ApJ...693L..76S,2009ApJ...699L.139W,2009arXiv0905.0340B}, we are still far from a complete understanding of the observed variation of SN Ia spectra. Detailed spectral studies are needed in order to limit the possible differences between low and high redshift objects, a basic
requirement for the use of SNe Ia as distance indicators, and could
potentially be used to further sharpen the standarizable candle
through secondary brightness indicators. 


During 2006-2007, {\nosniaspectra} spectra of SNe Ia were
obtained at the New Technology Telescope (NTT) and the Nordic Optical
Telescope (NOT) in a program designed for spectral identification of
objects detected by the Sloan Digital Sky Survey II (SDSS-II)
Supernova Survey \citep{1998AJ....116.3040G,2000AJ....120.1579Y,2008AJ....135..338F}. 
The SDSS-II Survey operated as a three-year survey
(2005-2007), aiming at finding a large number of intermediate-redshift
SNe Ia, to be used to estimate cosmological parameters. The search algorithm and the procedure for the spectroscopic observations have been
described in \citet{2008AJ....135..348S}. The first-year
photometry and spectroscopy have been presented in
\citet{2008AJ....136.2306H} and \citet{zheng08}, respectively.

The NTT/NOT SDSS spectra
provide a key opportunity to study SN properties. First, the SN
population is drawn from an interesting redshift range, where evolution
could be expected, yet the SNe are close enough to yield
a reasonably high S/N. Secondly, this
{\dataset} is large enough to allow statistical tests. This {\dataset} is described in detail in \citet{ostman09}.
%
%

We present quantitative measurements of {\nosniaspecmeas}
SN Ia spectra from the NTT/NOT samples with good
lightcurves and low to moderate host-galaxy contamination and compare
these with samples of nearby SNe Ia. We focus on potential evidence of
evolution, but also study correlations with lightcurve properties, such
as stretch and colour, as well as host galaxy properties like stellar mass and star formation rate.


The two main sources of systematic uncertainties in spectral studies
of SNe Ia are noise degradation and host galaxy contamination. These
effects complicate the search for potential spectral evolution, and would cause systematic errors since they will affect nearby and distant SNe differently. An unknown systematic bias could be misinterpreted as a sign of evolution, or
even obscure a real effect.

The analysis of spectral indicators presented here consists of several steps: 
\begin{itemize}
\item We first compare NTT/NOT spectra with \modifierat{nearby} data (Sect.~\ref{sec:results}). Deviating SDSS SNe, possible signs of \emph{evolution}, are collected into a \emph{deficit} subsample (named so since measurements are \modifierat{smaller} than average).
\item We then change focus and combine all data in order to search for correlations with global SN parameters (Sect.~\ref{sec:lccorr}). We use LC parameters (SALT and MLCS2k2) and host galaxy properties and search for the epoch ranges with the most significant correlations.
\item We finally try to understand the \emph{origin} of both the deficit subsample and the major correlations with global parameters (Sect.~\ref{sec:dis}). This is done through (i) \modifierat{studying} the spectral region around 4000-4500 {\AA} (rest frame),  (ii) a comparison of \emph{deficit} SNe with normal SNe and (iii) a short discussion of host galaxy correlations.
\end{itemize}

Through each step of this analysis, focus has been put on
minimising/mapping any sort of systematic error and/or observer
bias. The observations, data reduction and host galaxy subtraction
methods of the NTT/NOT spectra are presented in \citet{ostman09}, while comparisons of
indicator measurements will be presented here. Extensive Monte Carlo simulations were run in order to estimate errors caused by host galaxy subtraction or varying noise levels.




The full organisation of this paper is as follows: In Sect.~\ref{sec:indic} we introduce spectral indicators and in Sect.~\ref{sec:data} we
present the {\dataset}s used. In Sect.~\ref{sec:results} the indicator measurements, as a function of spectral epoch, are displayed for both the NTT/NOT and the reference SNe.
Sect.~\ref{sec:lccorr} contains studies of the correlations between
spectral indicators and lightcurve parameters as well as with host galaxy properties.
Sect.~\ref{sec:dis} contains a discussion (focused on evolution and the {\si} feature) and finally conclusions are given in
Sect.~\ref{sec:conclusions}.
We present further details of our Monte Carlo tests in
appendices~\ref{app:comphost} and~\ref{app:filter}.
\newnew{All SNe included in this study can be found listed in Appendix~\ref{app:tables}.}


\section{Spectral indicators}
\label{sec:indic}

\begin{figure}
  \centering 
  \includegraphics[angle=-90,width=\columnwidth]{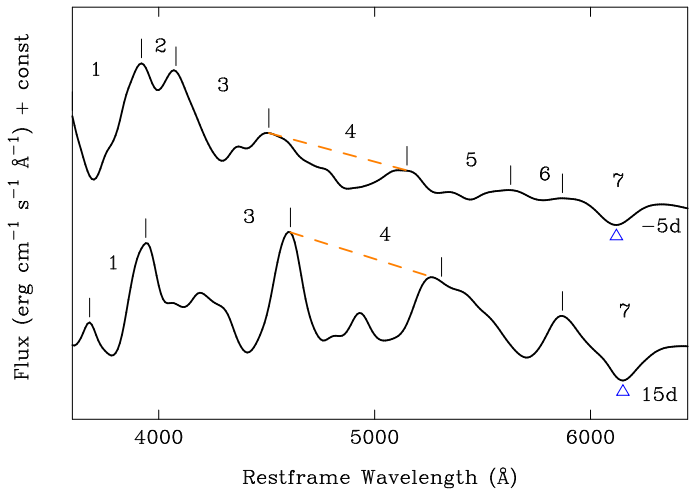}
  \caption{The feature regions used for the measurements of spectral
    indicators are shown for two template spectra at day $-5$ and $15$ relative to lightcurve peak \cite[templates from][]{2007ApJ...663.1187H}. The \emph{pseudo} continuum (dashed orange) and line minimum (marked with blue triangle) are shown for two features (feature 4 and 7, respectively). }
  \label{fig:spec_limits}
\end{figure}

\refcom{Elements in the SN ejecta will absorb photons originally emitted by radioactive material in the inner layers, thus causing the typical pattern of ``features'' visible in SN Ia spectra.}
In this analysis of spectral features we concentrate on seven regions corresponding to these studied by \citet{2004NewAR..48..623F} and \citet{2007AA...470..411G}. In Figure~\ref{fig:spec_limits}, the features are displayed for a typical spectrum at two different ages. 

Each feature is labelled after the ion that normally dominates
the absorption in this region (see Table~\ref{tab:pew}), but since
most absorption lines are blends of several lines and should not be
directly identified with physical properties, these regions will
simply be identified as features 1 to 7. 


With the term ``spectral indicator'' we refer to a measurement of a spectral feature of a SN Ia
spectrum. 
\refcom{Spectral indicators are always measured on rest-frame spectra, and all spectra presented here have been de-redshifted.}
In this paper we will discuss two indicators, pseudo Equivalent Widths (pEWs) and velocities. Feature 5, shaped like a 'w', has two minima; we use the redder of these as velocity indicator.

\subsection{Pseudo equivalent widths}
\label{sec:pewdesc}

\refcom{For astrophysical objects with a well-defined continuum, the equivalent width of an absorption feature can be measured easily. If, furthermore, the density structure is known and the absorption is caused by a single ion, this information can be used to deduce elemental abundances. }
For SNe Ia, the spectra are dominated by wide
  absorption features caused by mixed multiple absorption lines. The
  continuum can thus not be read directly from the observed data, and
  the physical interpretation of equivalent widths becomes non-trivial.
  Nevertheless, we can measure equivalent widths if an unambiguous (pseudo) continuum can be \emph{defined}. We do this following \citet{2004NewAR..48..623F} and \citet{2007AA...470..411G}. A lower and upper \emph{limit} is found at the peak-flux wavelength within lower and upper wavelength \emph{regions}. These regions, for the features used here, are given in Table~\ref{tab:pew}. 
The pseudo-continuum is defined as the straight line between the flux of these lower and upper limits, with the choice of peaks optimised so that the pseudo continuum is maximised while not intersecting the spectrum. 

With this pseudo-continuum a (pseudo) Equivalent Width (pEW) can be calculated (using the standard equivalent width formula):

\begin{equation}
  pEW = \sum_{i=1}^{N} \left( 1-\frac{f(\lambda_i)}{f_{c}(\lambda_i)} \right) \Delta\lambda_i,
  \label{eq:pew}
\end{equation}
where $f$ is the observed flux and $f_c$ is the pseudo-continuum.  The
sum is taken over all wavelength bins contained between the lower and
upper limit.   

\begin{table}
\caption{Feature boundaries (pEW)}
\label{tab:pew}
\begin{tabular}{p{0.9cm}p{2cm}p{2cm}p{2cm}}
\hline \hline
Feature &Dominating line &Lower region (\lang{centre} {\AA}) &Upper region (\lang{centre} {\AA}) \\
\hline
f1 & Ca~{\sc{ii}} H\&K & 3450 - 3800 & 3800 - 4100 \\
f2 & Si~{\sc{ii}} $\lambda$4000 & 3800 - 3950 & 4000 - 4200 \\
f3 & Mg~{\sc{ii}} $\lambda$4300 & 3850 - 4250 & 4300 - 4700 \\
f4 & Fe~{\sc{ii}} $\lambda$4800 & 4300 - 4700 & 4950 - 5600 \\
f5 & S~{\sc{ii}} W     & 5050 - 5300 & 5500 - 5750 \\
f6 & Si~{\sc{ii}} $\lambda$5800 & 5400 - 5700 & 5800 - 6000 \\
f7 & Si~{\sc{ii}} $\lambda$6150 & 5800 - 6100 & 6200 - 6600 \\
\hline
\end{tabular}
\end{table}


The pEW definition has the advantage of not having to fit any function
to the data (most features are clearly non-Gaussian) as well as being
insensitive to multiplicative differences between spectra (assuming
the multiplied factor does not change drastically over the range of
the feature).
However, pEWs are not insensitive to additive flux differences.

As for equivalent widths, the \emph{statistical} error is given by
\begin{equation}
  \sigma_{pEW}^2 = \sum_{i=1}^{N} \left( \frac{\sigma_{f}^2(\lambda_i)}{f_c^2(\lambda_i)} + \frac{f^2 (\lambda_i) }{f_{c}^4 ( \lambda_i )} \sigma_{f_c}^2 (\lambda_i)  \right) \Delta\lambda_i^2.
  \label{eq:dpew}
\end{equation}
It consists of two parts, the first is obtained from the error
spectrum, $\sigma_{f}$, while the second propagates the uncertainty
from the choice of pseudo-continuum, $\sigma_{f_c}$.

To avoid subjectivity in the pEW measurements, all steps were automated, e.g. the level of filtering was
determined through lookup tables (see below and Appendix B) and boundaries for the pEW are determined using computer algorithms. 
%
The automated code was validated by measuring the pEWs on the
same data as \citet{2007AA...470..411G}. The same indicator trends were obtained when using the same input spectra.\fixme{Does this need to be further quantified?}\footnote{As a further check on these algorithms the automatic measurements were manually revised and the outcome compared with our basic results. No major deviations were seen.}

In addition to the statistical error, there are several sources of
systematic uncertainties. 
Host-galaxy contamination can both change
the shape of the feature and induce an additive flux change. 
Differential slit loss effects, which
although being multiplicative can have large effects on the
host-galaxy subtraction process, are included  in contamination errors.
These systematic pEW uncertainties
are discussed below.




\paragraph{Noise-filtering uncertainties}

\modifierat{Filtering or smoothing is necessary in order to identify the end points of spectral features.}
However, the optimal filter parameters will
change with Signal-to-Noise (S/N) ratio; noisy data need stronger smoothening to
reduce the impact of random fluctuations in choosing 
the feature endpoints, while the same filter
strength will dilute information in high S/N data. Over or under filtering can introduce a measurement bias.
This is a particular concern when making redshift comparisons, since
distant SNe  have (in general) lower S/N than nearby
SNe.

Extensive Monte Carlo (MC) simulations were \modifierat{carried out} to make an optimal
choice between filtering methods and their parameters. The simulations
are described in detail in Appendix~\ref{app:filter}. We conclude that
a standard boxcar smoothing performs well, provided that the
boxcar width is modified depending on the noise level of the spectrum
and what feature is being studied. The simulations were used to find the optimal boxcar width for each S/N.
For each feature and S/N level we also obtain an uncertainty from the simulations which is added in quadrature to the systematic pEW error.

\paragraph{Host-galaxy contamination uncertainties}

The host galaxy contamination is the single largest source
of systematic errors for pEW studies. Unsubtracted host light can both cause an error through a flux offset and, if the underlying galaxy is changing with wavelength, through shifting pEW feature boundaries. 
Monte Carlo simulations were performed to estimate the uncertainties
due to the host-galaxy subtraction. When host galaxy contamination could be estimated using photometry, both the error and a possible bias is retrieved as a function of contamination (this is the case for all NTT/NOT spectra as will be discussed below). 
\modifierat{For spectra where no host-galaxy is subtracted (reference spectra and low contamination SDSS spectra) we add  the uncertainty expected for an
uncorrected 10\% galaxy contamination. }
%
See Appendix~\ref{app:comphost} for a more detailed description of these simulations.

\begin{figure}
  \centering 
  \includegraphics[angle=-90,width=0.9\columnwidth]{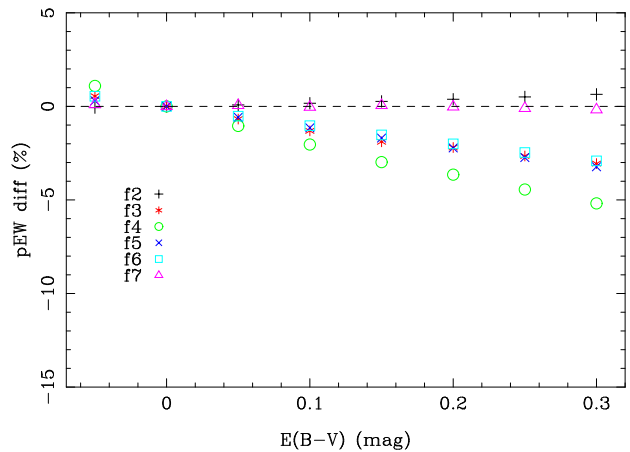}
  \caption{Percent change in pEW for features 2 to 7 as a function of E(B$-$V). The measurements are based on the \citet{2007ApJ...663.1187H} SN template at peak brightness and assuming \citet{cardelli89} type dust with $R_V=1.7$.
}
  \label{fig:ebvpew}
\end{figure}

\paragraph{Reddening}
Pseudo-equivalent widths are also affected by un-corrected reddening by 
host-galaxy dust. \new{In Figure~\ref{fig:ebvpew} we show how pEWs, as measured on a template spectrum, change as dust-like extinction is applied to the template. All features gradually
\emph{decrease} with colour excess, $E(B-V)$. Changes are
smaller than $10\%$ for $E(B-V)\le 0.3$ mag. As expected, wider features change more with extinction than narrower ones.}
%


\subsection{Line velocities}

The position of absorption and emission features can also be used to probe SN properties. We study the wavelength minima of the features defined above. \refcom{As reference minima we use the \emph{rest} wavelength of the ion that each feature was named after, these are given in Table~\ref{tab:vel} (3rd column).}

\begin{table}
\caption{\refcom{Feature minima wavelengths}}
\label{tab:vel}
\begin{tabular}{p{0.9cm}p{2.6cm}p{3cm}}
\hline \hline
Feature &Dominating line ($\sim\lambda$ observed) & Rest wavelength ({\AA}) \\
\hline
f1 & Ca~{\sc{ii}} H\&K & 3945.12 \\
f2 & Si~{\sc{ii}} ($\lambda$4000) & 4129.73\\
f3 & Mg~{\sc{ii}} ($\lambda$4300) &  4481.20\\
f4 & Fe~{\sc{ii}} ($\lambda$4800) & 5083.42\\
f5 & S~{\sc{ii}} W     & 5536.24\\
f6 & Si~{\sc{ii}} ($\lambda$5800) &  6007.70\\
f7 & Si~{\sc{ii}} ($\lambda$6150) &  6355.21\\
\hline
\end{tabular}
\end{table}

These are converted into velocities through the relativistic Doppler formula,
\begin{equation}
  v_{\rm abs} = c ~ \frac{ (\lambda_m/\lambda_0)^2 - 1}{ (\lambda_m/\lambda_0)^{2} + 1},
  \label{eq:velocity}
\end{equation}
where $\lambda_0$ is the laboratory wavelength of the ions creating the feature and $\lambda_m$ is the measured wavelength in the rest-frame of the host galaxy (SNe without host galaxy redshifts are thus excluded from velocity studies).

As for pEWs, it is difficult to give a direct physical interpretation
of line velocities for SNe Ia since most spectral features
consist of blends of ions. \refcom{Also, different ions dominate features at different epochs and thus shift the minimum position. Nonetheless we use the same reference wavelength, with the consequence that measurements are not guaranteed to be the velocity of an ion, but are rather a general measurement that can be compared between different SNe. In practice we only study each feature during epochs where no drastic changes happen to the minima shape.} Some of the features lack clear minima, like feature 4, and are thus harder to evaluate. {\sisix} (f7), typical of SNe Ia, most often yield unambiguous measurements \refcom{and is thus usually the best estimate of the ``true'' expansion velocity}.

All spectra are binned using bin widths constant in velocity,
$c\Delta\lambda/\lambda=2000$ km s$^{-1}$. Binned error spectra were
calculated through weighted error averages in each bin.
 
The flux minima for the different features were obtained in the binned
spectra, i.e. without interpolation to a finer wavelength grid and
without fitting a general shape to the region. Although a sub-bin
fitting procedure would, in principle, give a finer determination of
the minimum, this is not viable with the noise level in our sample.

\paragraph{Velocity error estimates}

Velocity measurements are less sensitive to most major systematic
uncertainties, but noisy or heavily contaminated spectra can still
have substantial uncertainties. These were studied using similar Monte
Carlo simulations as for pEWs, see Appendix~\ref{app:comphost} and
\ref{app:filter}. For noisy or contaminated data, the dispersion
is non-negligible, but leads to no major bias, provided host-galaxy
subtractions are performed on host contaminated spectra.

Systematic uncertainties obtained from the simulations are added to each velocity measurement depending on the S/N and the estimated contamination. These are added in quadrature with a 200 km s$^{-1}$ peculiar velocity error.\fixme{THE STATISTICAL ERROR STILL MURKY. IS THIS PEC VEL ERR GOOD?}


\section{{\Dataset}s}
\label{sec:data}

Our sample consists of SNe Ia observed with  the NTT and the NOT as a part of the SDSS-II
Supernova Survey \citep{ostman09}. The set consists of
{\nosniaspectra} SN Ia spectra of {\nosneia} different
objects. SDSS SNe are labeled after their SDSS Supernova ID number, see \citet{ostman09} for the respective IAU names.

\newnew{Observations made at the NTT were performed using the ESO Multi-Mode Instrument (EMMI) and have a wavelength coverage from 3800 to 9200 {\AA}, a wavelength dispersion of 1.74 {\AA} per pixel, and a spatial resolution of 0$\farcs$166 per pixel before binning}. A binning of 2x2 was used.
%
\newnew{The NOT spectra were obtained using the Andalucia Faint Object Spectrograph and Camera (ALFOSC) with grism 4.}
%
NOT spectra have a wavelength range from 3200 to 9100 {\AA}, a wavelength dispersion of 3.0 {\AA} per pixel, and a spatial resolution of 0$\farcs$19 per pixel. See \citet{ostman09} for detailed information.

Lightcurve properties such as stretch, colour and maximum absolute
magnitude, as well as the spectral epochs, were obtained with the SALT
lightcurve fitter \citep{guy05}. The spectral \emph{epoch} is defined 
with respect to the peak of the $B$-band lightcurve.

After applying lightcurve quality cuts, requiring photometric
  observations both prior and post maximum brightness, we are left
  with {\nosniaspectralc} spectra. Out of these, {\nobefcont} spectra
  have both good host-galaxy subtraction and are of sufficient quality for
  spectral features to be identified. Finally, we apply a host-galaxy
  contamination cut of $<60$\% in the $g$-band, motivated by Monte
  Carlo simulations, which leaves us with {\nosniaspecmeas} spectra.\footnote{This limit is somewhat arbitrary: many subtractions of higher contaminated spectra succeed, but the risk of a subtraction significantly failing increases above $60$\% host contamination. See Appendix~\ref{app:comphost}.}
A list of all NTT/NOT spectra used in this analysis is given in Table~\ref{tab:sdsspec}.

The SDSS NTT/NOT spectra are compared to a low-redshift reference
SN sample which consists of three subsets, data from the Harvard-Smithsonian Center for Astrophysics (CfA), the Supernova Cosmology Project (SCP99)
and the Online Supernova Spectrum Archive (SUSPECT). Since the NTT/NOT spectra cover the spectral epochs between
{\sdssminepoch} days and +{\sdssmaxepoch}, we have only studied reference spectra up to epoch
30.

The CfA sample consists of {\nospeccfa} spectra of {\noobjcfa} SNe Ia
from \citet{2008AJ....135.1598M}.
The SCP99 {\dataset} contains {\nospecscp} spectra of {\noobjscp} SNe observed by the Supernova
Cosmology Project
in 1999 that were studied by \citet{2007AA...470..411G}.
The SUSPECT {\dataset} collects publicly available SN spectra\footnote{\texttt{http://bruford.nhn.ou.edu/suspect/.}}, we use {\nospecsuspect} spectra of {\noobjsuspect} Type Ia Sne. 
A list of all spectra in our reference sample is given in
Table~\ref{tab:snspectra}. The table also contains the source of the
lightcurve parameters as well as the original spectroscopic reference for SUSPECT spectra.
These lightcurve parameters are lacking for some SNe and these objects are thus 
excluded from analysis where such information is required.

\begin{figure*}[htb]
  \centering
  \includegraphics[angle=0,width=\halfcol\columnwidth]{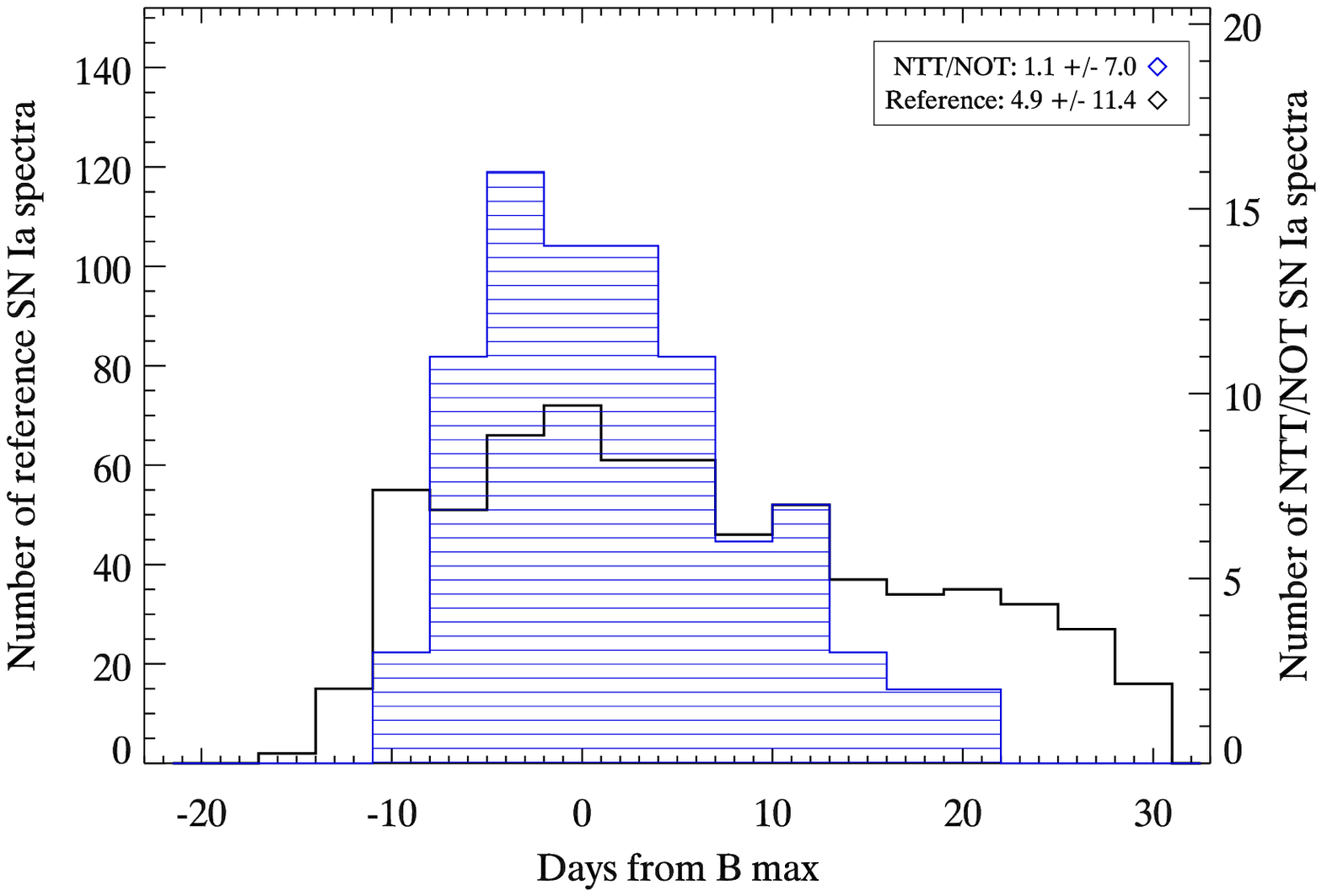}  
  \includegraphics[angle=0,width=\halfcol\columnwidth]{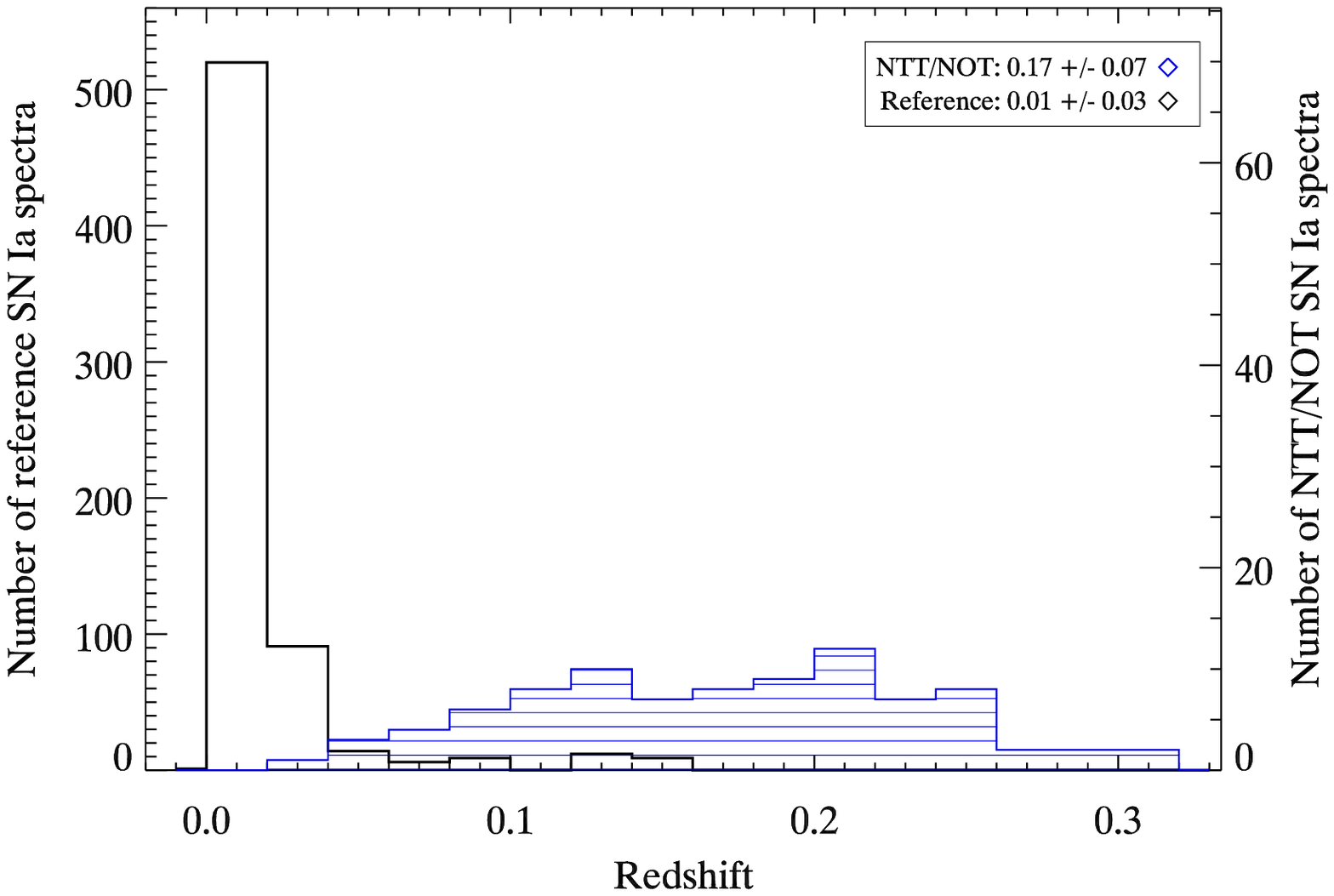}  
  \includegraphics[angle=0,width=\halfcol\columnwidth]{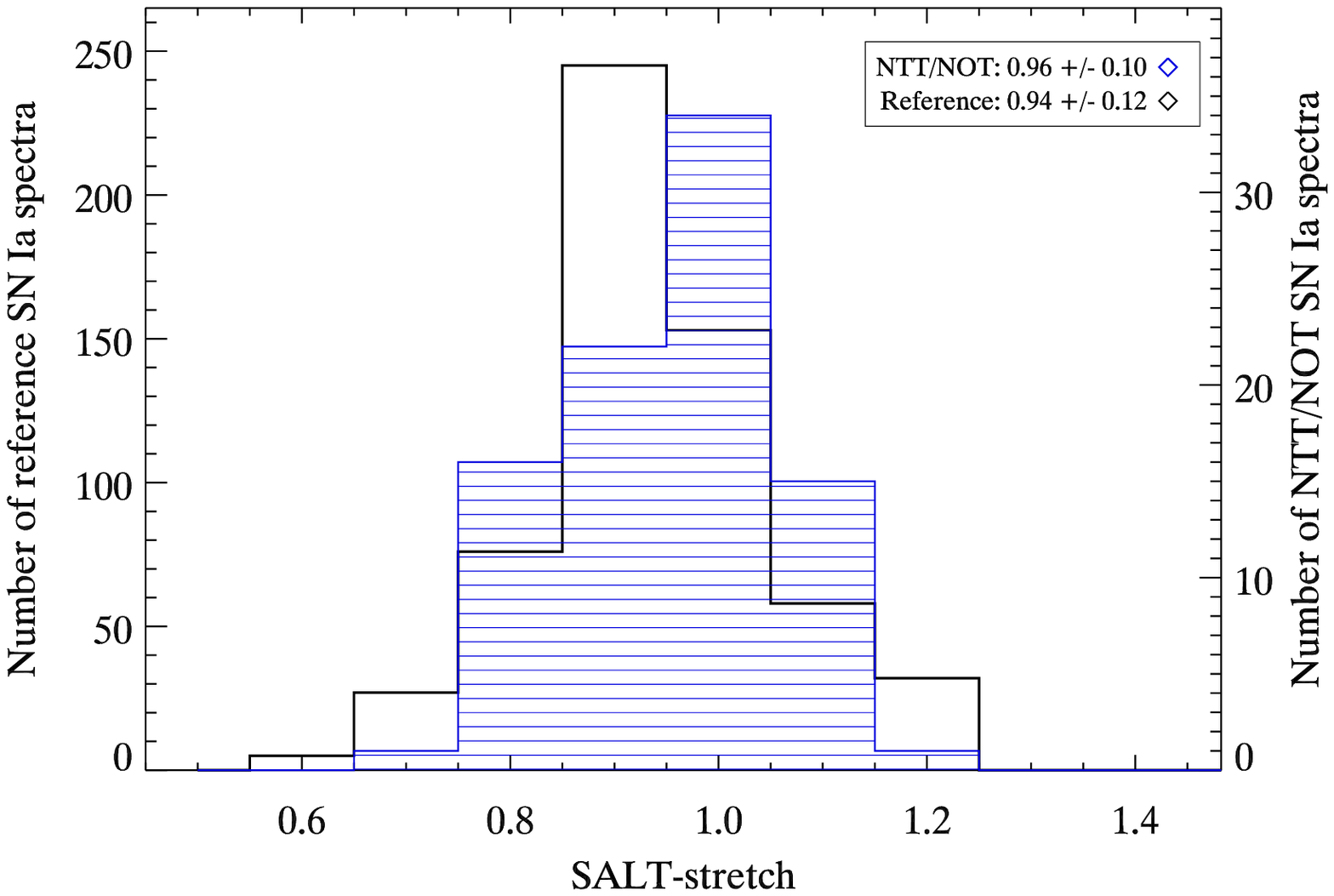}  
  \includegraphics[angle=0,width=\halfcol\columnwidth]{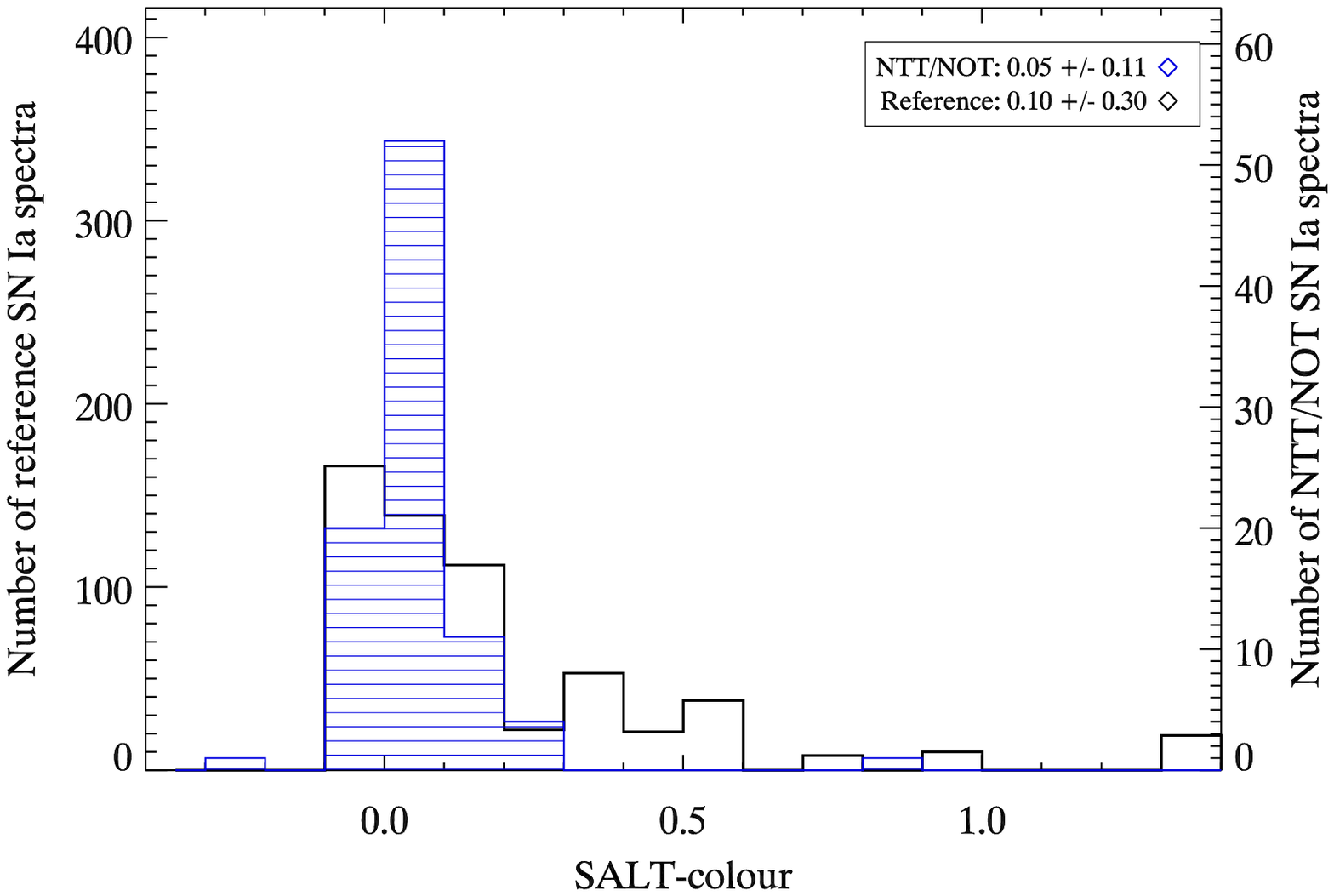}  
  \caption{Distribution of epoch, redshift, SALT stretch factor and SALT
    colour for the SN spectra used in our study. The epoch is defined here
    as the number of days in rest frame from $B$ band maximum
    brightness. The white histogram is for the reference sample while
    the striped histogram is for the NTT/NOT spectra used here. Legends show mean and Gaussian $1\sigma$ levels for the subsets.}
  \label{fig:histograms}
\end{figure*}

We present in Figure~\ref{fig:histograms} the distribution of epoch, redshift, SALT stretch and SALT colour for the NTT/NOT sample together
with the reference sample used in this paper.

The NTT/NOT sample has significantly larger redshifts than the
comparison sample, a median value of {$\bar z_{SDSS}=$ \distmediansdssredshifts}
compared to {$\bar z_{ref}=$\distmedianrefredshifts}. The 
spectral epoch distribution is also somewhat
different: the NTT/NOT spectra are more centred around the lightcurve
peak, while the comparison sample includes earlier and later epochs.
This is a natural effect arising from the differing magnitude limits of the SN searches.
\new{The distribution of SALT stretch is fairly \modifierat{constant between the samples} with a
Kolmogorov-Smirnov (KS) probability of {\distksallstretch}, thus showing that we can not reject the assumption that they belong to the same distribution.} The median value for the NTT/NOT sample and the reference sample are {\distmediansdssstretch} and {\distmedianrefstretch}, respectively.
However, the SALT-c (colour) distributions are significantly
different as can been see by a visual inspection, with a tail of red
colour SN in the reference sample. This is not surprising since local SNe Ia
can be detected even with a few magnitudes of extinction. The median value for the NTT/NOT sample and the reference sample are {\distmediansdsscolour} and {\distmedianrefcolour}, respectively.

While the SALT lightcurve fit output is used to make the initial
analysis, we have also obtained MLCS2k2 \citep{2007ApJ...659..122J}
lightcurve fits. MLCS output consist of a lightcurve shape-dependent
parameter $\Delta$ and the $V$-band extinction, $A_V$. The $A_V$ parameterisation assumes that any reddening not corrected for by the $\Delta$ parameter can be
described by a Milky Way-like extinction law. These fits
are used to validate
results found using SALT and study lightcurve fit dependant
effects. We use the MLCS lightcurve fits of all reference SNe
contained in the \citet{2009ApJ...700.1097H} \dataset~together with
MLCS fits of the NTT/NOT SDSS SNe using the SNANA fit package \citep{2009PASP..121.1028K}
and employing the same quality cuts and settings as in
\citet{2009ApJS..185...32K}. 
The only exception to this procedure was that we used $R_V = 1.7$, this change was made to comply with \citet{2009ApJ...700.1097H} but has very small effect on our analysis.

\refcom{In \citet{ostman09} three potential peculiar SNe Ia from the NTT/NOT \dataset~are presented: two of SN1991T-type and one SN2002cx like. In this paper we can not confirm any additional 'SN1991T', 'SN1991bg' or 'SN2002cx' SNe. \citet{2010arXiv1006.4612L} find, in their luminosity limited sample, that 77\% of all Type Ia SNe are normal, 18\% SN1991T-like, 4\% SN1991bg-like and 1\% SN2002cx-like. While 'SN1991bg' and 'SN2002cx' either would have escaped detection entirely, due to their low luminosity, or have been singled out based on lightcurve properties, it is likely that a fraction of the SNe used here are really of the 'SN1991T'-type. That these are not identified can be explained by two effects: (i) The S/N is often good enough to classify a SN as Ia, but not high enough for a strict subclassification \citep{ostman09} (ii) 'SN1991T' like SNe are often mistaken for normal if no early spectra exist \citep[an 'age bias',][]{2010arXiv1006.4612L}. We will later, in Section~\ref{sec:dis}, show that a fraction of $\sim20$\% of the SDSS SNe have shallower features and would represent 'SN1991T' like SNe at later epochs well.}

The NTT/NOT spectra have well documented uncertainties \citep{ostman09}. However, most of the spectra in the reference sample lack error estimates. For such
cases, a constant flux error of 5\% of the average spectral flux was
used to compute the uncertainties in spectral indicators.

\subsection{Host galaxy subtraction}
\label{sec:hostsub}

A large fraction of the SDSS NTT/NOT SNe have significant host galaxy
contamination which could affect spectral indicators. In particular, 
comparisons with virtually host-free spectra for local
reference SNe could lead to systematic differences that may 
be confused with evolution with redshift.
Great care was taken to remove the host galaxy light efficiently, given 
the data available, as well as understanding remaining errors.

\refcom{A large fraction of the NTT/NOT SNe were not observed in paralactic angle, and are thus affected by differential slit losses. In \citet{ostman09} we calculate slit loss for all spectra, as functions of estimated seeing conditions and centring wavelength. Because of the uncertainties associated, mainly regarding centring, we choose not to apply calculated slit losses directly to the spectra, but rather to incorporate corrections into the host subtraction pipeline, where we also account for reddening. Considering the typical centring wavelength (6500 {\AA}) and the limited wavelength range studied (4000-6500 {\AA}), slit loss and reddening exhibit similar differential attenuation and can be fitted together.}

 A detailed description of the host galaxy subtraction can be found in
\citet{ostman09}.
In brief, the galaxy SED which is subtracted is estimated by minimising the
difference between the observed spectrum and a combination of a SN
template and a set of galaxy eigenspectra. 
To the SN template, a second degree polynomial is multiplied to account for reddening (e.g. due to host galaxy dust extinction) and differential slit loss effects.
The minimisation can be described with the formula
\begin{equation}
f_{\rm{fit}}(\lambda) = a_0 s(\lambda) \cdot f_{\rm{SN}}(\lambda) + \sum_{i=1}^3 a_i g_{i}(\lambda),
\end{equation} 
where $f_{\rm{SN}}$ is the SN template, $g_i$ the galaxy eigenspectra, $s$ the second degree polynomial and $a_i$ weights which are fitted in the subtraction.
The SN templates in the fit were the Hsiao templates \citep{2007ApJ...663.1187H} within epochs
$\pm$5 days from the SN spectral epoch as obtained from the
lightcurves. Models of the peculiar SNe 1991bg and 1991T \citep{2002PASP..114..803N} within the same epoch interval were also included in the fit, but no new clear cases of these subtypes were found.
Three galaxy eigenspectra from SDSS \citep{2004AJ....128..585Y} were
used for the galaxy SED in the fit.
The second degree polynomial $s(\lambda)$ was locked to have $s\equiv1$ at $\lambda=6600$
{\AA} and $s<1$ for all other wavelengths (thus only having one degree of freedom). The wavelength of the $s$
function peak is chosen to match the wavelength where most
spectra were centred on the slit. It should be noted
that the slit loss function is asymmetric around the centring wavelength, but the fit during
subtraction is only made between 4000 and 6000 {\AA} and thus
the behaviour of $s$ at longer wavelengths will not affect the fit.
The polynomial is only multiplied with the SN SED, and not
with the galaxy. This was done since galaxies, not being point
sources, are significantly less affected by slit loss.

Several modified versions of this host subtraction were tried: Other sets of eigencomponent spectra, more eigencomponent spectra, free polynomial (instead of restricted) and slit loss applied to the host galaxy. For individual spectra one of these alternative methods might achieve better fits, but globally they were all either less stable or as good but with significantly more parameters to fit.

\begin{figure*}[htb]
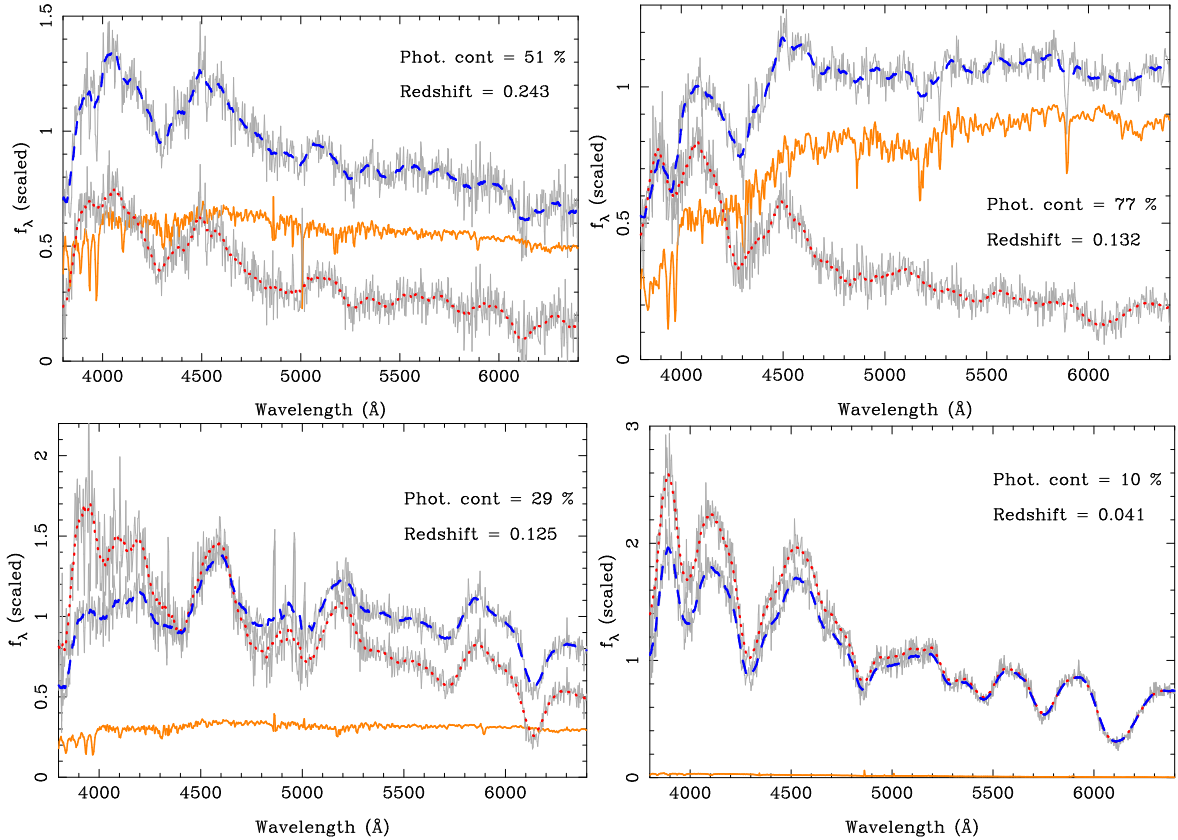

	\centering
        \includegraphics[width=\mycol ,angle=-90]{15705fig7.ps}
        \includegraphics[width=\mycol ,angle=-90]{15705fig8.ps}
        \includegraphics[width=\mycol ,angle=-90]{15705fig9.ps}
        \includegraphics[width=\mycol ,angle=-90]{15705fig10.ps}
        
        \caption{ Display of sample host galaxy subtractions of SDSS
        SN16637, 12907, 17886 and 13894 (clockwise from top left, \refcom{restframe epochs -1, 0, -4 and 9}). All
        flux values (y-axis) have been scaled. Wavelength (x-axis) is rest-frame wavelength. \new{The grey line shows the unsmoothed original spectrum, while the dashed blue line shows the smoothed version.}
        Dotted red line show final subtracted \emph{slit
        loss/reddening corrected} spectrum and orange solid line show
        best fit galaxy. Note that because of the slit loss/reddening correction the galaxy and subtracted spectrum do not sum to the raw
        spectrum. \new{The galaxy contamination estimated from photometry (g-band) and redshift have been written in each panel.}
}
\label{fig:multfit}
\end{figure*}

\begin{figure*}[htb]
	\centering
        \includegraphics[width=\mycol,angle=-90]{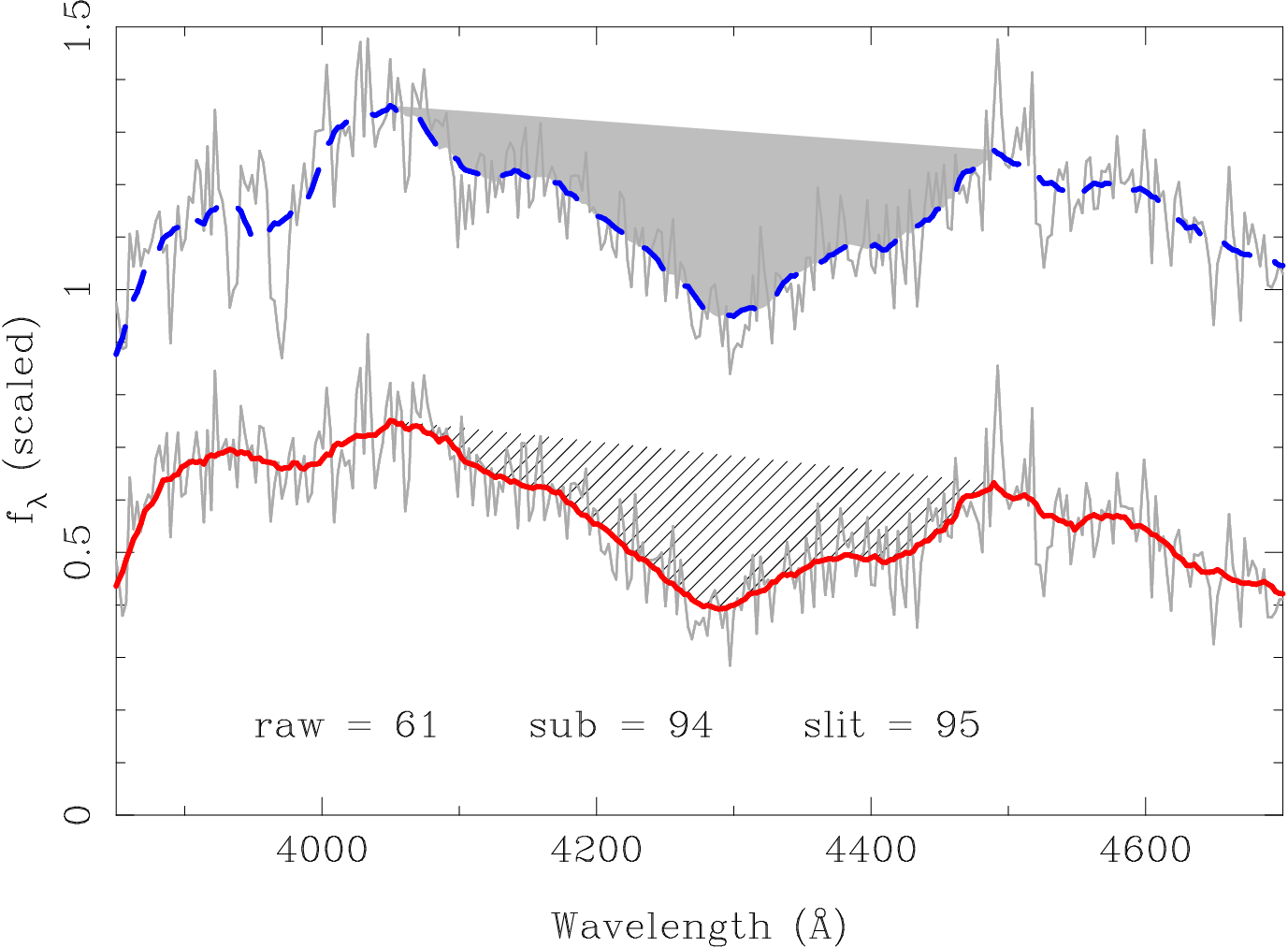}
        \includegraphics[width=\mycol,angle=-90]{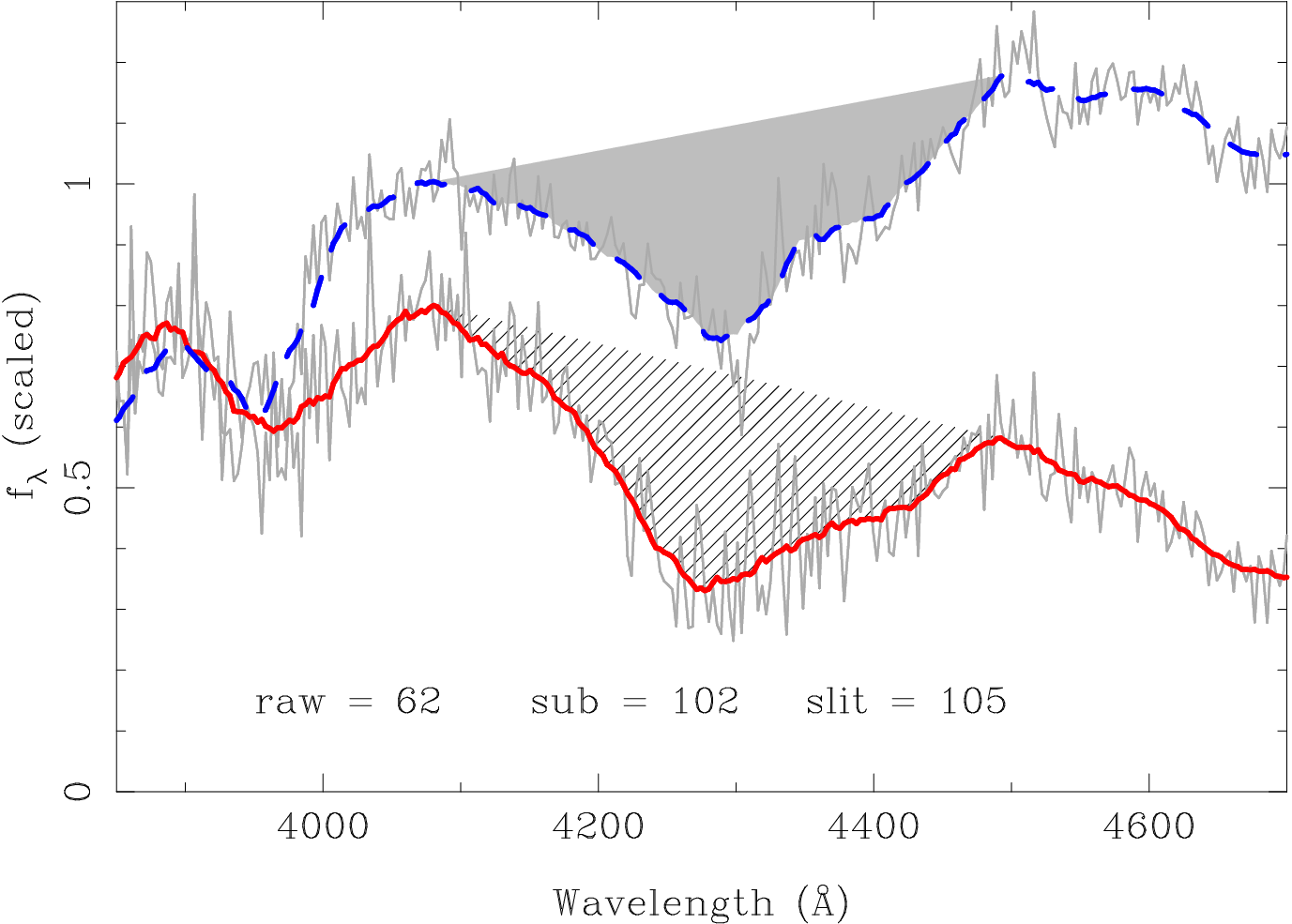}
        \includegraphics[width=\mycol,angle=-90]{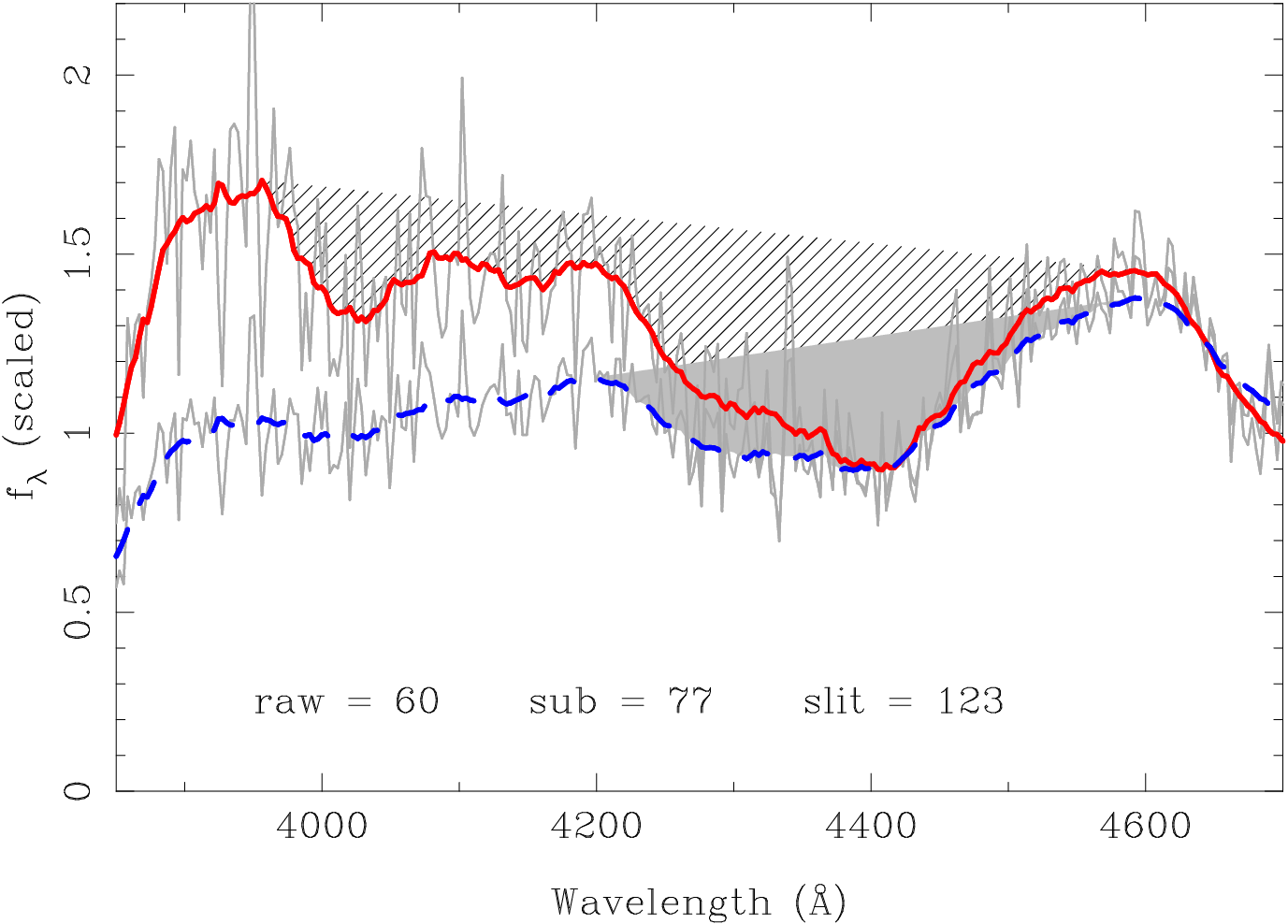}
        \includegraphics[width=\mycol,angle=-90]{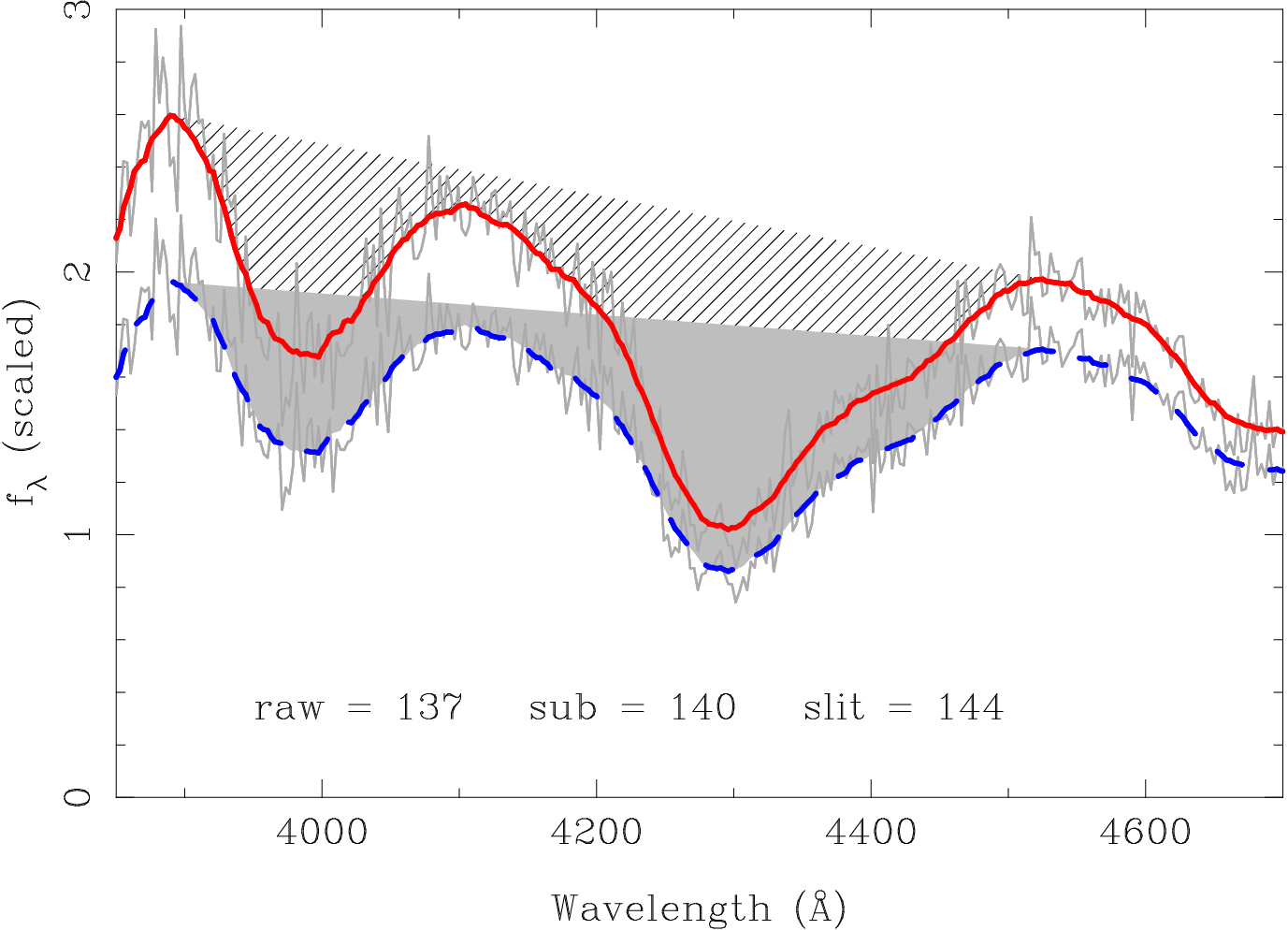}
        \caption{ Expanded view of the feature 3 region for the same spectra
        displayed in Figure~\ref{fig:multfit}, showing the effects of
        host subtraction and slit loss/reddening correction on
        pEW. Wavelength (x-axis) is rest-frame wavelength. The blue dashed line is the raw spectrum, the red solid line the
        final subtracted and corrected spectrum. The marked regions
        show where pEW is calculated, the calculated pEWs ( in {\AA}) are written in
        panels for raw spectra (``raw''), subtracted spectra (``sub'') as well as
        subtracted \emph{and} corrected spectra (``slit''). These sample spectra show
        host subtraction change pEWs both through additive offsets and
        changed feature limits.}
\label{fig:subsample}
\end{figure*}

Figure~\ref{fig:multfit} shows subtraction samples of SNe with contamination ranging from very high ($77$\%) to very low but including reddening/slit loss.
For moderate contamination, 10 to 60\% galaxy light in the $g$-band,
the multiplicative host-galaxy subtraction works well. \emph{Within these
limits the best fit galaxy SED is subtracted from the observed
spectrum before any measurements are done.} An error associated with host contamination uncertainties was calculated for each indicator
measurement. In Figure~\ref{fig:subsample} we show examples of how host subtraction affects pEW measurements.
For spectra with very low contamination, no subtraction is done and the error estimated for 10\% unsubtracted light added.
Spectra with very high
contamination are excluded from the analysis.

In Appendix~\ref{app:comphost} we describe 
simulations designed
to estimate uncertainties and evolution detection limits. We have performed several tests in order to determine the
stability of our results. These include comparisons with synthetic spectra
and comparisons between different host galaxy subtraction methods.

The low-z reference sample spectra were \emph{not} host subtracted. These are sufficiently local to allow subtraction of most host galaxy contamination during data reduction. Visually, they do not appear to contain significant host galaxy light.

\section{Results: Comparing the reference and NTT/NOT samples}
\label{sec:results}

\modifierat{After host subtraction all spectra are processed through the automated indicator measurement pipeline.}
The error bars of the measurements are \modifierat{symmetric} geometric sums of the statistical uncertainty of the measured indicator and the noise-filtering and host-galaxy subtraction systematic errors.

All spectral epochs used are rest frame epochs and are thus corrected for time dilation.

\subsection{The reference sample}

\begin{figure}
  \centering 
  \includegraphics[angle=0,width=\columnwidth]{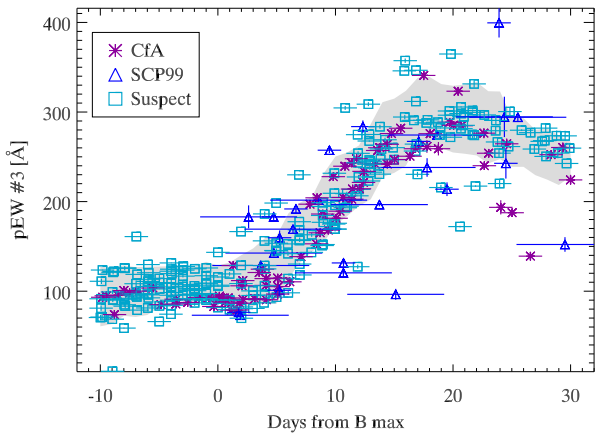}
  \caption{pEW values for the reference data for f3 vs epoch. The
  shaded region is the same one sigma contour as shown in the left
  panel of Figure~\ref{fig:pew_evo_vsepoch}. The different symbols
  show the measurements that were used to construct the grey region,
  with different symbols denoting the different subsets of the
  reference sample.}
  \label{fig:pew_bandsample}
\end{figure}

The reference set measurements were
combined into  \refcom{1-$\sigma$} contours to facilitate statistical comparisons. 
The contour is calculated for each day as the weighted mean and
uncertainty of the indicator (pEW/velocity) for $\pm$3 days. The broad epoch
interval is used to make a smooth curve (stable with respect to
outliers). 
As a justification for the definition of the band, the
measurements underlying the band for pEW for feature 3 are shown in
Figure~\ref{fig:pew_bandsample}.  

\begin{figure}
  \centering 
  \includegraphics[angle=0,width=\columnwidth]{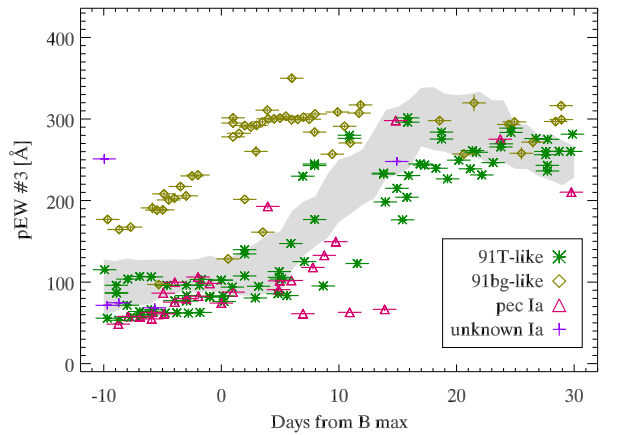}
  \caption{\modifierat{Peculiar SNe compared with normal.} The shaded region shows the one sigma contour constructed from the normal SNe Ia in the reference sample \emph{and} the NTT/NOT spectra. The symbols show the measured pEWs for the SN 1991T-like, SN 1991bg-like and peculiar SNe Ia in the reference sample.}
  \label{fig:pew3pecs}
\end{figure}

We confirm the displacement of the unusal SNe Ia compared to the overall trend, as shown in Figure~\ref{fig:pew3pecs} for the reference sample. Peculiar SNe are not included in the reference sample, but it is in practise impossible to make a strict definition regarding which SNe should be considered as ``normal''. The number of observed spectra per SN also varies, thus giving artificially high weight to certain objects.\emph{We thus do not expect the reference sample to completely match the NTT/NOT SNe detected in a rolling SN survey.} The fraction of SNe of different SN Ia subtypes are different, as well as the distribution of lightcurve color. We will return to possible consequences of this when discussing evolution in Section~\ref{sec:dis}.


\subsection{Pseudo-equivalent widths}
\label{sec:pewtwosamples}

\begin{figure*}
  \centering
  \includegraphics[angle=0,width=\halfcol\columnwidth]{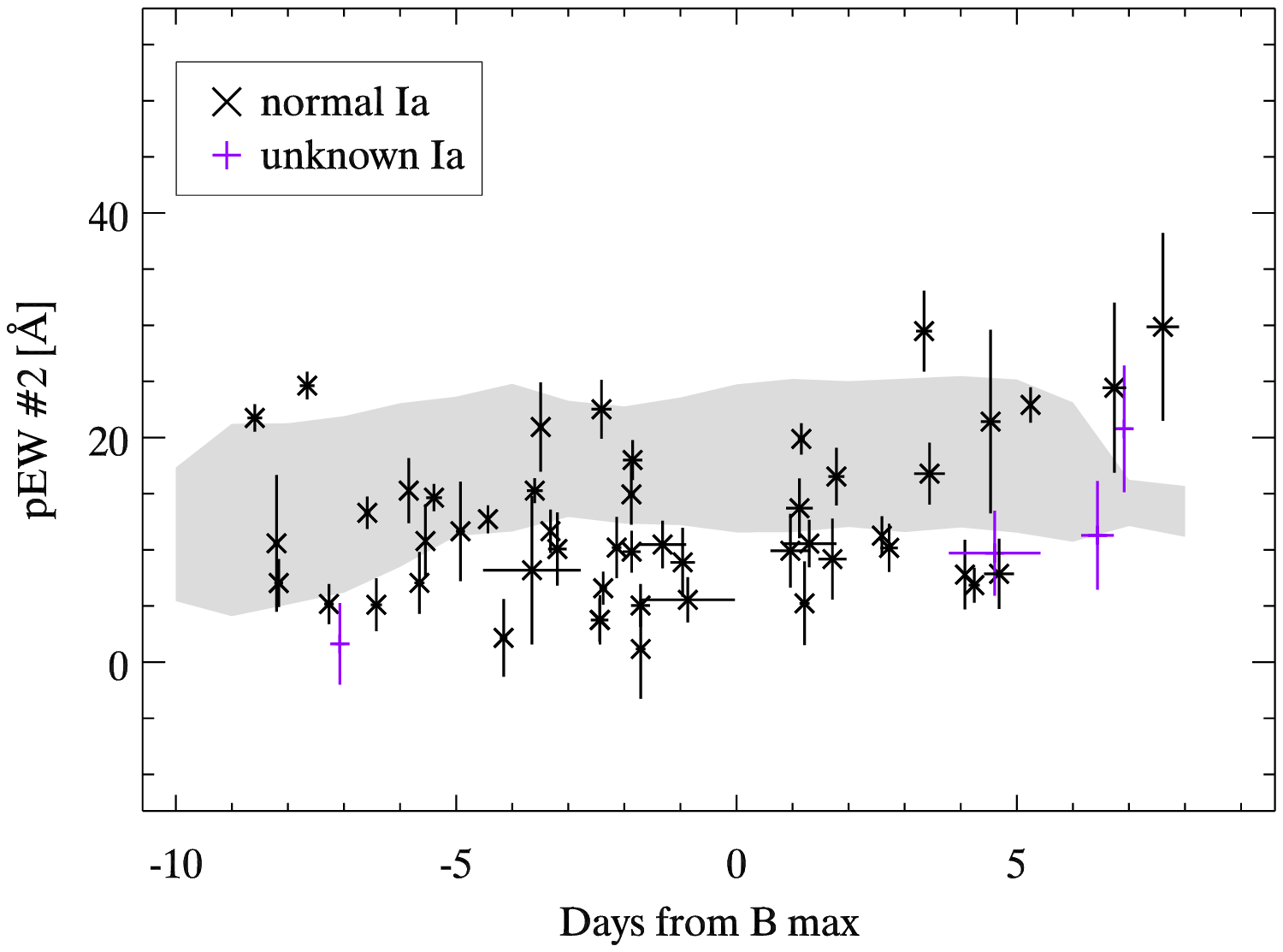}
  \includegraphics[angle=0,width=\halfcol\columnwidth]{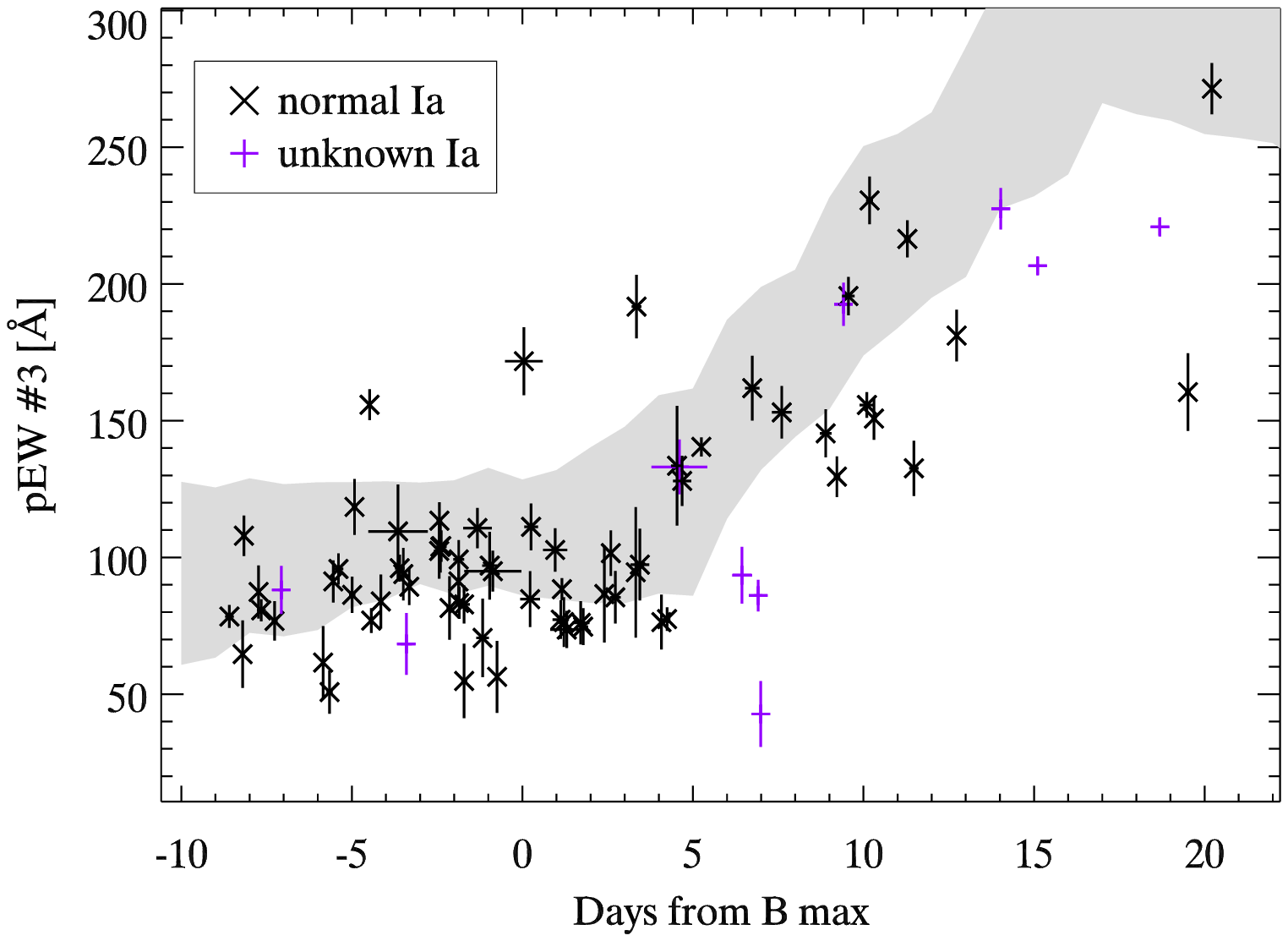}  
  \includegraphics[angle=0,width=\halfcol\columnwidth]{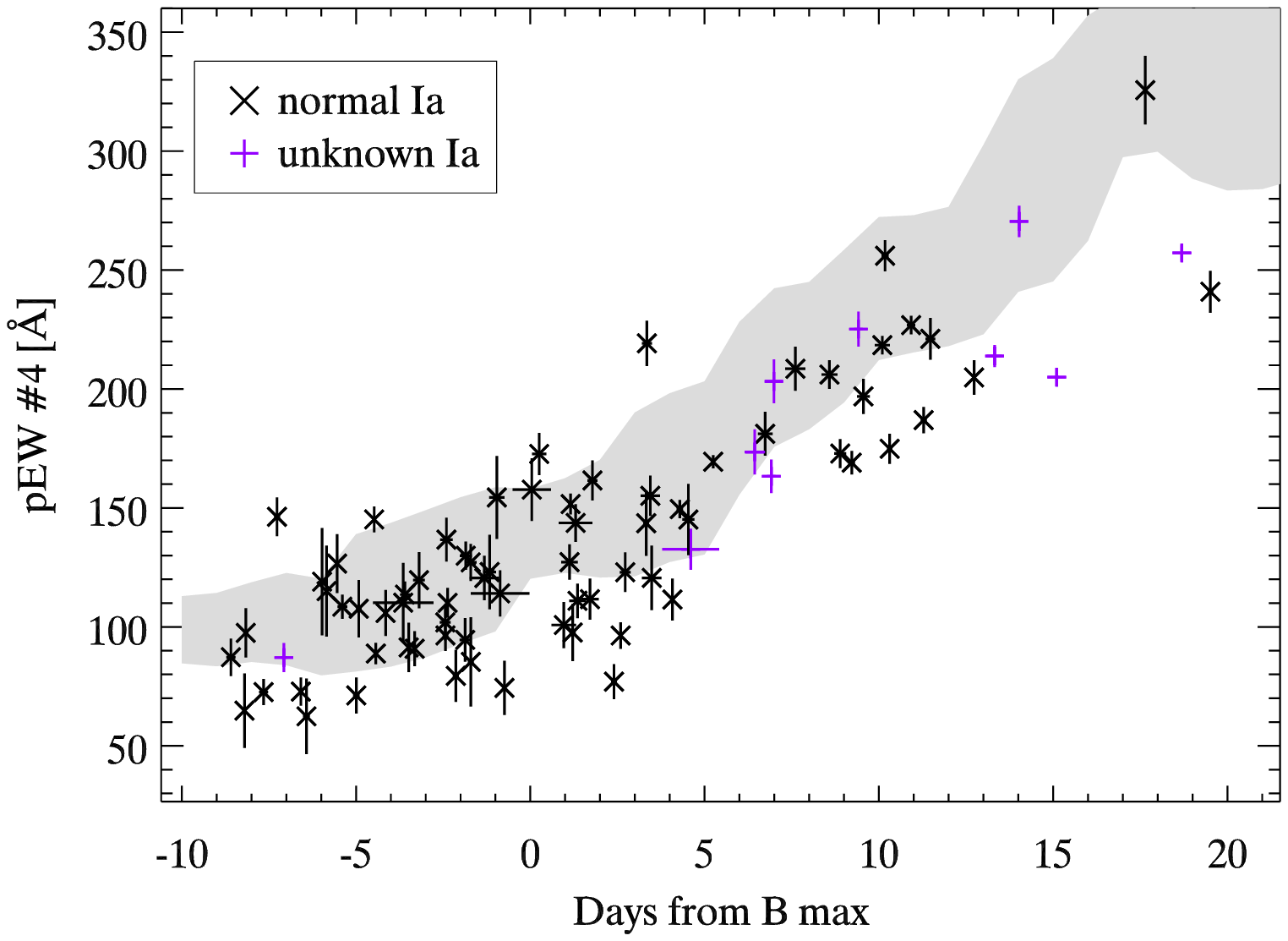}    
  \includegraphics[angle=0,width=\halfcol\columnwidth]{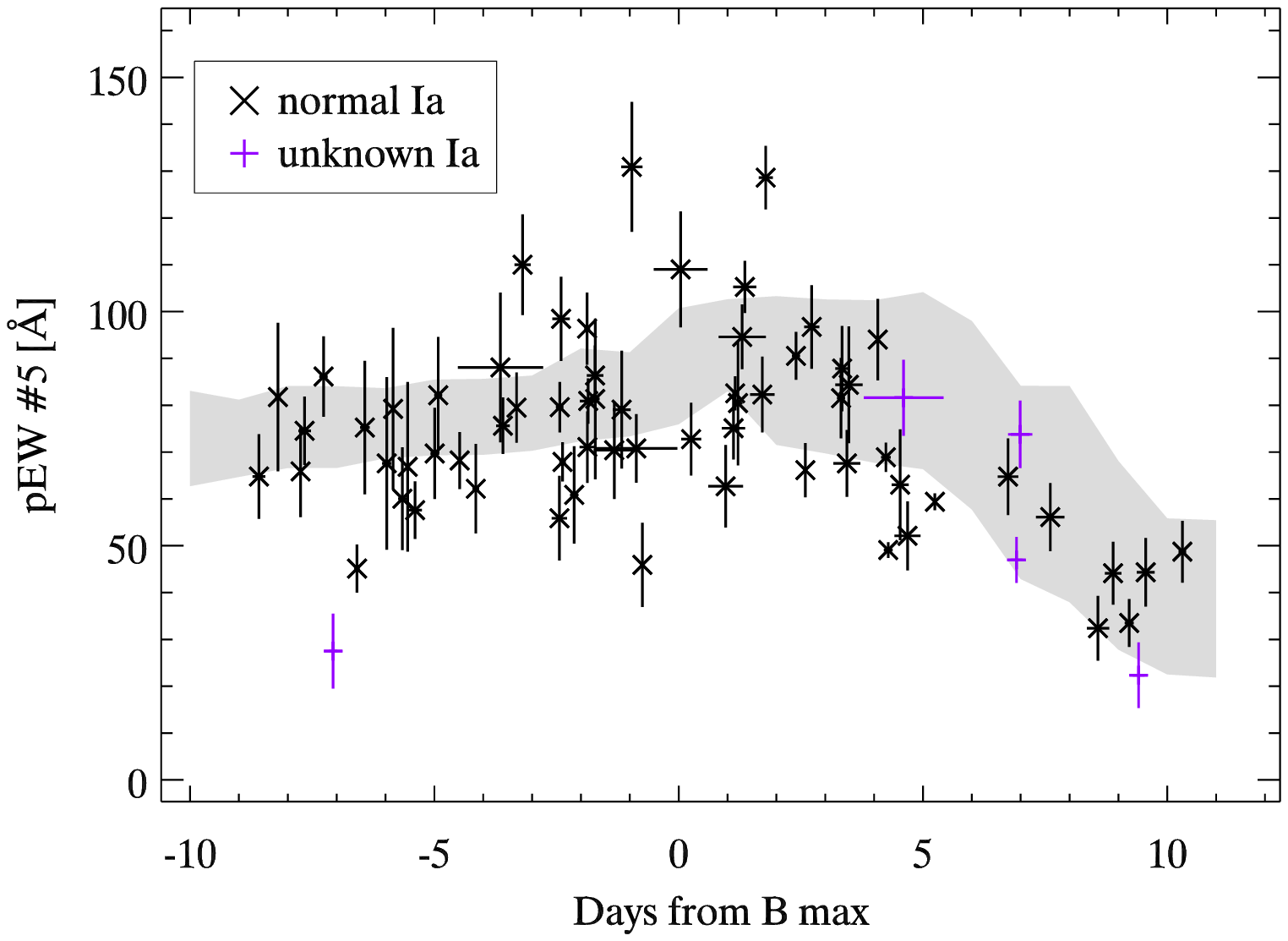}
  \includegraphics[angle=0,width=\halfcol\columnwidth]{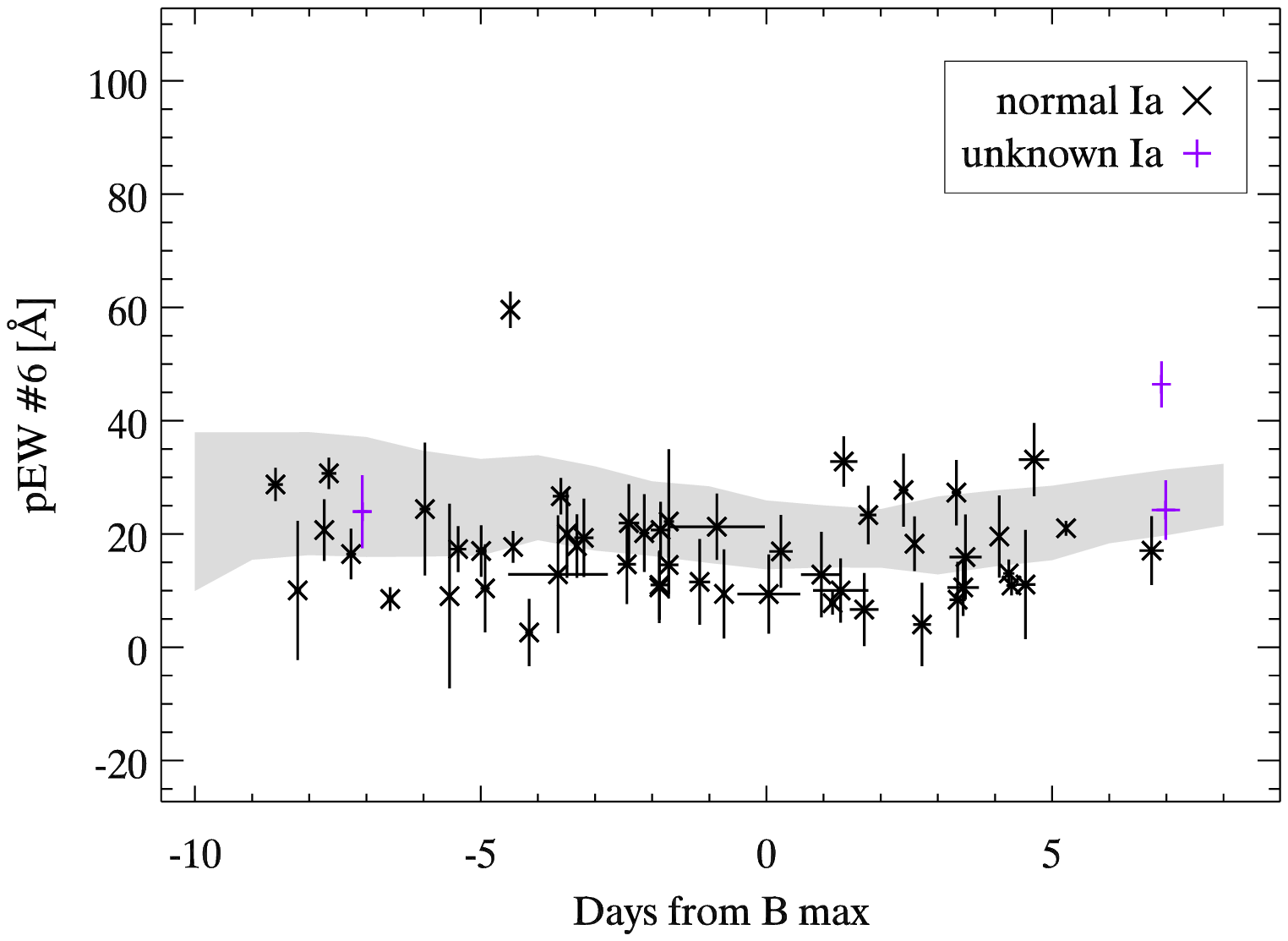}
  \includegraphics[angle=0,width=\halfcol\columnwidth]{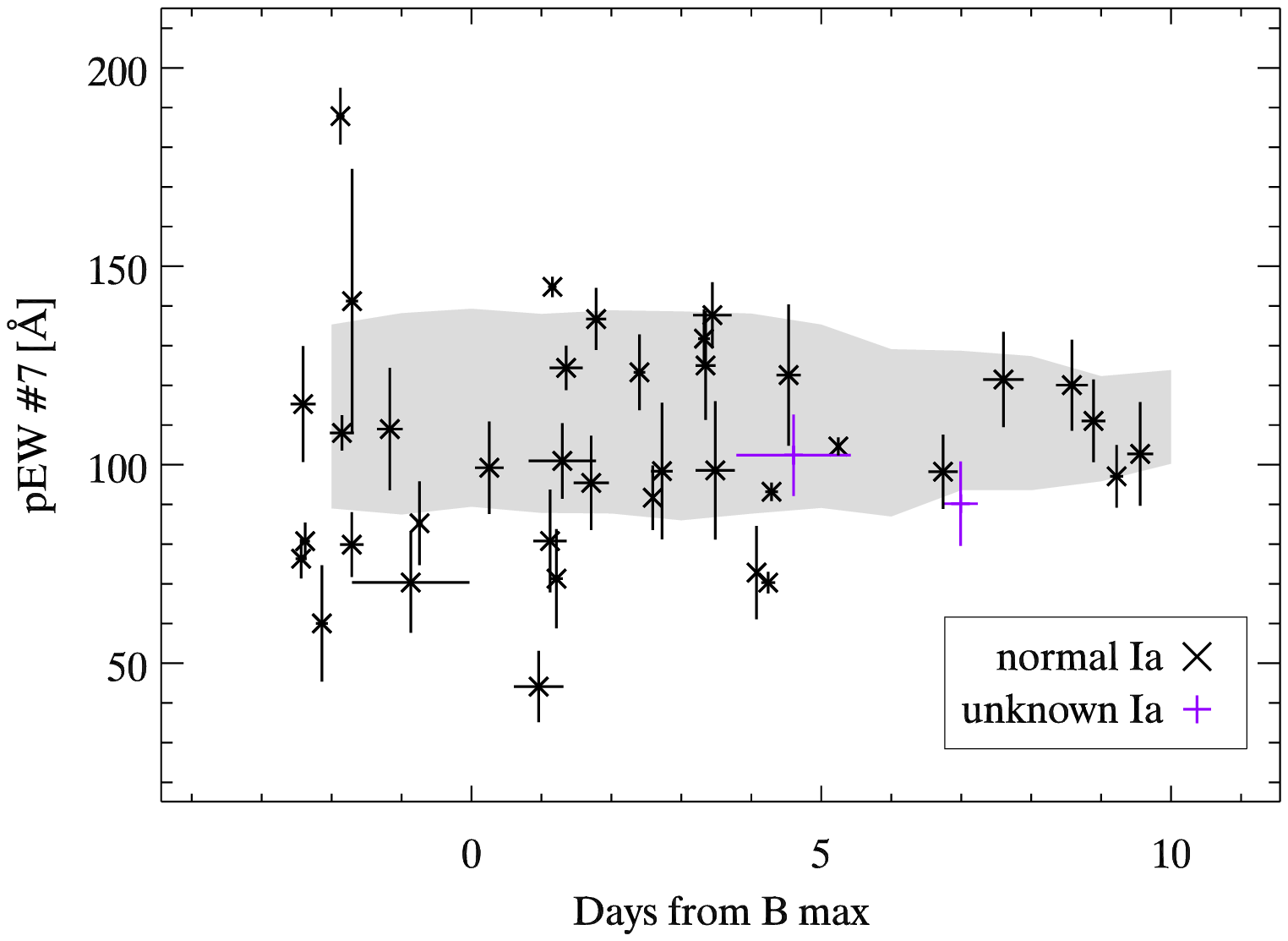}
  \caption{Study of pseudo-equivalent widths for feature 2 to 7 vs epoch. The shaded band shows the one sigma contour for
  the normal SNe Ia in the reference sample. The points
  show the measurements on the NTT/NOT spectra, where the error bars
  include both statistical and systematic errors. The different
  symbols show two categories: spectra identified as of the normal SNe Ia subtype by SNID \citep[SuperNova IDentification;][]{2007ApJ...666.1024B} or spectra identified as SNe Ia but of unknown subtype.}
  \label{fig:pew_evo_vsepoch}
\end{figure*}

Figure~\ref{fig:pew_evo_vsepoch} shows the measured pEW values of
features 2 to 7 for the NTT/NOT spectra as a function
of epoch.  These are compared to the corresponding 1 $\sigma$ contour for the
normal SNe Ia in the reference sample. Features are only measured for epochs when they can be clearly defined.

\begin{figure*}[htb]
  \centering
  \includegraphics[angle=0,width=\halfcol\columnwidth]{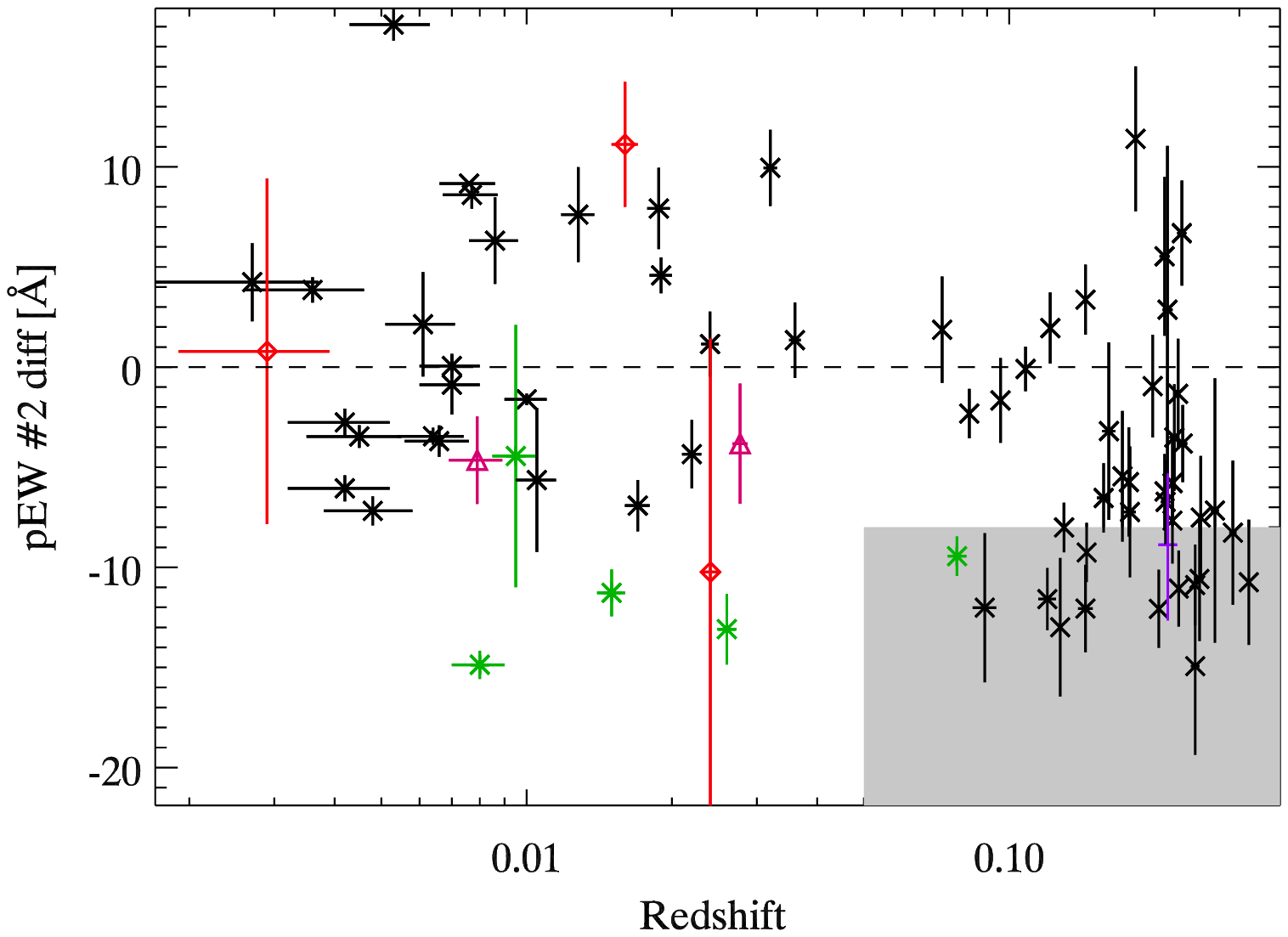}
  \includegraphics[angle=0,width=\halfcol\columnwidth]{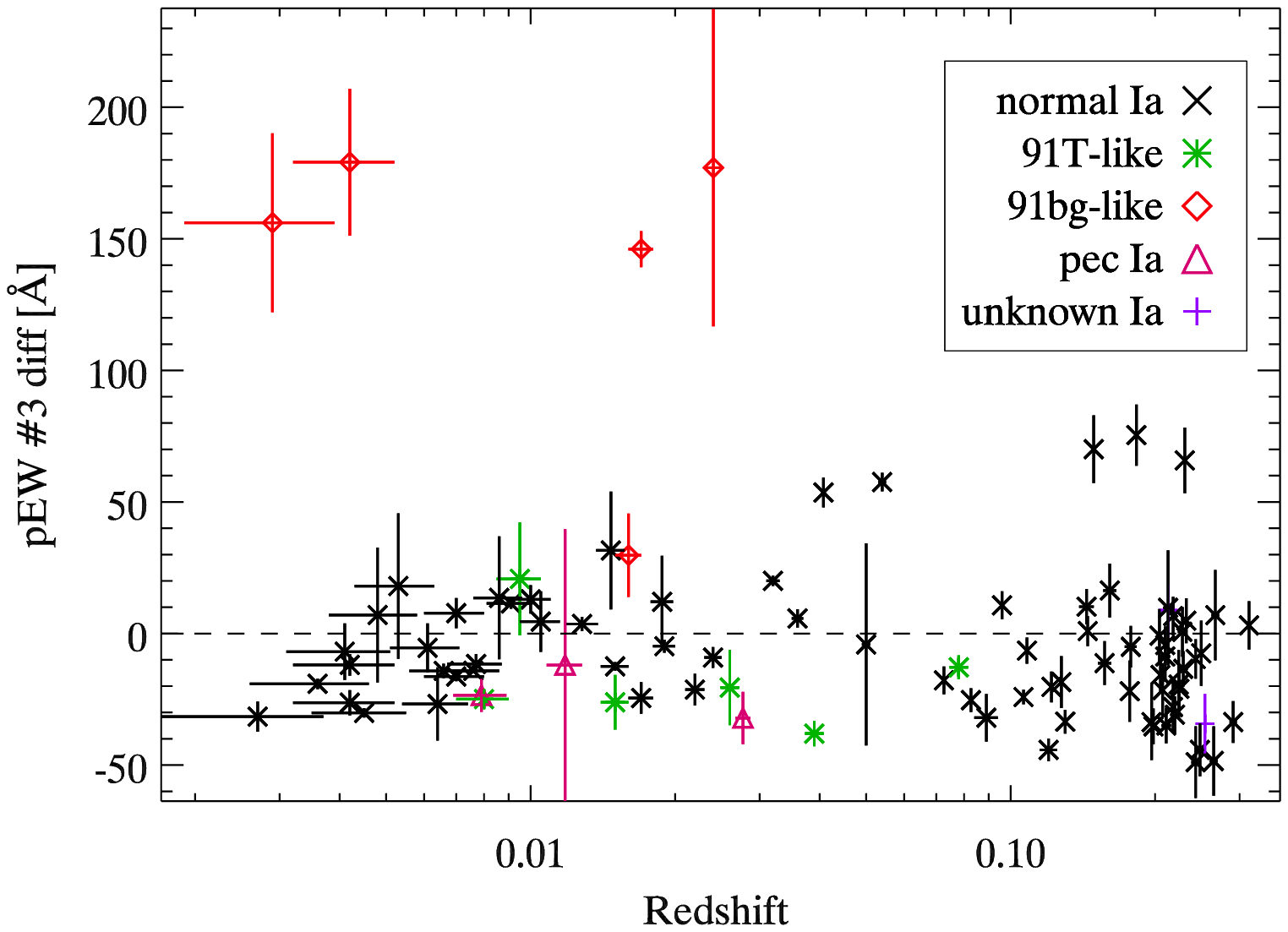}
  \includegraphics[angle=0,width=\halfcol\columnwidth]{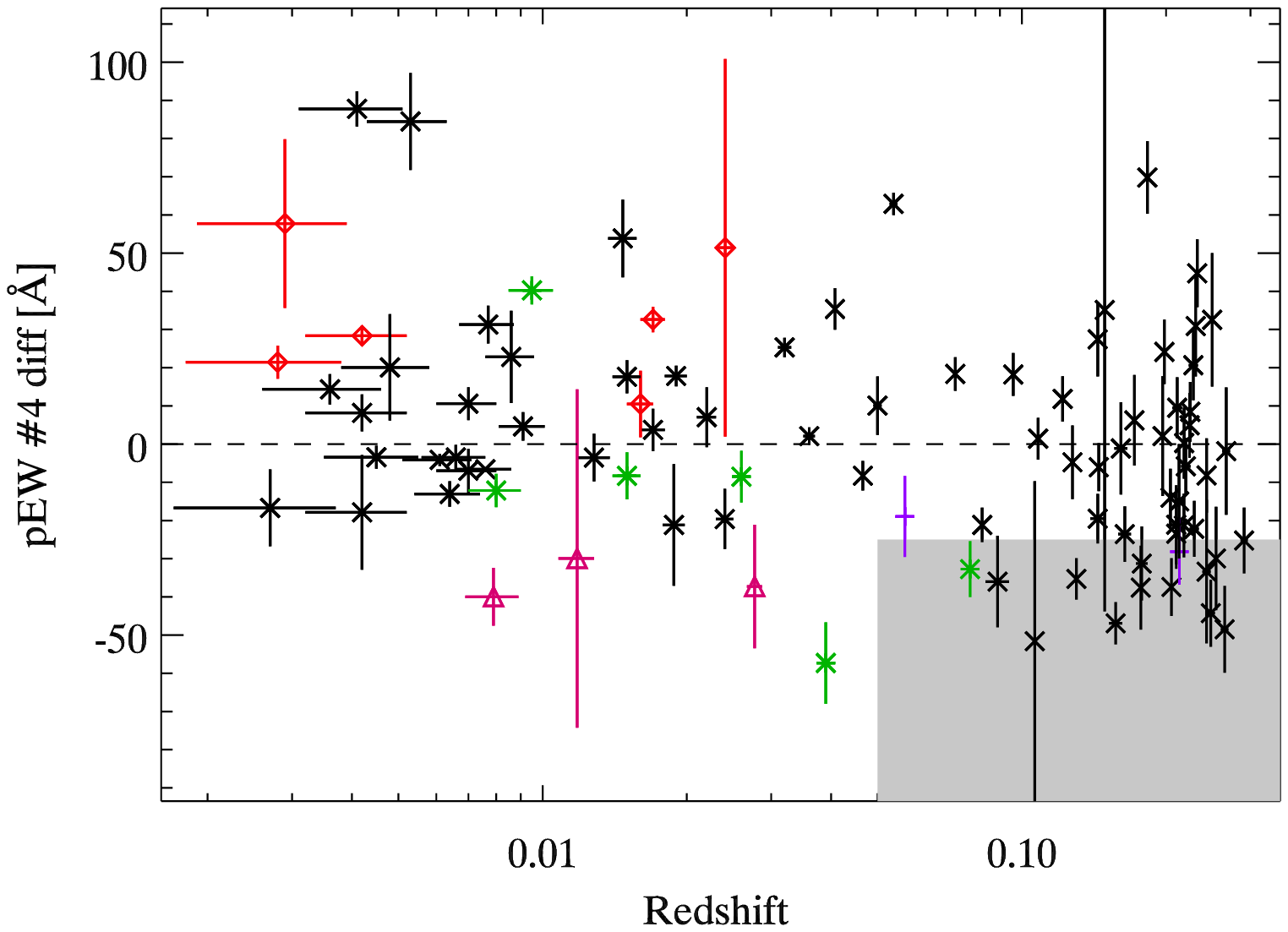}
  \includegraphics[angle=0,width=\halfcol\columnwidth]{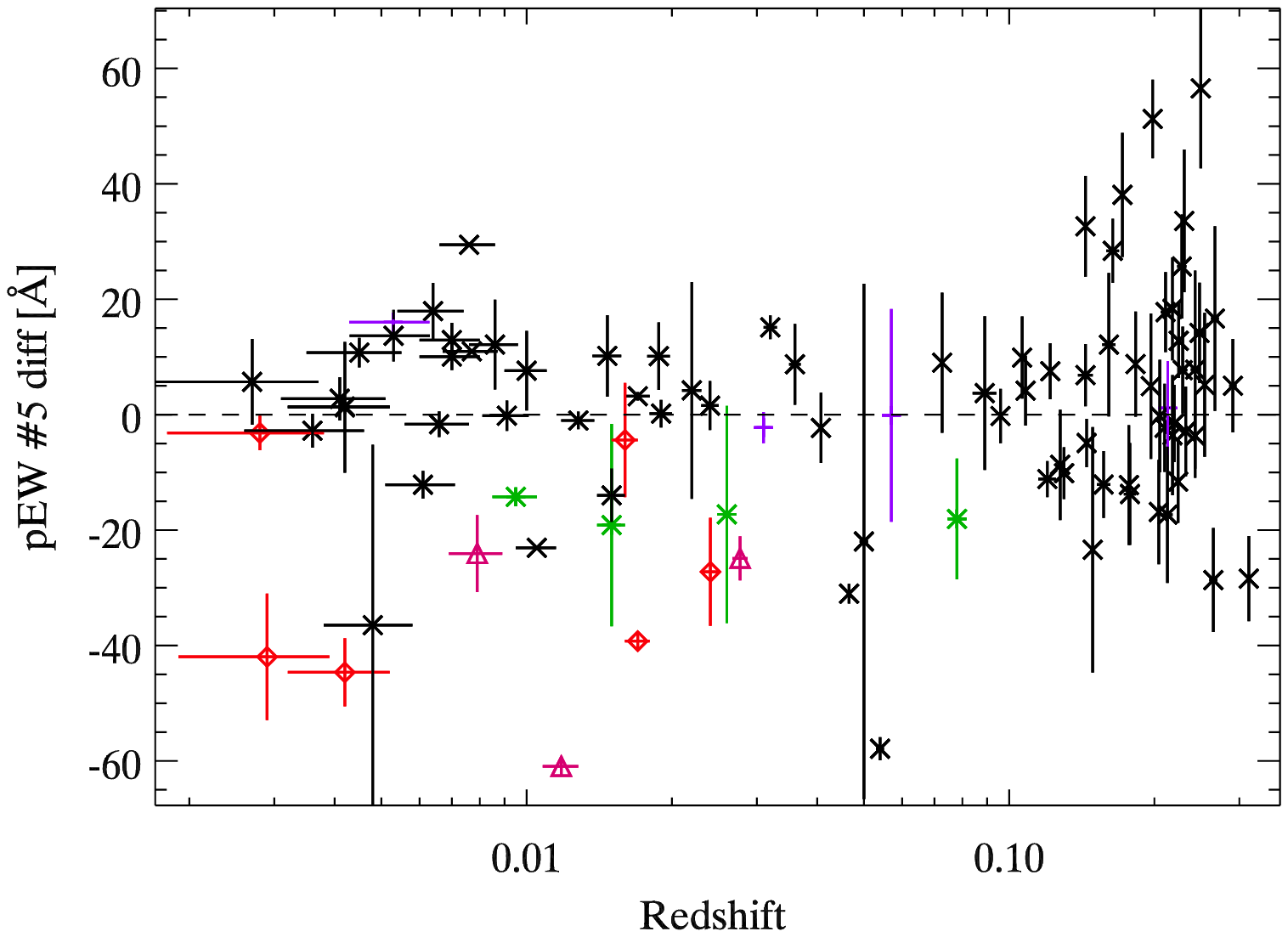}
  \includegraphics[angle=0,width=\halfcol\columnwidth]{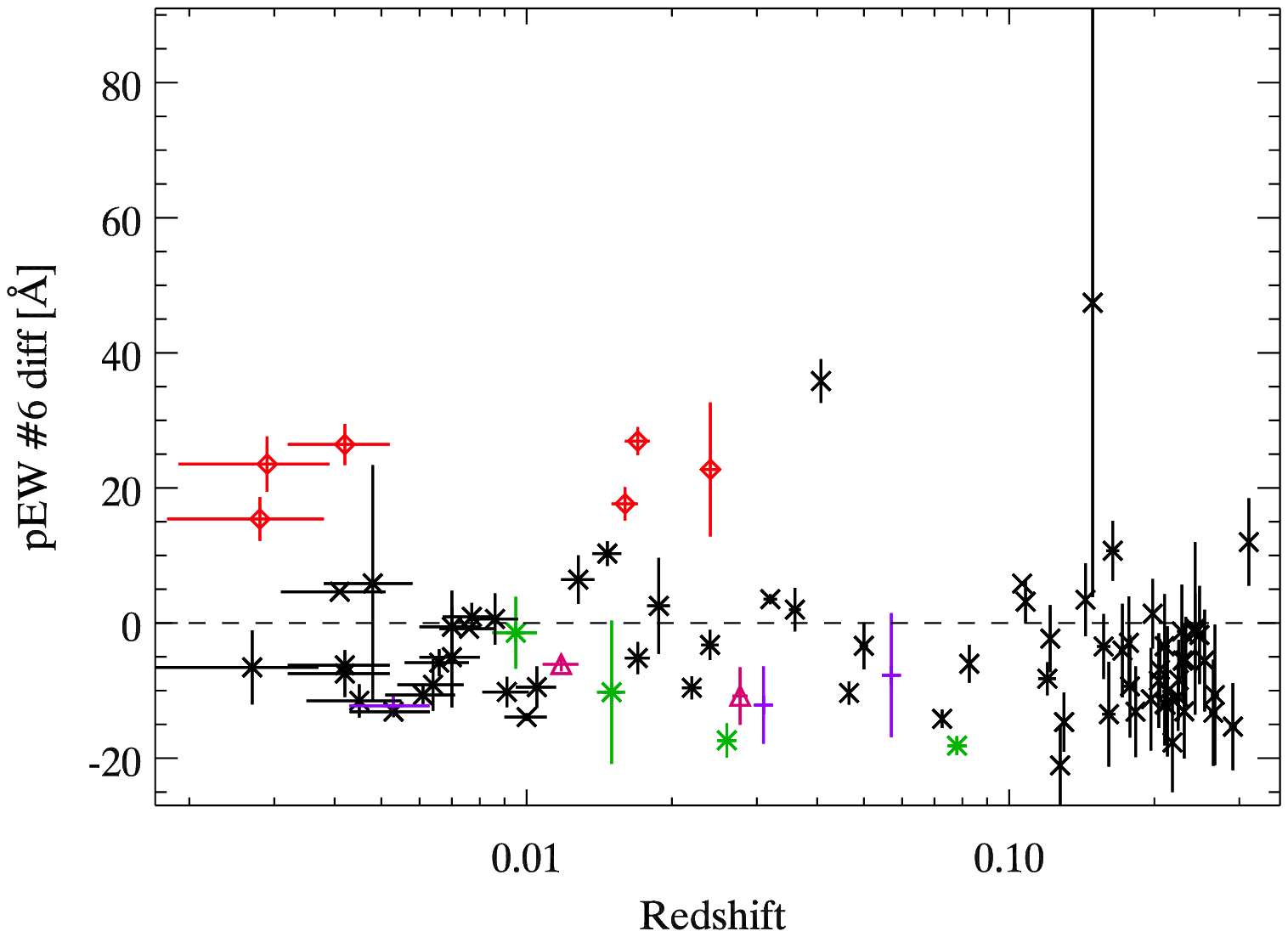}
  \includegraphics[angle=0,width=\halfcol\columnwidth]{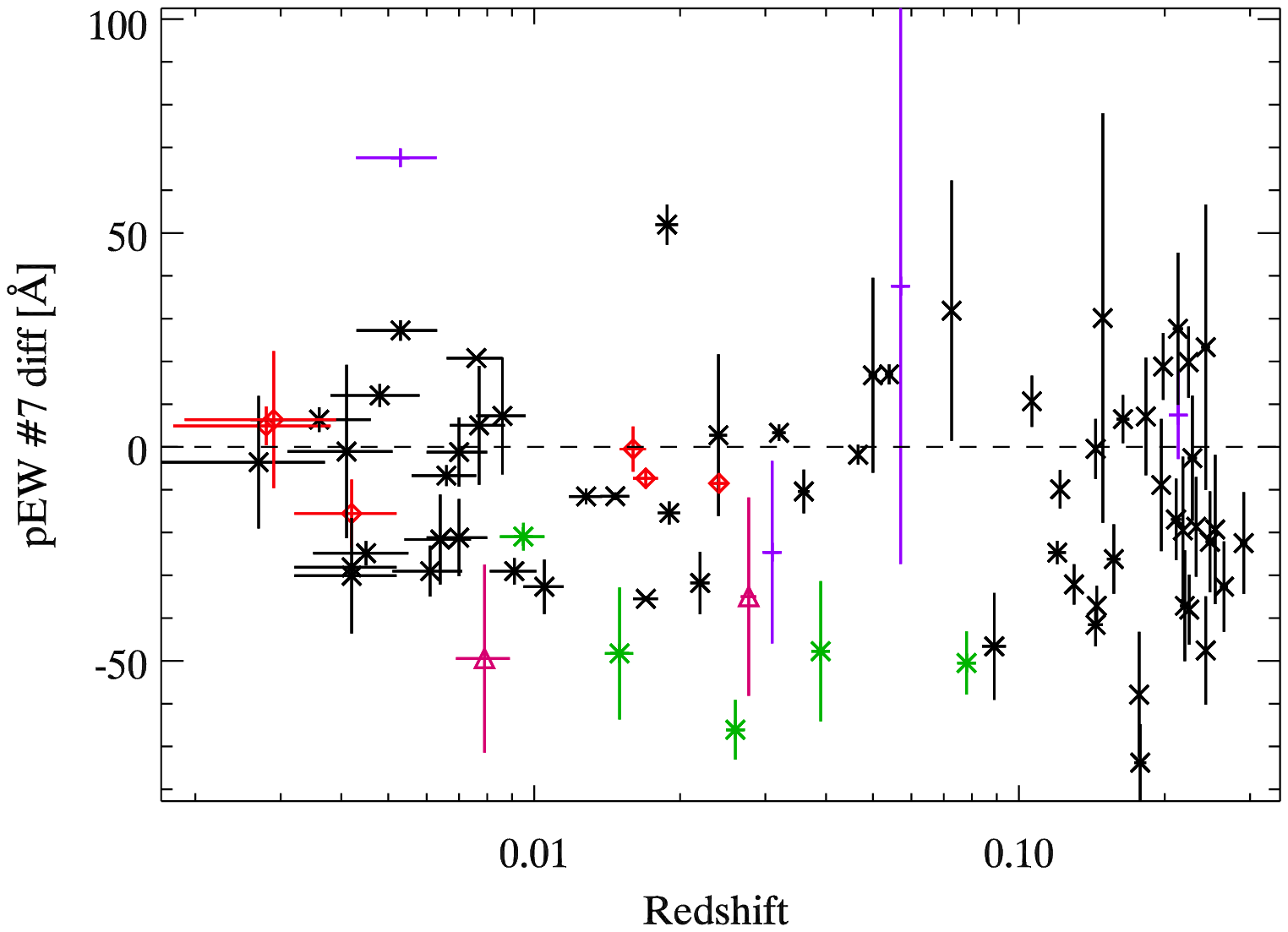}
  \caption{ Comparison of the pseudo-equivalent width measurements
    around lightcurve peak for features 2 to 7 vs redshift. In all figures
    the average spectroscopic evolution \emph{among non-peculiar low-z reference SNe} has been subtracted. It is thus the \emph{pEW-difference} (compared to normal low-z SNe) that is plotted on the y-axis. SNe are divided according to subtype classification, with the colour scheme following the legend in the upper right plot. In the two upper left panels (f2 and f4) the shaded region show how the \emph{pEW-deficit} sample is defined.}
  \label{fig:pew_peak_pec}
\end{figure*}

To be able to study the correlations of spectral indicators with
different parameters we want to remove the epoch dependence. This is
done by fitting a function describing the epoch evolution in pEW and
then subtracting it from the measurements.\footnote{\new{Indicators that vary little or gradually with epoch are fitted with a linear function. Indicators that show sudden changes (like pEW f3) are fitted using a logistic function: $f(t)=A/(1+e^{(t_{br}-t)/\tau})+B$.}}. 
This \emph{epoch independent} quantity ($\Delta pEW$) is shown in Figure~\ref{fig:pew_peak_pec}
for \emph{all} SNe Ia NTT/NOT and reference spectra (including peculiar
types) as a function of redshift for epochs around maximum
brightness.

 We now examine Figure~\ref{fig:pew_evo_vsepoch} and Figure~\ref{fig:pew_peak_pec} for significant differences between distant and local SNe. The NTT/NOT measurements 
generally match the \refcom{1-$\sigma$} contour of the reference sample within
uncertainties.  There are, however, some regions where the NTT/NOT pEW
measurements appear, on average, \emph{lower than the reference set
average}. These differences appear most significent for pEW f2 and pEW f4, this is most easily seen in Figure~\ref{fig:pew_peak_pec} (top left and mid left, shaded region).

\emph{In order to study the origin of these differences we collect all SNe with \emph{either} f2 \emph{or} f4 pEW measurements below
those of normal SNe in the reference sample into a \emph{pEW-deficit}
sample} (thus $\Delta pEW$-f2 $< -8$ {\AA}  OR $\Delta pEW$-f4 $< -25$ {\AA}). This subset is examined in detail in Section~\ref{sec:dis},
where we \new{for example} discuss the effects of different lightcurve parameter
distributions. 
\new{The limits in pEW for the \emph{deficit} sample are arbitrary; we do not expect them to precisly single out a physically distinct subset of SNe. It is rather to be seen as a starting point for a discussion of possible differences between local and distant SNe.}


\subsection{Comparing the two samples - line velocities}

\begin{figure*}
  \centering
  \includegraphics[angle=0,width=\halfcol\columnwidth]{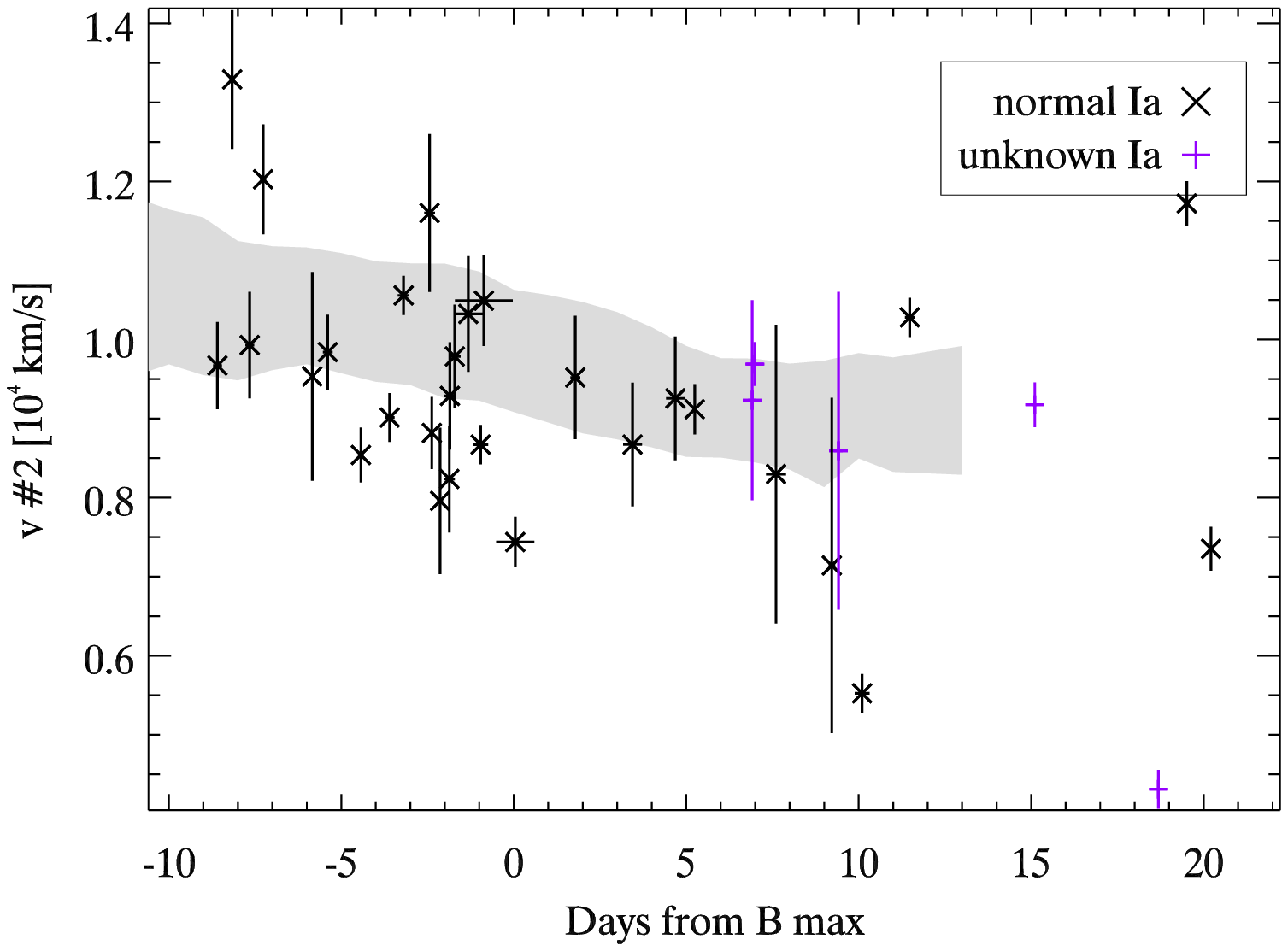}  
  \includegraphics[angle=0,width=\halfcol\columnwidth]{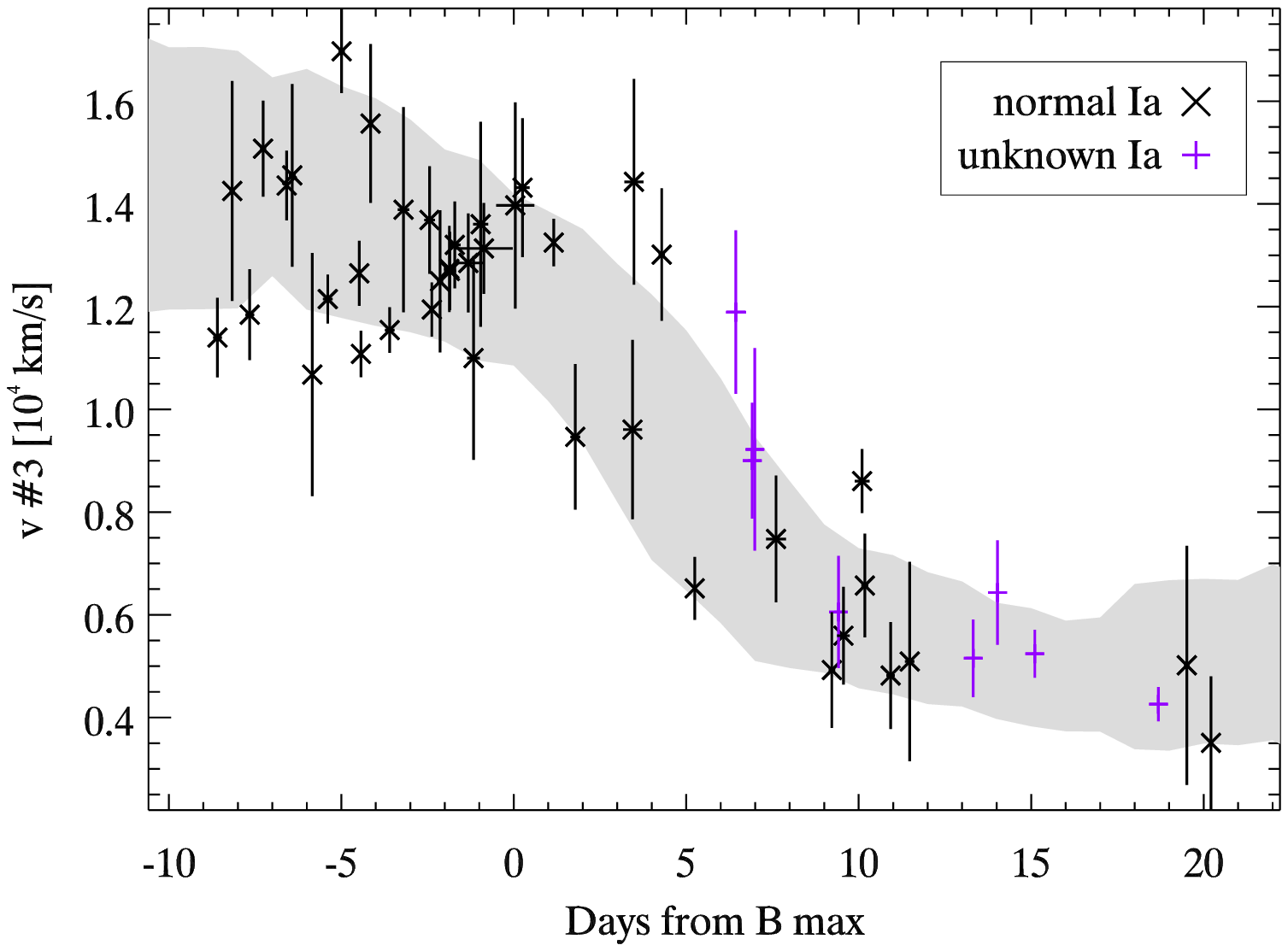}  
  \includegraphics[angle=0,width=\halfcol\columnwidth]{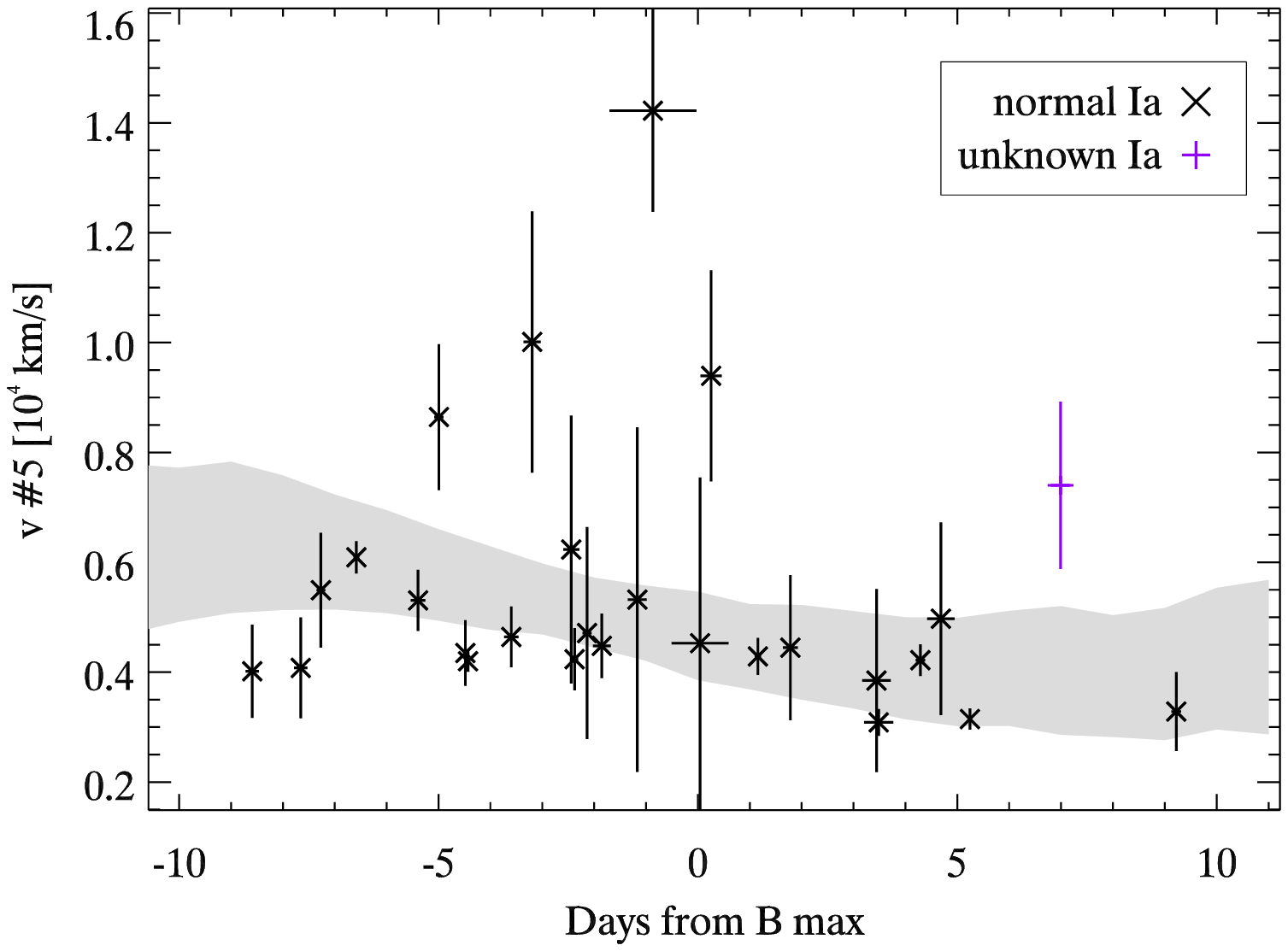}
  \includegraphics[angle=0,width=\halfcol\columnwidth]{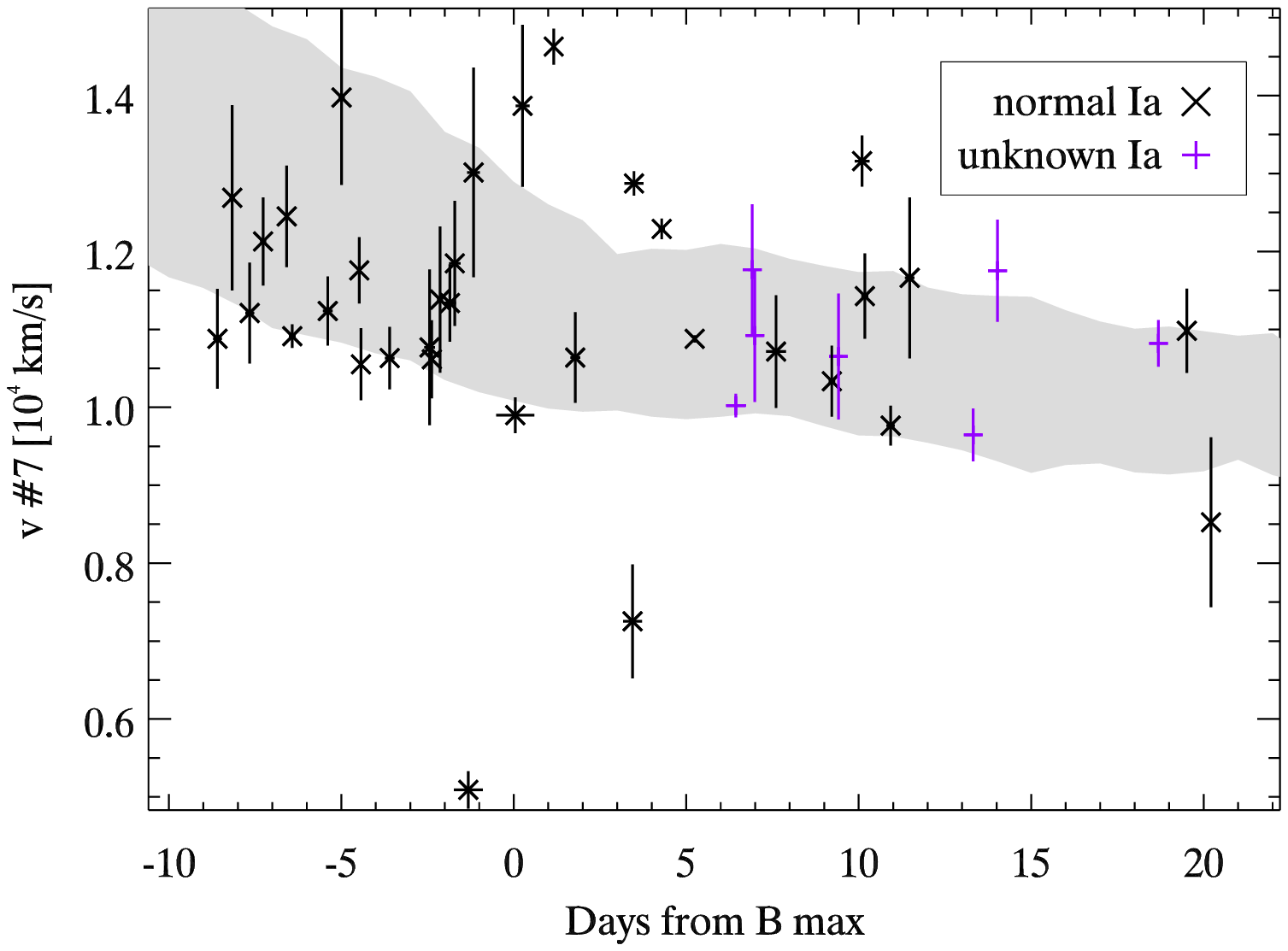}
  \caption{Comparisons of line velocities (f2, f3, f5, f7) between the reference sample and the higher redshift SDSS spectra. The shaded band shows the one sigma contour for
  the normal SNe Ia in the reference sample. }
  \label{fig:vel_evo_vsepoch}
\end{figure*}

Figure~\ref{fig:vel_evo_vsepoch} shows some of the more stable line
velocities for the NTT/NOT sample together with the \refcom{1-$\sigma$} contour for the normal SNe Ia in the reference
sample. Only measurements of SNe with redshifts measured from galaxy lines are included, i.e. the subset of objects with redshifts measured from SN features are excluded here. No signs of sample differences are detected.

\new{The epoch evolution of the velocity of f7 {\sisix} (with epoch) have been extensively studied \citep{2005ApJ...623.1011B,2009ApJ...699L.139W,2010Natur.466...82M}. However, the NTT/NOT sample does not contain enough SNe with multiple spectra to measure velocity changes.}

\subsection{Summary: Comparing the reference and NTT/NOT samples}
The samples are generally consistent, possibly deviating in a subset of NTT/NOT SNe with pEW measurements below these for normal SNe Ia in the reference sample. These were collected in a \emph{pEW-deficit} sample.
%


\section{Results: Correlations with SN parameters}
\label{sec:lccorr}

The collected sample (both low- and high-z SNe) was used to search
for correlations between spectral indicators and global properties of
Type Ia SNe. Since many of the features evolve with epoch, \modifierat{we study the epoch corrected pEW- and velocity-differences, $\Delta pEW$ and $\Delta v$, as introduced above.}

The correlation is calculated taking into account the estimated uncertainties of the indicators.

\subsection{Correlation statistics}

As the basic measure of correlation between measurements $R_i,S_i$, $(i=1..n)$,
we use Spearman's rank
correlation coefficient, 
\begin{equation}
r_S = {\sum_i{(R_i - \bar R)(S_i - \bar S)} 
        \over
        { \sqrt{\sum_i{(R_i-\bar R)^2}}\sqrt{\sum_i{(S_i-\bar S)^2}} }}.      
\label{eq:spearman}
\end{equation}
 This is similar to the Pearson correlation
coefficient, i.e., a non-parametric measure of correlation, but
relies on ranked variables. This method is preferable for variables not following
a Gaussian distribution. Spearman's rank correlation is also less sensitive
to outliers. The output coefficient range from $-1$ to 1, with $-1$ being
perfect negative correlation, 0 no correlation and 1 being perfect
positive correlation.
The significance of a correlation $r$ from $n$ elements can be estimated roughly using Student's t distribution of dimension ($n-2$) and $t=r\sqrt{(n-2)/(1-r^2)}$ \citep{numerical}. For example, rank correlation $0.6$ corresponds to less than 1 \% chance of being random if $n\gtrsim18$. In our analysis we mark any correlation $r>0.6$ where $n\gtrsim15$ for further study. More realistic confidence analysis should be done using permutation tests. This is done when using flexible ranges below, where we account for the fact that we probe a large number of correlations (and thus expect statistical fluctuations to cause some large $|r|$). We first present the basic correlation coefficients for spectra close to lightcurve peak.

\subsection{Correlations with lightcurve parameters}

\paragraph{Correlations around lightcurve peak}

We have searched for correlations with SALT stretch and colour as well as absolute magnitude, $M$, for spectra within $\pm$3 days from maximum brightness. The absolute magnitude is corrected for stretch and colour and calculated assuming a fiducial cosmology. Table~\ref{tab:peakcorr} lists the
calculated Spearman coefficients. The statistical
significance in standard deviation from the null correlation
hypothesis is shown in parenthesis. At least 50 measurements were used
in each correlation estimate.

\begin{table*}
\begin{center}
\caption{Correlations at peak brightness: $r_S$, Spearman correlation coefficient (number of standard deviations from null). At least 50 measurements used for each entry.}
\label{tab:peakcorr}
\begin{tabular}{|c|ccc|ccc|}
\hline
\hline
f & pEW - s & pEW - c & pEW - M & velocity - s & velocity - c & velocity - M\\
\hline
1&-&-&-&$+$0.19 (0.9)&$-$0.02 (0.1)&$+$0.05 (0.3)\\
2&$-$0.73 (4.8)&$+$0.19 (1.2)&$+$0.19 (1.3)&$-$0.00 (0.0)&$+$0.22 (1.3)&$+$0.01 (0.1)\\
3&$+$0.13 (0.9)&$+$0.34 (2.4)&$+$0.12 (0.8)&$+$0.16 (1.0)&$+$0.08 (0.5)&$-$0.15 (1.0)\\
4&$-$0.26 (1.8)&$+$0.42 (3.0)&$-$0.05 (0.4)&$-$0.36 (2.2)&$+$0.26 (1.6)&$+$0.13 (0.8)\\
5&$-$0.40 (2.9)&$-$0.18 (1.3)&$+$0.15 (1.1)&$+$0.19 (1.1)&$-$0.18 (1.1)&$-$0.13 (0.8)\\
6&$-$0.36 (2.4)&$+$0.02 (0.2)&$-$0.07 (0.4)&$-$0.02 (0.1)&$+$0.18 (1.1)&$-$0.14 (0.8)\\
7&$-$0.55 (3.8)&$+$0.15 (1.0)&$-$0.03 (0.2)&$+$0.07 (0.4)&$+$0.22 (1.3)&$-$0.02 (0.1)\\
\hline
\end{tabular}
\end{center}
\end{table*}

\begin{figure*}[htb]
  \centering
  \includegraphics[angle=0,width=\halfcol\columnwidth]{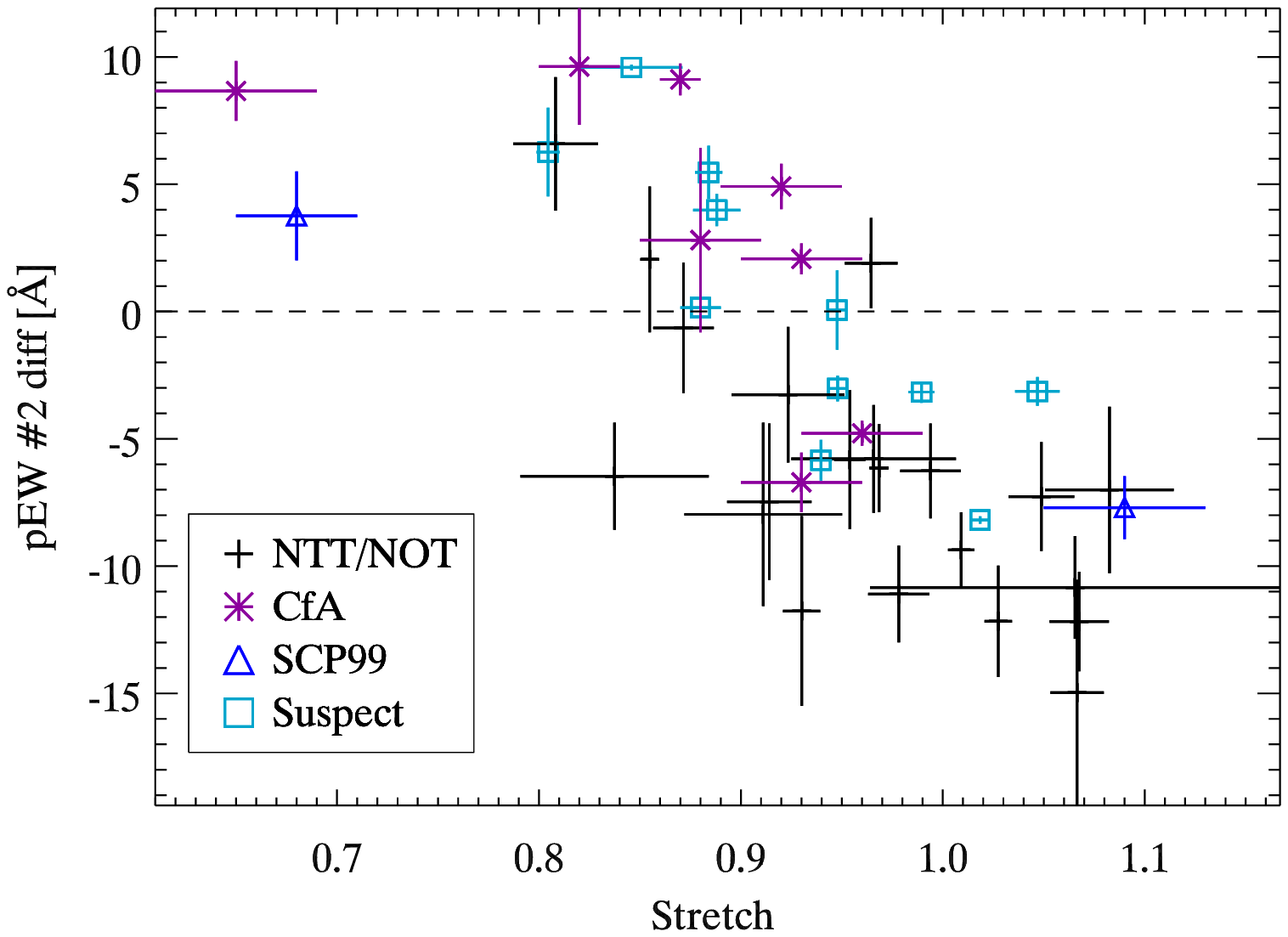}  
  \includegraphics[angle=0,width=\halfcol\columnwidth]{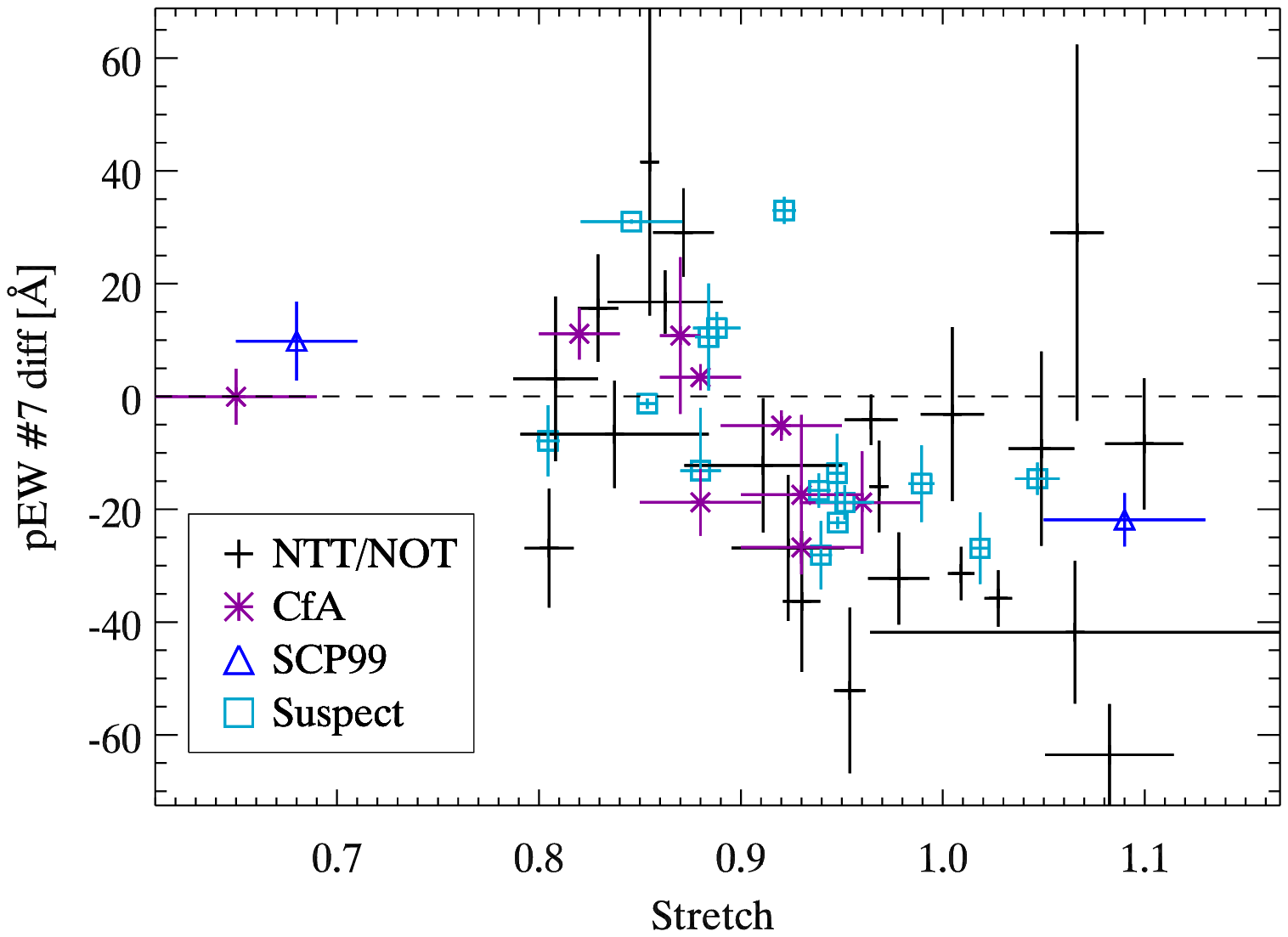}  
  \includegraphics[angle=0,width=\halfcol\columnwidth]{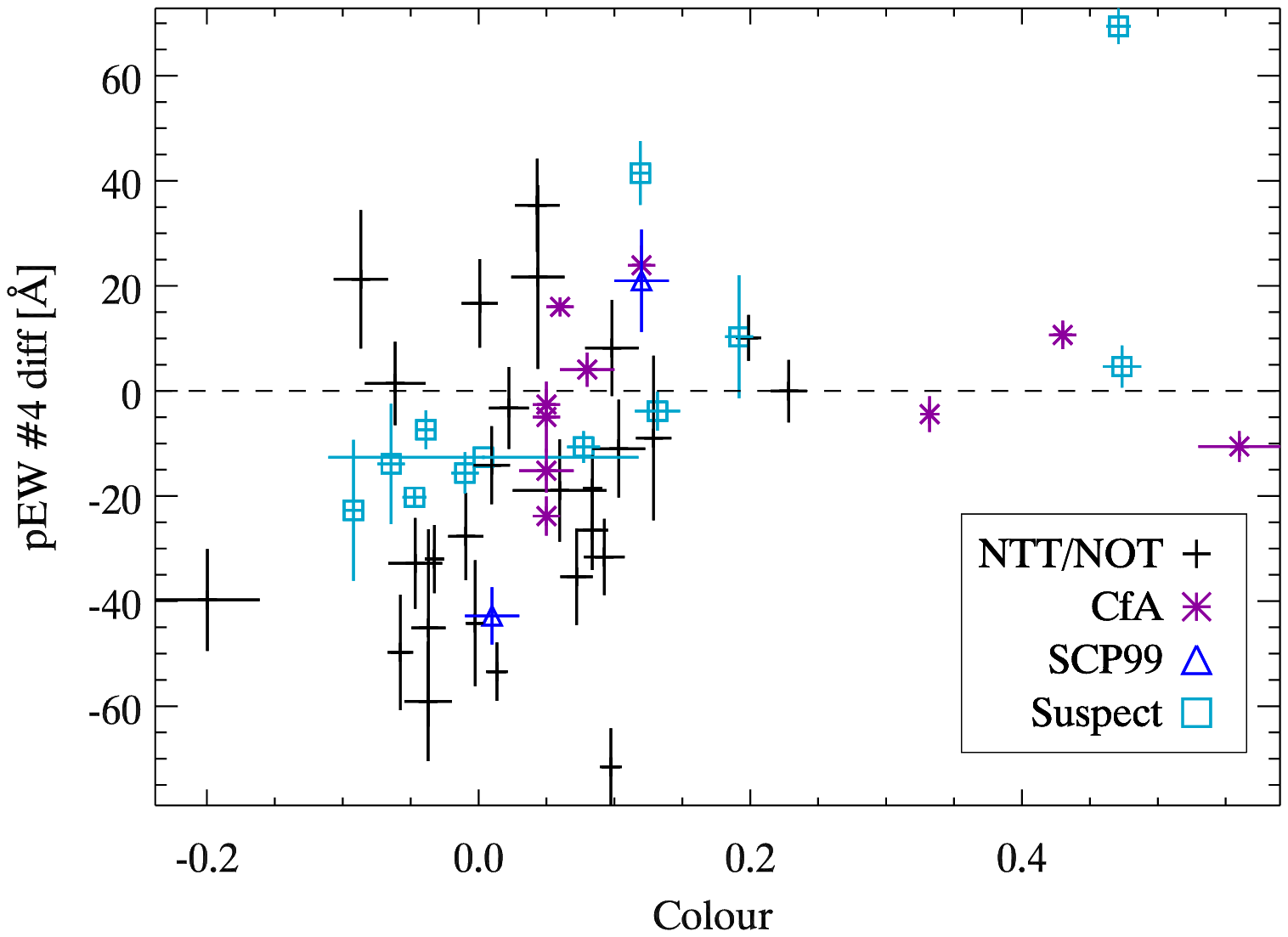}  
  \includegraphics[angle=0,width=\halfcol\columnwidth]{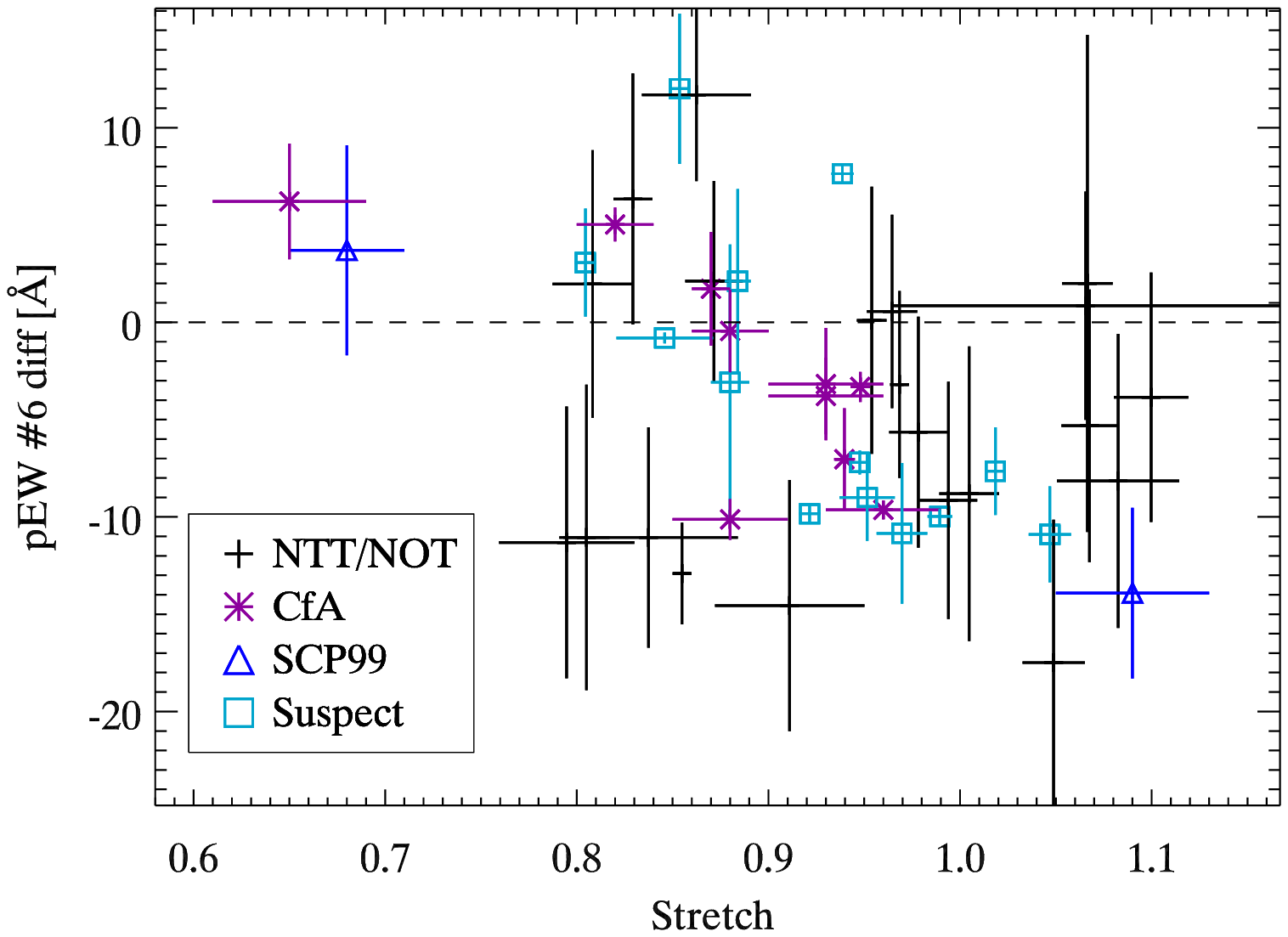}              
  \caption{These correlation studies show the measured pEW minus the expected pEW for the corresponding epoch for the full sample of normal SNe Ia vs SALT lightcurve parameter (stretch, colour) for epoch ranges around peak ($\pm3$ days). Indicators chosen are the ones showing largest correlation for this epoch range.
  }
  \label{fig:maxcorr}
\end{figure*}

Some pseudo-equivalent widths around peak luminosity do show strong
correlation with lightcurve parameters. These include f2 and f7
correlating with stretch and f4 showing a correlation with lightcurve
colour. In Figure~\ref{fig:maxcorr} we show these strong correlations,
together with the f6 correlation with stretch. For the latter, all SNe
except the low S/N NTT/NOT objects show a strong correlation; this feature is
too small to probe among noisy data. The correlation of the depth of
this feature with lightcurve width has been reported earlier, see
e.g. \cite{2008MNRAS.389.1087H}.

Also noticeable is that most correlations have the same
\emph{direction} (sign): SNe with wide lightcurves (large stretch) have weaker
pEW values and redder SNe (large colour) have in general larger
equivalent widths. While these \modifierat{colour} correlations are not strong they are
consistent and \emph{opposite} in direction to what
would be expected from an application of pure \citet{cardelli89} type
extinction (see Figure~\ref{fig:ebvpew}).

Any correlation between spectral indicators and peak magnitudes
\emph{corrected for stretch and colour} would be of great interest,
since this could be used to ``sharpen'' Type Ia SNe as standard
candles. Most spectral indicators correlate only weakly with absolute magnitude (after correction for stretch and colour). The  most significant correlation is for pEW f4; this is, however, a weak correlation of only moderate significance ($\sim 3 \sigma )$.

We find that none of the velocities are strongly correlated with
stretch, colour or absolute magnitude around peak brightness. The kinetic energy, as sampled by line velocities, thus appear to be independent from the optical luminosity.

\paragraph{Flexible epoch ranges}

\begin{figure*}[htb]
  \centering
  \includegraphics[angle=0,width=\halfcol\columnwidth]{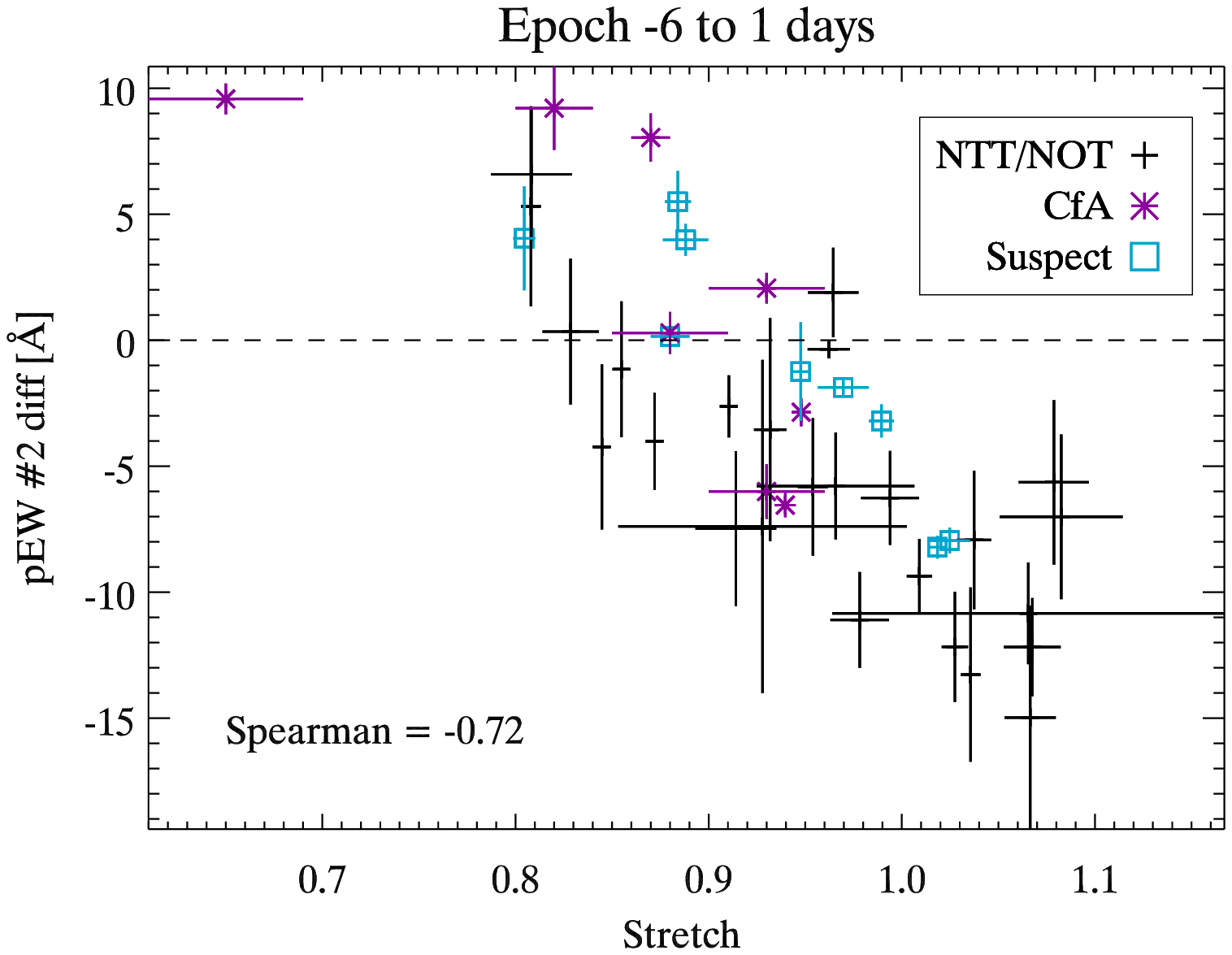}  
  \includegraphics[angle=0,width=\halfcol\columnwidth]{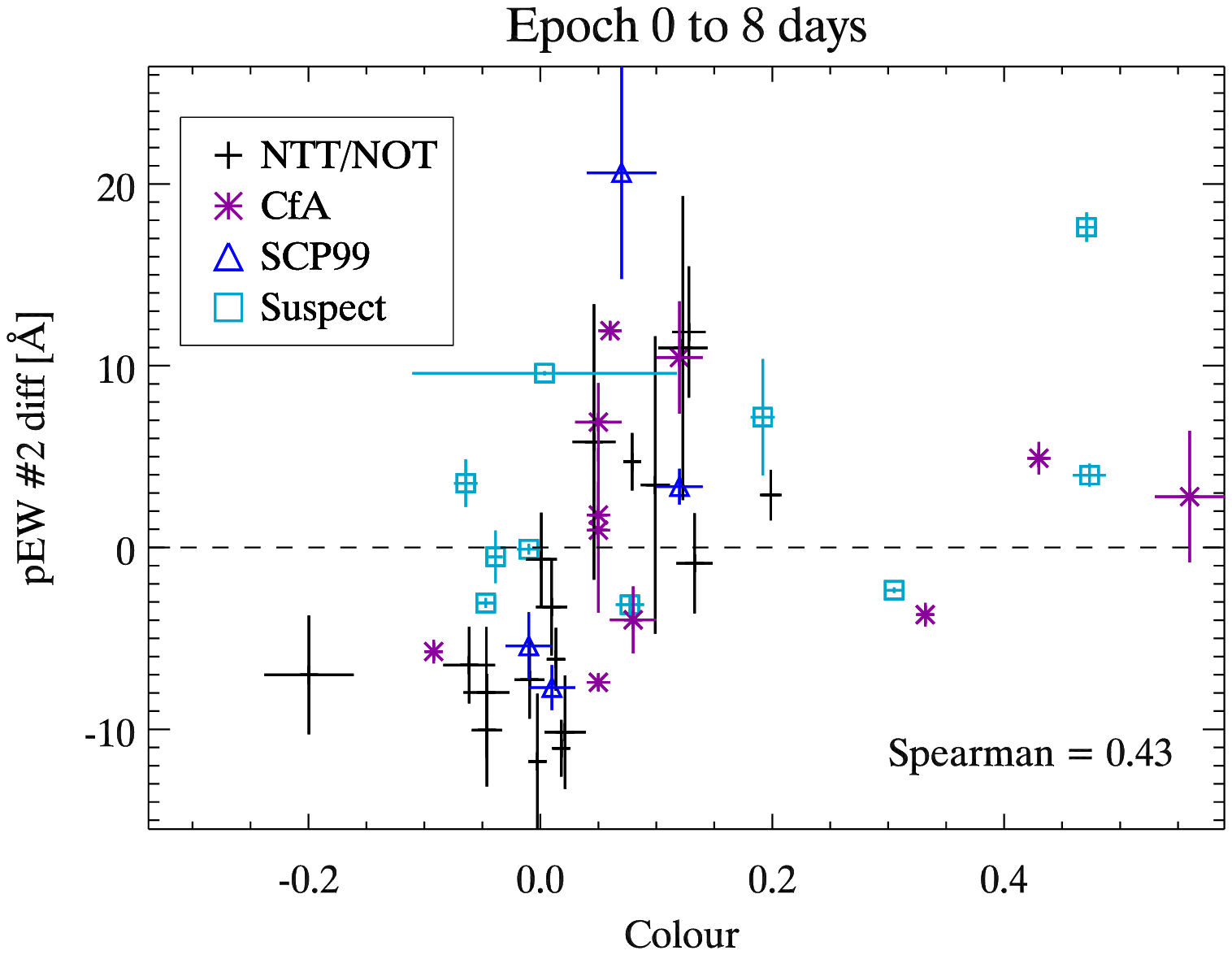}  
  
  \caption{These correlation studies show $\Delta pEW$, the measured pEW minus the
expected pEW for the corresponding epoch (for the full sample of normal
SNe Ia), vs SALT lightcurve parameter (stretch, colour) for
different epoch intervals. These are given above each panel. The Spearman
correlation coefficient is given in each plot. MC simulations show
that the chance that these correlations are random consequences of a
search through epoch ranges is one or less in 1000.  }
  \label{fig:corr}
\end{figure*}

There is no {\em a priori} reason to expect a fixed epoch range
around lightcurve peak to be the epoch range where spectral indicators
correlate best with lightcurve parameters. In Figure~\ref{fig:corr}
we present a sample of the bracketed epoch ranges where the most significant
correlations were found through a blind search involving all
indicators. Before analyzing the results we will describe the blind search in detail, as well the MC studies performed in order to determine how significant the search output is.

For a given set of epochs, an indicator measurement
(e.g. pEW f2) and a global property (e.g. stretch), we loop
through all epoch ranges (containing at least 15 measurements) and
save any correlations with $|r_S|>0.4$. A combination is thus defined by (spectral indicator, global property, min epoch, max epoch) e.g. ('pEW f2','stretch',-8,0).
The number of such combinations, for each indicator, range from 500-1000 (if using all SNe and an indicator well defined at all epochs) to 50-100 (if using a subset of SNe and an indicator not existing at all epochs).

For any real correlation found we also expect ``neighbouring'' epoch
ranges to be correlated. If, for
example, the epoch range $0-8$ yields a strong correlation we would
also expect epoch ranges like $1-7$ and $2-9$ to show correlation. We thus rank correlations between indicators and global
properties through the \emph{number of epoch ranges with
$|r_S|>0.4$}. When we discuss a correlation in an epoch range,
this is thus only one in a series of neighbouring correlating epoch
ranges.

While this epoch bracketing is necessary in order to find the epoch
ranges where indicators are sensitive to global parameters, this
method increases the probability of finding random correlations. For
any set of three parameters, it will always be possible to find
\emph{some} correlation between two of these through restrictions of
the third. For any correlating parameters we thus have to find the
probability of finding such correlation(s) in random data. This is
done through Monte Carlo simulations where we retain the epoch
and indicator values and randomise the global property. We then search
for correlations among epoch ranges exactly as for real data. This
process is repeated 1000 times and the number of strongly correlating
epoch ranges is saved for each. This result can be used to find
the probability of finding as many strongly correlating epoch ranges
by chance.
For the correlations presented below, we find either zero or one out of 1000 iterations to yield as many correlated epoch ranges. We thus find that the probability that these correlations are completely random is equal to or smaller than 0.001.

The strongest correlation found was for the pEW for feature 2 and
stretch when probing epochs right before maximum brightness. Depending
on the epoch interval used, the Spearman correlation coefficient is
about 0.6-0.7. It is a stable correlation in the sense that the coefficient remains large when the epoch interval is perturbed. A correlation between
lightcurve width and feature 2 has been previously discussed by
\citet{2008A&A...477..717B} and \citet{2008AA...492..535A}.

The pseudo-equivalent width for feature 2 also seems to be correlated
with the fitted SALT colour parameter, as shown in the left panel
of Figure~\ref{fig:corr}, but mainly in the epochs just \emph{after}
peak. While SNe with SALT-c $\gtrsim 0.2$ do not appear to correlate
with pEW, those below this rough limit seem to do. This
could be interpreted as a sign of two different sources of reddening,
where e.g. the highly reddened supernovae are dust extincted while
most supernovae get their colour from some intrinsic property which is
correlated with the strength of feature 2. 

\modifierat{As was shown in Section~\ref{sec:pewdesc}, dust absorption according to \citet{cardelli89} would create a small pEW change in the \emph{opposite} direction. }

\begin{figure*}[htb]
  \centering
  \includegraphics[angle=0,width=\halfcol\columnwidth]{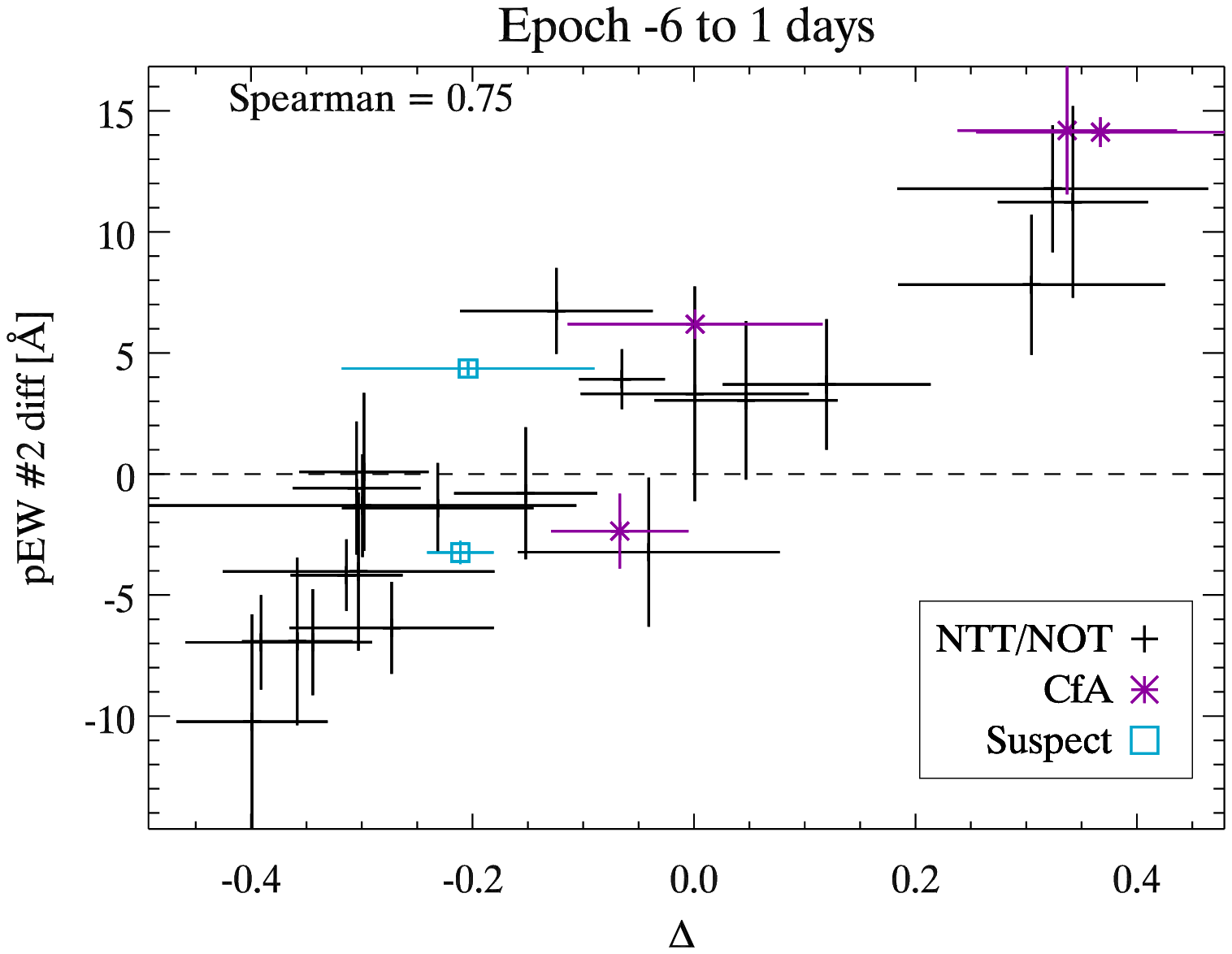}  
  \includegraphics[angle=0,width=\halfcol\columnwidth]{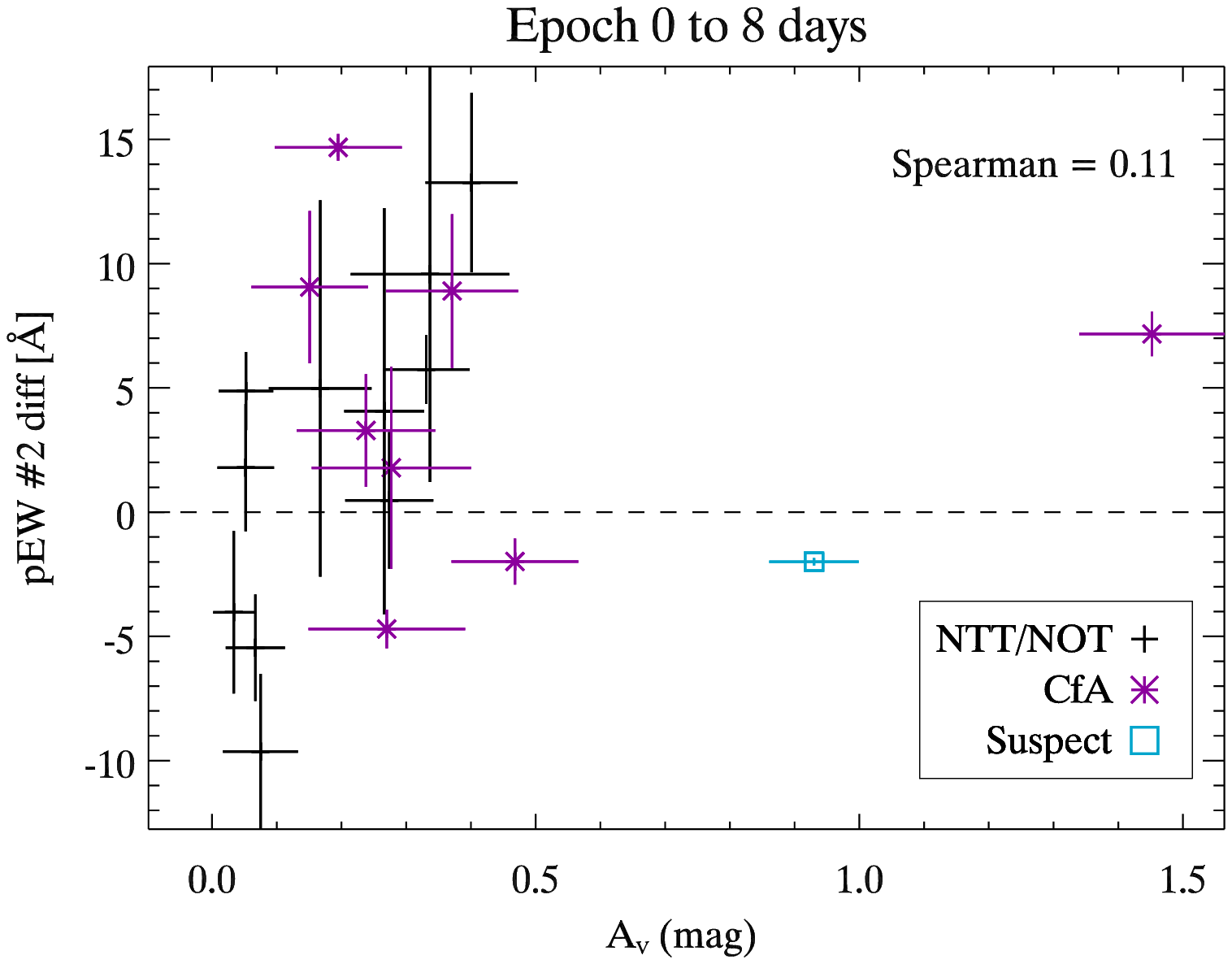}  
  \includegraphics[angle=0,width=\halfcol\columnwidth]{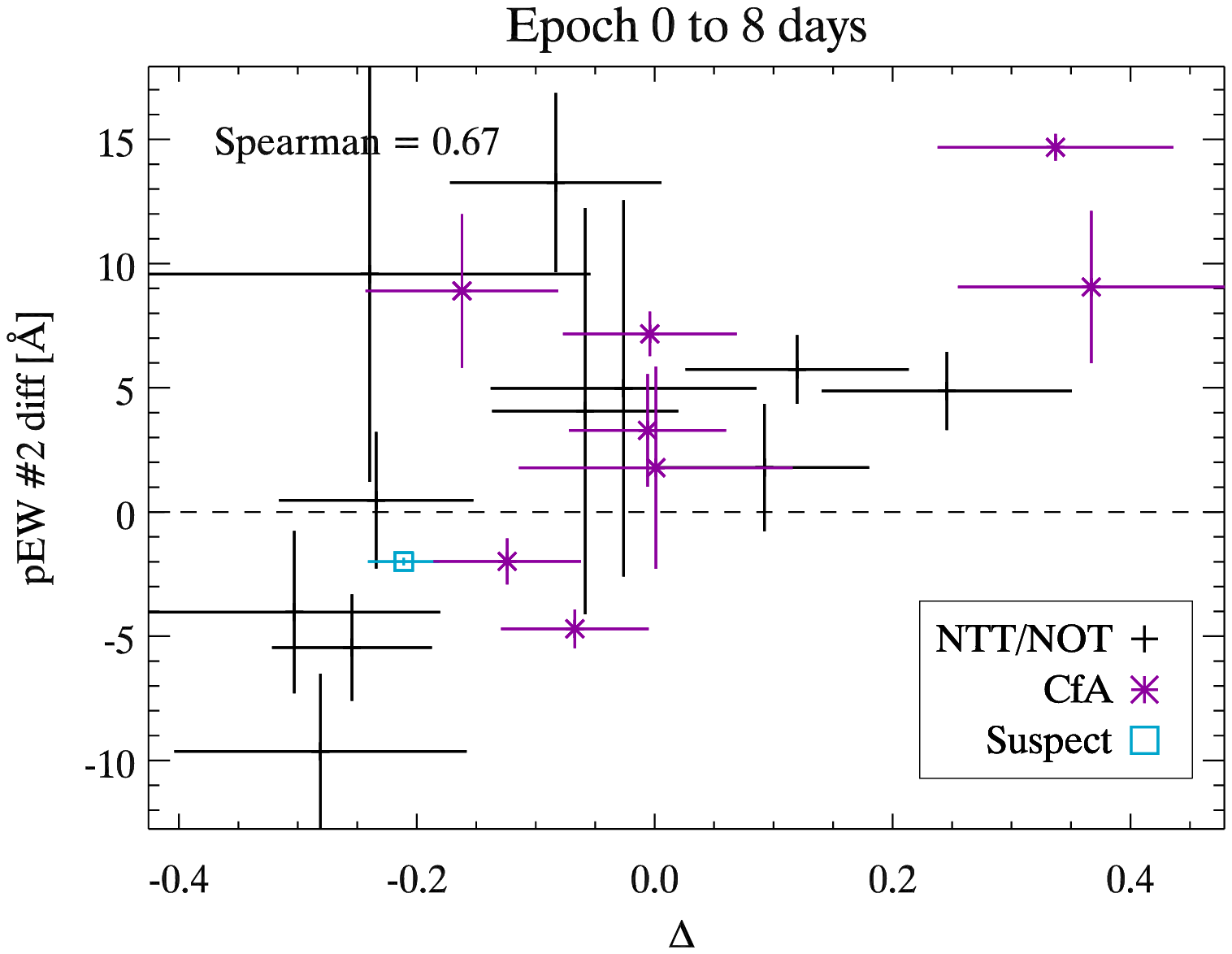}              
  \caption{These correlation studies show $\Delta pEW$, the measured pEW minus the expected pEW (for the corresponding epoch for the full sample of normal SNe Ia), vs MLCS2k2 lightcurve parameter ($\Delta$, $A_V$) for different epoch intervals. These are given above each panel. The Spearman correlation coefficient is given in each plot. The epoch ranges were chosen to correspond to those used in Figure~\ref{fig:corr}. For the post lightcurve peak epoch range (0-8) we include correlations with both $\Delta$ and $A_V$.
  }
  \label{fig:mlcscorr}
\end{figure*}

To probe the origin of these lightcurve correlations the same
epoch ranges and indicators as displayed in Figure~\ref{fig:corr} were
examined using MLCS fit parameters. These results can be seen in
Figure~\ref{fig:mlcscorr}. Correlation with lightcurve shape
($\Delta$) is strong, while correlations with $A_V$ are less clear.
The correlation with $\Delta$ in the epoch range 0-8 is significant,
while for the same epoch range, we find a correlation with 
SALT colour but only a weak correlation with stretch.

The origin of these correlations is further discussed in Section~\ref{sec:dis}, where we focus on feature 2, {\si}.

\subsubsection{Summary: Correlations with lightcurve parameters}
Pseudo-equivalent widths, as measured in this sample, do correlate
with lightcurve properties. We recreate strong \modifierat{linear} correlations between
f2-\si~both with SALT stretch and MLCS $\Delta$.
Correlations with SALT colour were
also found. In general we see weak correlations between most pEWs
around lightcurve max and both stretch and colour in the sense that
wider, bluer SNe have small equivalent widths.

\subsection{Host-galaxy properties}
\label{sec:hostprop}

It is well documented that star forming, late type galaxies host
brighter SNe with wider lightcurve shape
\citep{1996AJ....112.2391H}. Since correlations between lightcurve
shape (stretch) and spectral indicators seem to be present in the data
analysed, we expect the host-lightcurve correlation to propagate to a
correlation between spectral and host galaxy properties. Recent
studies have also found indications of correlations between host
galaxy properties, like mass and metallicity, and supernova absolute
magnitude that does \emph{not} appear to be captured by lightcurve shape
or colour
\citep{2008ApJ...685..752G,2010ApJ...715..743K,2010arXiv1003.5119S,2010arXiv1005.4687L}. It
is thus of great interest to investigate if spectral indicators correlate with
host galaxy properties, especially beyond what is
related to  lightcurve stretch.

All SN host galaxies were studied using the stellar formation code
PEGASE. A description of this process can be found in \citet{smith10},
and a comparison with lightcurve properties and Hubble diagram
residuals in \citet{2010arXiv1005.4687L}. Here we use the estimated host
galaxy type, host mass (in units of $M_{\odot}$) and specific star
formation rate (sSFR; defined as the star formation rate per stellar
mass, $yr$) and compare with spectral indicators. Host type is
defined based on specific star formation rate: type zero indicate no
star formation, type one moderately star forming ( $-12.5 < \log(sSFR)
< -9.5$) and type two star forming ($\log(sSFR) > -9.5$). This
parameter thus largely overlaps with sSFR, with the important exception
that hosts with no star formation (Type 0) are not displayed in the
plots over sSFR values.

As previously, this search for correlations was performed using various epoch intervals, but with only SDSS NTT/NOT SNe. Once again, the largest $|r|$ were found when
probing the second feature ({\si}). When maximising the
correlation for host mass and specific SFR, we obtained the epoch
interval between 0 to 8 days past peak (which coincides with the
epochs where a strong colour correlations is also seen). 

\new{We also minimize the KS probability for indicator measurements from SNe
in different host galaxy types to originate from the same distribution.
We do this through first compairing SNe from higly star-forming hosts
(type 2) with remaining SNe (from host types 0 and 1), and then compairing
SNe from non star-forming hosts (type 0) with star-forming (type 1 and
2).}
 
\begin{figure*}[hbt]
  \centering
  \includegraphics[width=\halfcol\columnwidth]{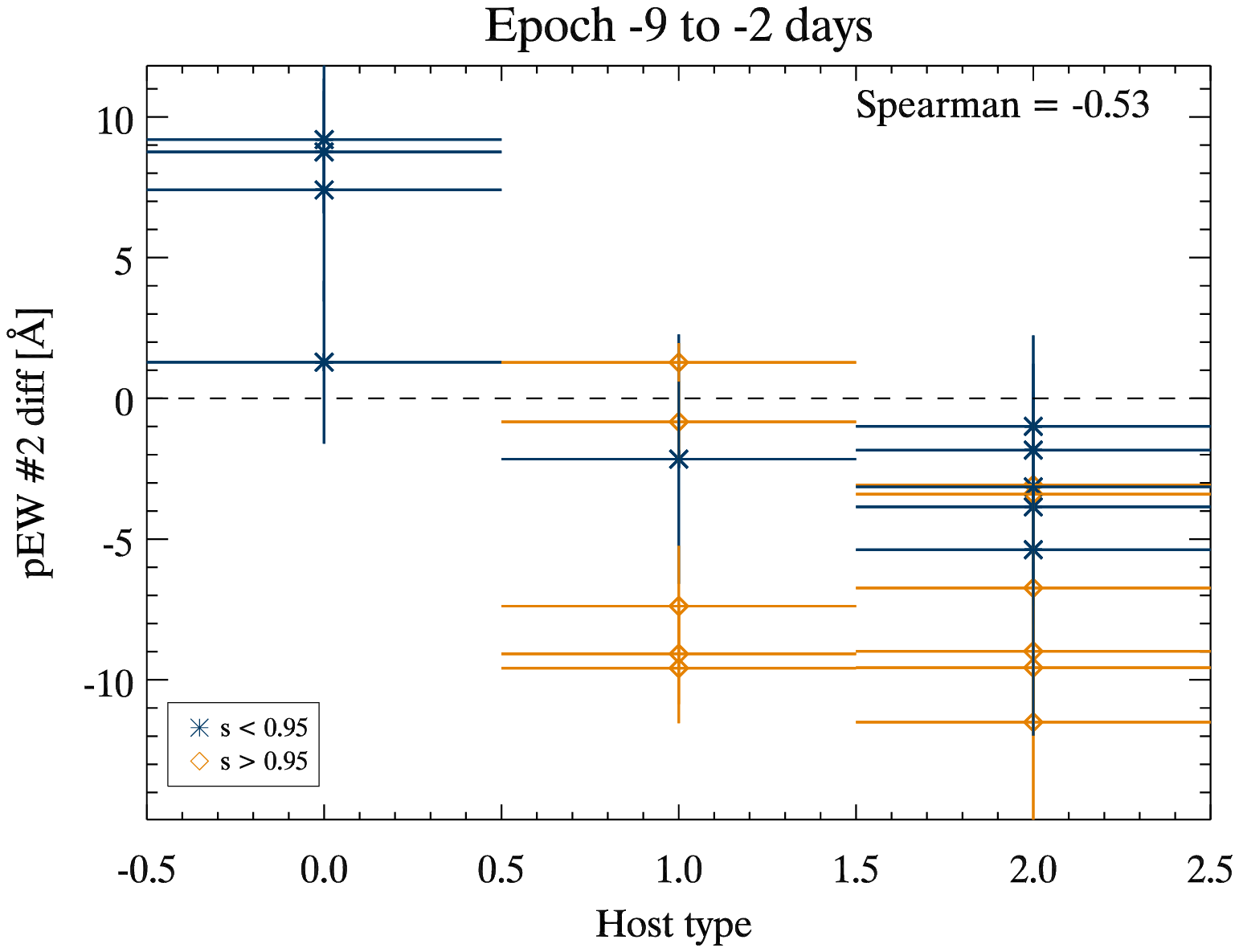}  
  \includegraphics[width=\halfcol\columnwidth]{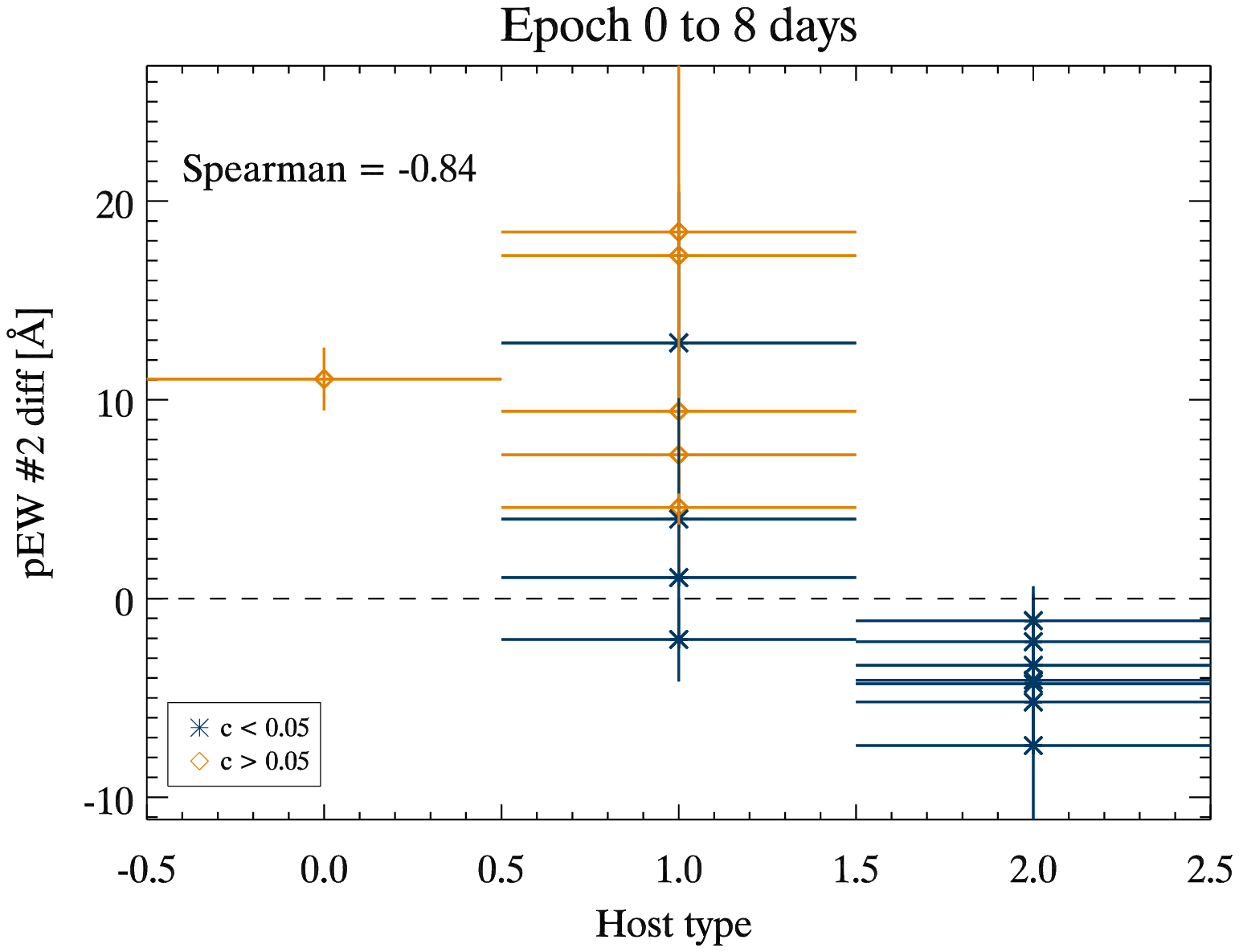}  
  \includegraphics[width=\halfcol\columnwidth]{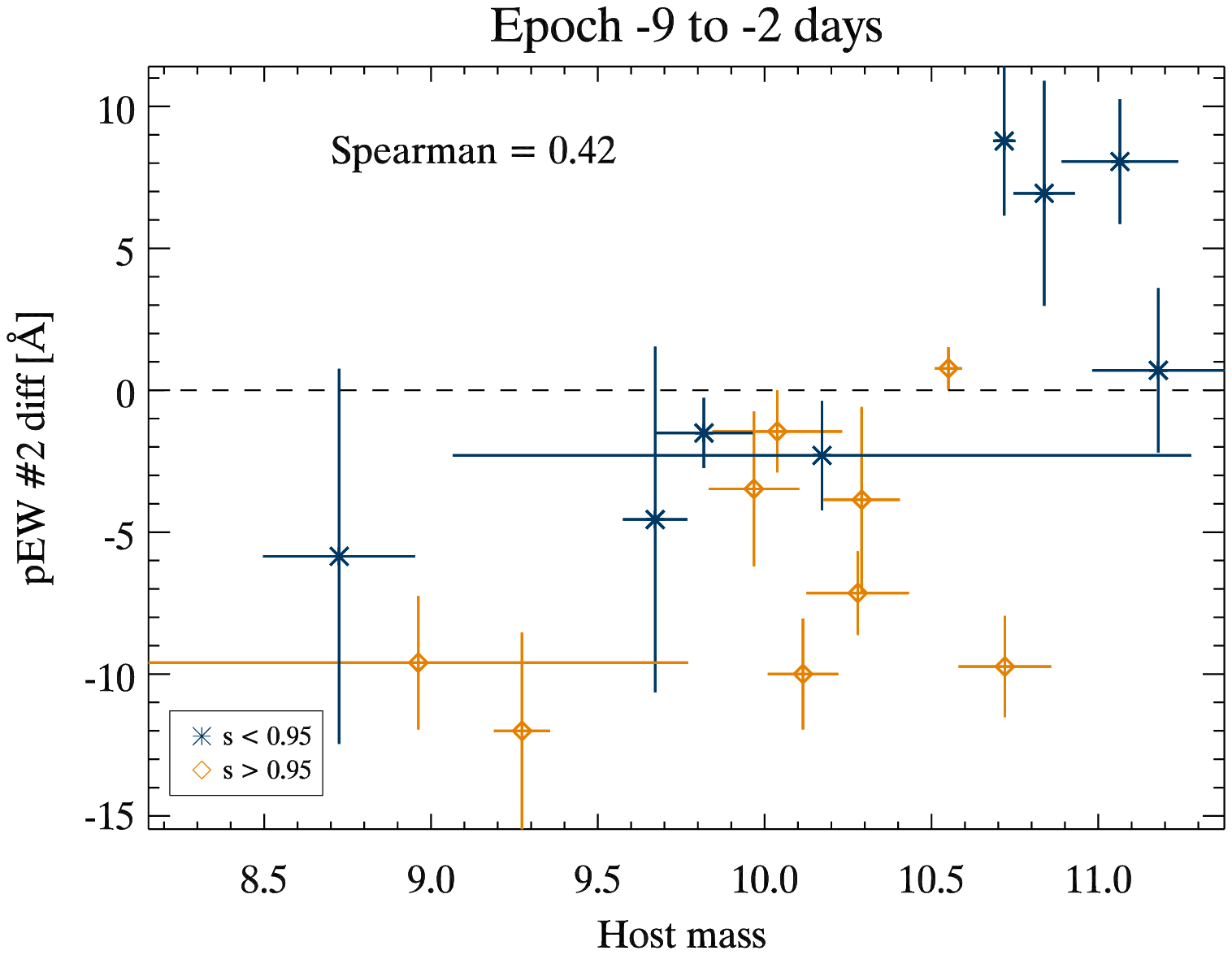}   
  \includegraphics[width=\halfcol\columnwidth]{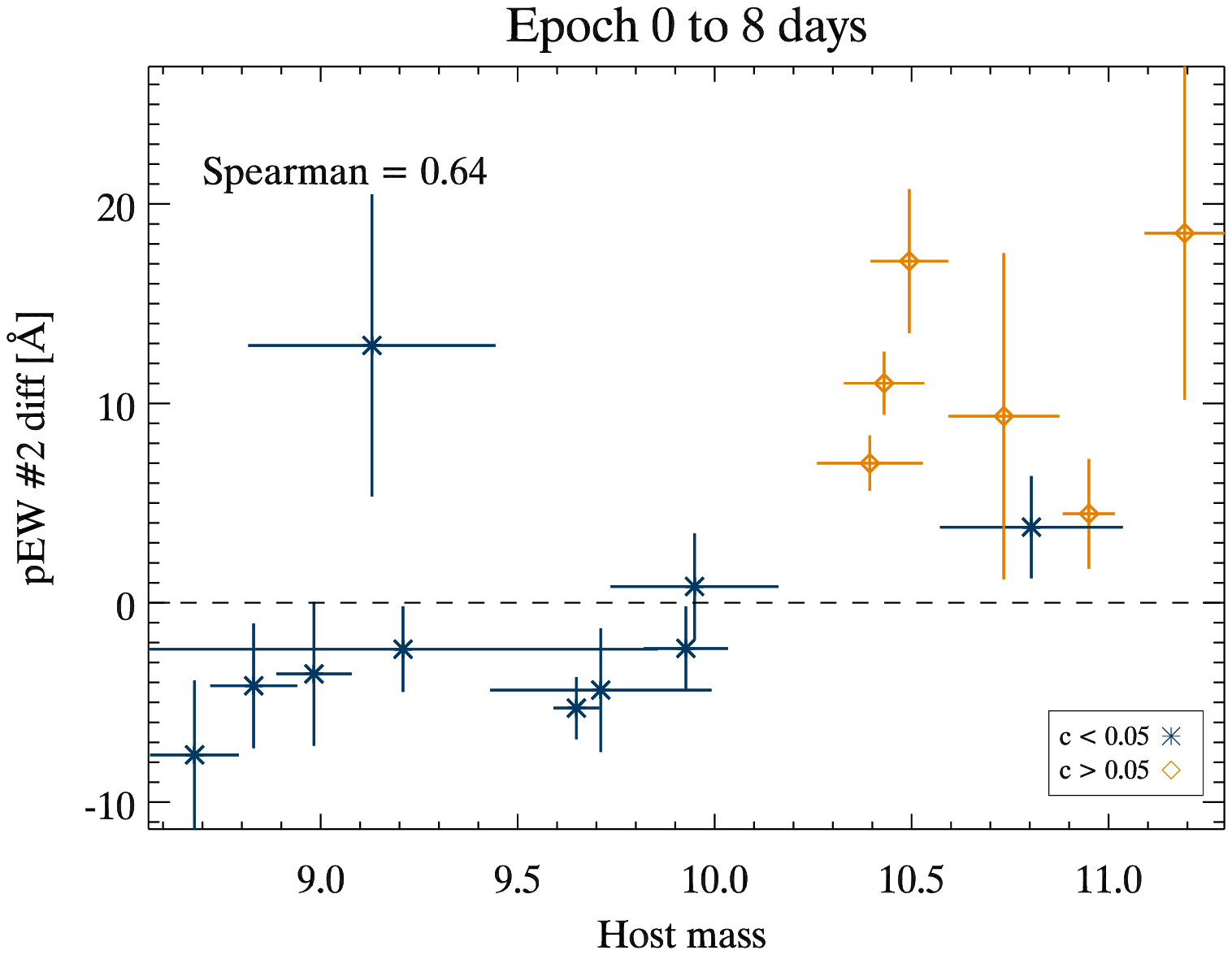}
  \includegraphics[width=\halfcol\columnwidth]{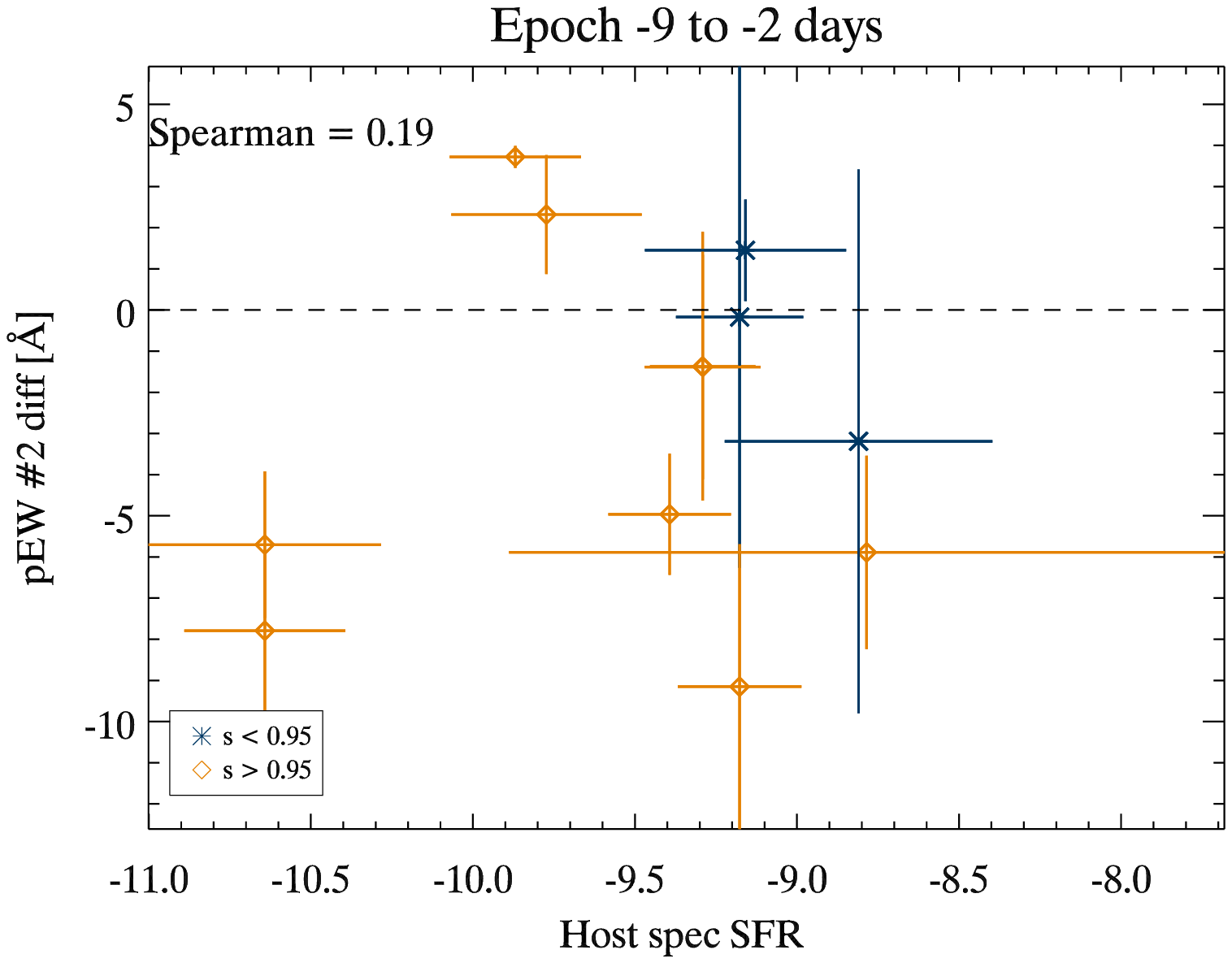}   
  \includegraphics[width=\halfcol\columnwidth]{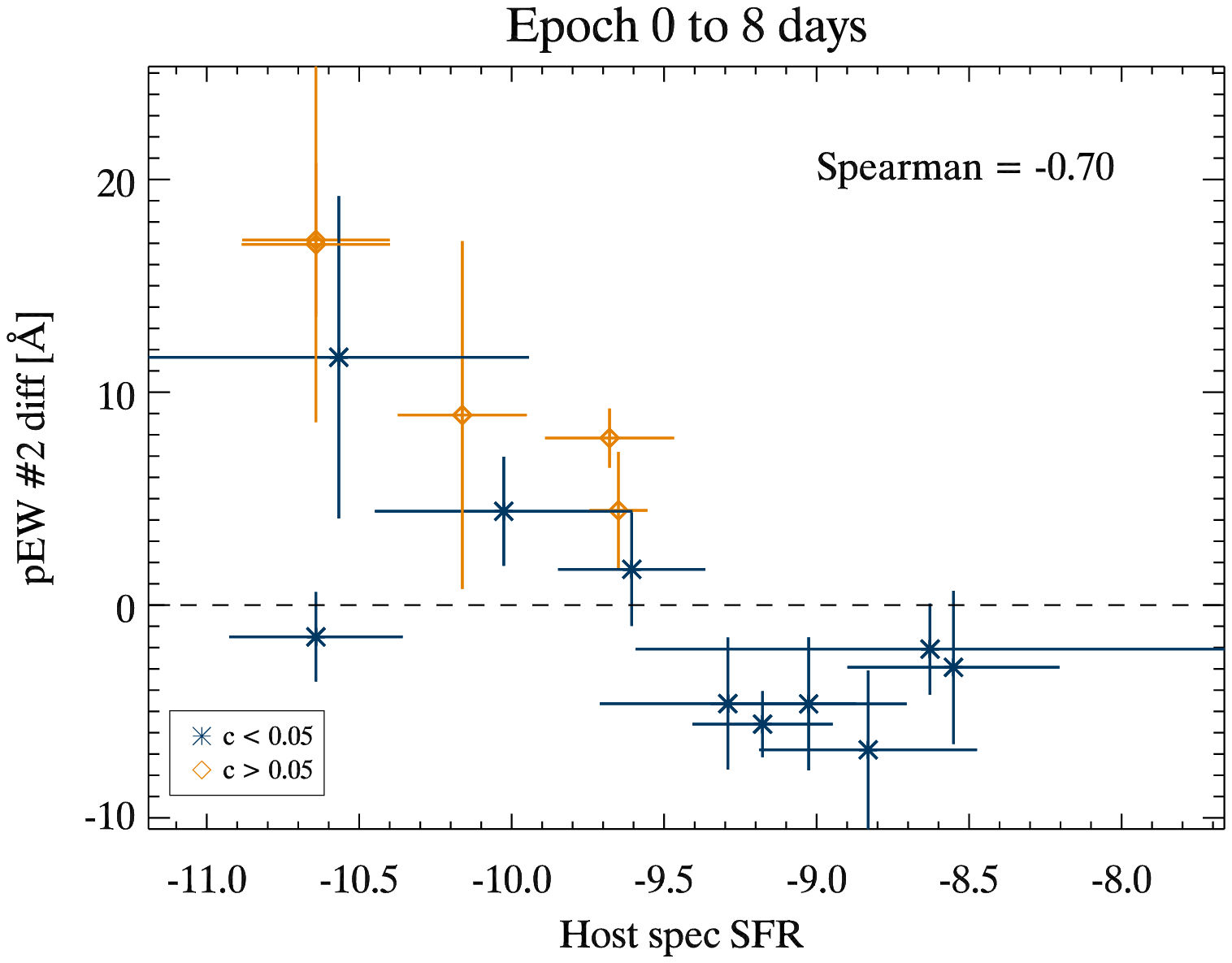}     
  \caption{ Pseudo equivalent width of feature 2 compared with host galaxy properties. \emph{Left panels:} Pre peak epochs (-9 to 2) with colours marking low/high lightcurve stretch SNe. \emph{Right panels:} Post peak epochs (0 to 8) with colours marking low/high lightcurve colour SNe. \emph{Top:} Host type vs pEW (0 not star forming, 1 moderately star forming, 2 highly star forming. \emph{Middle:} Log of Host galaxy mass ($M_\odot$) vs. pEW  \emph{Bottom:} Log specific start formation rate (log sSFR,~$yr^{-1}$) vs pEW.
 } 
  \label{pic:hosttype}
\end{figure*}

The epoch interval with \emph{least} probability of distributions of pEWs from all host types being the same is epoch $-9$ to $-2$. 
These correlations are shown in Figure~\ref{pic:hosttype}.

\paragraph{Before lightcurve peak}

The largest $|r|$ connection between lightcurve width and pEW was
found to be for f2 during the epochs right before lightcurve peak. In
the left panels of Figure~\ref{pic:hosttype} we compare this feature
with host galaxy properties. As can be expected, assuming a relation
between stretch and host galaxy type, we see that actively star forming
galaxies have lower pEW f2 values than passive galaxies. A possible
alternative explanation is that SNe in passive galaxies (Type 0) form
a separate sub group: These all have large pEW f2 values, lower than
average lightcurve widths and are only found in the very most massive host
galaxies (as can be seen in the mid left panel of
Figure~\ref{pic:hosttype}). More statistics is needed to determine
whether such a subgroup exists, or if a continuous trend with host type
or mass is present. 

\paragraph{After lightcurve peak}

\modifierat{We also found pEW f2 after lightcurve maximum to be related to host galaxy properties (right panels of Figure~\ref{pic:hosttype}). Lightcurve width and pEW f2 are correlated but not as strongly as before maximum. Instead we see a tentative correlation with SALT colour.}

Most significant in this epoch range is what seems to be
a linear correlation between sSFR and pEW f2. Alternatively, as for
the epochs before peak, this could be explained using a subgroup of
SNe, in this case consisting of blue SNe with small pEW f2 values and
high specific star formation rate. This relationship is emphasised if we include
host mass information; all SNe in this group have \modifierat{low} host
masses. Note that all SNe in actively star forming hosts (Type 2) have smaller than average SALT-c lightcurve colours ($c<0.05$), thus suggesting little dust
extinction.\fixme{Evaluate this? Need to include mcls?} Since we see a correlation with host galaxy mass, a random (uncorrelated) star formation rate would mean a correlated \emph{specific} star formation rate (since this is the ratio between SFR and host mass). This clearly needs further study; we would expect SNe in small, star forming hosts to be more extincted.

\subsubsection{Summary: Correlations with host galaxy properties}

Host galaxy properties and spectral indicators, mainly pEW f2,
are clearly connected. 
Of special interest is whether further
subgroups among normal SNe can be identified. Our results could be
interpreted as a hint of the presence 
of \emph{two subgroups}. One consisting of low stretch SNe with wide
pEW f2, hosted by passive massive galaxies, and another consisting
of blue SNe in actively star forming, low-mass galaxies.\fixme{Find the average corrected abs mag for these, and
compare with the rest? Did a quick check and did not find significant
- need larger numbers or better fits.}


\section{Discussion}
\label{sec:dis}

\modifierat{The discussion is split into a further examination of correlations in the {\si}~region (Section~\ref{sec:rest4000}), a search for signs of evolution with redshift (\ref{sec:evodisc}), a discussion of host galaxy properties (\ref{sec:hostdisc}) and finally we revisit some systematic effects (\ref{sec:sysdisc}).} This division does not mean that these topics are separate, they are rather closely related.

We use composite (average) spectra as a tool for searching for physical differences between subsets of SN Ia spectra. Composite spectra can be normalised/created in a number of ways. \modifierat{This can make it hard to interpret the difference between two composite spectra in terms of physics} or compare composite spectra created using different methods.\footnote{For example, \citet{2001AJ....122..549V} show how continuum measurements on Quasar composite spectra change depending on combination method.} 

The composites shown here were created through scaling all spectra to have average $f_\lambda = 1$ in the studied region before combining. The uncertainties of the composites were then investigated through jackknife techniques. This simple procedure is sufficient since our purpose is to compare different subsets of spectra, not provide a ``true'' supernova composite.


\subsection{Rest frame 4000-4500}
\label{sec:rest4000}

This region roughly correspond to features two and three.
The absorption feature at $\sim4050$, here called f2 and usually attributed to
Si~II, has been shown to correlate with both luminosity and the
stretch parameter in the sense of wider more luminous SNe having
smaller equivalent width
\citep{2008A&A...477..717B,2008AA...492..535A}. The same trend,
using partially the same data, is seen clearly in this study
using both SALT stretch (left top panel of Figure~\ref{fig:corr}) and
MLCS $\Delta$ (left top panel of Figure~\ref{fig:mlcscorr}).
\modifierat{We can further show that (i) the trend is \emph{continous} with SALT-s/MLCS-$\Delta$ and (ii) strongest before lightcurve peak.}

Including the broader absorption region around $4200$~{\AA} \modifierat{(here called f3, attributed to MgII amongst other ions)},
\citet{2007AA...470..411G} found a correlation between the ``breaking
point'' of this feature and lightcurve width and
\citet{2009ApJ...693L..76S} found tentative signs of evolution, where
high redshift objects have smaller equivalent
widths. \citet{2009arXiv0905.0340B} used SNfactory data to look for
the spectral regions most correlated with peak brightness; the best
such was the ratio F(6420{\AA})/F(4430{\AA}).

Finally one of the major spectral differences between normal and
SN 1991bg-type SNe is the dominant Ti absorption at these
wavelengths, even though these subluminous SNe are not included in
this analysis.

This region is thus important both in order to understand and model
supernova explosions as well as to make SNe better standard candles
for use in cosmology. We will here try to probe the origin of the
\emph{pre-peak correlation with stretch} and the \emph{post-peak
correlation with colour} through comparisons of composite spectra.

\begin{figure}
  \centering
  \includegraphics[angle=-90,width=\columnwidth]{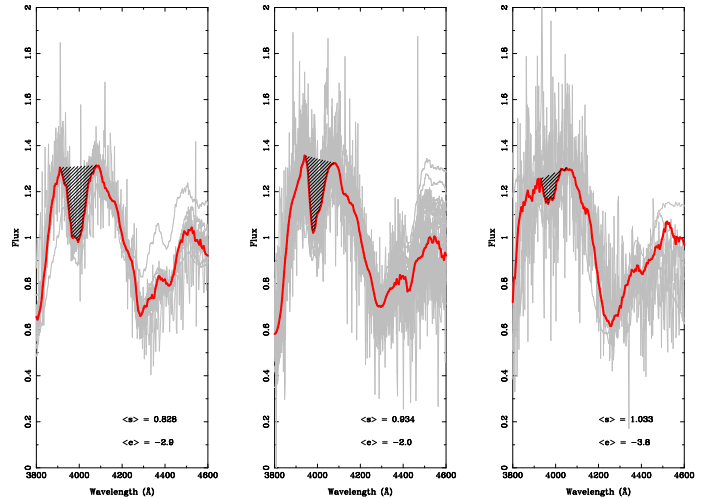}  
  \caption{ Composite spectra constructed from all peak spectra (NTT/NOT as well as reference SNe),
  divided according to stretch. \new{The average lightcurve stretch, s, is printed in each
  panel, increasing from left to right. The average epoch, e, is also included
  for each composite. The plots show both the composite (thick red)
  and individual (grey) spectra.}
}
  \label{pic:stretchcomposite}
\end{figure}

\paragraph{Pre-peak stretch composite} All spectra contained in the epoch region between -6 and 1 (the
 region where the correlation with stretch was strongest) were
 included. \modifierat{Multiple spectra of single SNe were pre-combined.} The sample was first divided into
 three stretch bins (low, intermediate and high). These composites can
 be seen in Figure~\ref{pic:stretchcomposite}. Both the width and
 depth of the {\si} feature grow smaller with increasing stretch
 (exemplified by the almost nonexistent absorption for overluminous
 SN 1991T-like objects). \refcom{Around the epoch of peak luminosity, the photosphere temperature and composition seem to be such that the ${}^{56}$Ni yield, through the deposited energy, strongly affects \si~ionisation and abundance, thereby causing a correlation between lightcurve width and \si~depth.}

As is discussed in \citet{2009ApJ...693L..76S}, changes in SN
populations \citep{2007ApJ...667L..37H} will be seen as a change in
average spectra with redshift. \citet{2008A&A...477..717B} argues that
this pEW-lightcurve width correlation could be used to correct SN Ia
lightcurve \emph{instead} of using the lightcurve width. Our results,
using a larger sample, support this conclusion. However, since
photometry will be obtained for most SNe anyway and high S/N spectra
in a small epoch range is needed, the practical use is limited.

\paragraph{Post-peak colour composite} Correlating spectroscopic indicators with lightcurve SALT colour
yielded a statistically significant correlation between colour and
{\si} pEW (using spectra obtained a few days after lightcurve
peak). This correlation grows stronger if events with a colour above
$\sim0.2$ are excluded. The need to remove highly reddened events
could be explained if these are caused by a separate effect compared
with moderately reddened SNe (e.g. circumstellar absorption). A
correlation between spectral indicators and colour would be of great
interest, as this would show that at least some part of the
colour-luminosity relation seen in Type Ia SNe originates in the SNe and
not in any extinction external to the explosion.

\begin{figure}
  \centering
  \includegraphics[angle=-90,width=\columnwidth]{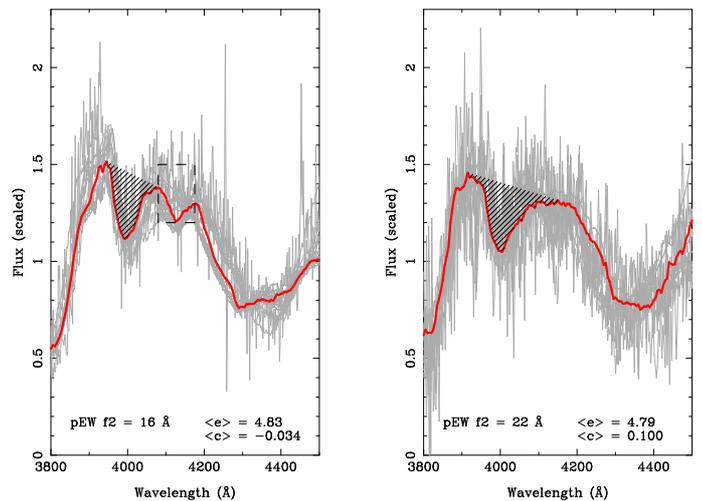}  
  \caption{Composite spectra constructed from all spectra with epochs
in the range (3-8). The left panel composite is based on spectra with colour
below average, the right panel one is based on spectra with colour above
average. The plots show both composite (red) and individual
(grey) spectra. The average SALT colours and epochs are printed for each composite, as well as the pEW for the indicated region. \new{The dashed square marks a region where a small feature seems to exist only in the low colur sample \refcom{(SALT c $<0.03$)}, either because it is not resolved \emph{or} because it is very rare among the higher colour SNe \refcom{($0.03<$ SALT c $<0.3$)}.}}
  \label{pic:colcomposite}
\end{figure}

In Figure~\ref{pic:colcomposite} we explore composite spectra constructed based on colour. Excluding very reddened eventes ($c>0.3$) the average SALT colour for spectra with rest frame epochs in the (0-8) range is 0.03. 
As with stretch one composite spectra was created out of all spectra with colour below average and one out of all spectra with colour above average. The mean SALT colour values for the low colour composite is $-0.03$ and for the high colour composite $0.10$.

\modifierat{The most interesting difference in Figure~\ref{pic:colcomposite} is} a resolved small
absorption at $4150${\AA} among the bluer SNe. This feature can not be seen in
the corresponding redder colour composite. Various authors have
suggested absorption by C~II, Cr~II, Co~II or Fe~III in this region, e.g. \citep{2005AJ....130.2278G,2008PASP..120..135B,2010ApJ...713.1073S}
(as well as D. Sauer, private communication)\fixme{Is this how to do
it? Also confirm with daniel.}. The average pEW will be smaller for
the blue composite since the peak at 4070 {\AA} provides a bluer
feature bound; this bound will be set $\sim$4120 {\AA} for the red
(high colour) version. The average epochs for the two composites are
very similar, and the effect is thus likely not created by epoch
variations.\footnote{Feature 3 also show some correlation with colour
in the same epoch range. Since these features share a border this correlation can very well be caused by flux change around $4100$ {\AA}.} We finally note that the blue edge of the $3900$ {\AA} peak is slightly weaker in the lower than average colour composite (left panel Figure~\ref{pic:colcomposite}).

Extinction by dust could, especially in the presence of noise,
cause a pEW correlation with SALT-c or MLCS-$A_V$ as follows: Highly
extincted, i.e., highly reddened SNe, have a large fraction of the
flux around $4000$ {\AA} absorbed. The relative S/N levels will thus be
smaller for redder SNe, which could make it impossible to resolve the
small $4150$ {\AA} absorption. This effect would be more significant for SNe
with already low S/N.

\new{While it is clear that the composites differ with respect to the 4150 {\AA}
absorption, we cannot further distinguish between: (i) random small
sample effects (ii) lower S/N preventing resolution of the small dip and
(iii) a difference in the nature of the explosion correlated with SALT
colour. The last alternative is intriguing, and strengthens the interest
in using spectral indicators to disentangle the dependence of supernova
properties on colour.}

\subsection{Evolution}
\label{sec:evodisc}

Evolution of SNe Ia is a ``thorny'' issue; it is hard to make firm predictions or document that SN properties have ``evolved''. From a theoretical point of view,
changing metallicity (and other parameters changing with time, like galaxy ages)
\modifierat{could} have an impact on SN Ia progenitor systems, but the
details remain largely unknown. Previous work on SN Ia 
indicates that their
properties appear to vary depending on the surrounding environment
\citep{2008ApJ...685..752G,2010arXiv1003.5119S,2010arXiv1005.4687L} and
redshift \citep{2007ApJ...667L..37H}. So far these effects are  interpreted as due to population evolution \cite[a larger fraction of ``prompt''
high stretch objects at higher redshifts,][]{2005A&A...433..807M}. 

Such evolution would not per se imply any fundamental problems for SN cosmology. However, if spectral evolution \emph{not} captured in lightcurve shape fits would be detected, this could create a cosmological parameter bias since these spectral changes are likely to affect luminosity. We thus need to show that we understand such changes.

Both the ESSENCE \citep{2008ApJ...684...68F} and SNLS \citep{2009A&A...507...85B,2009ApJ...693L..76S} collaborations have reported indications of evolution, or systematic differences in redshift binned composite spectra. Both groups use samples where the average lightcurve width increases with redshift. Based on the lightcurve shape correlations found above, we would thus expect decreasing equivalent widths with redshift. This agrees qualitatively with the differences found by both groups.

\subsubsection{The pEW-deficit subsample} 

As was discussed in Section~\ref{sec:pewtwosamples}, the deviations between the reference and NTT/NOT SNe can be attributed to a set of SNe with lower pseudo-equivalent widths than the non-peculiar reference SNe. These were collected in a \emph{pEW-deficit} sample. 
We will here study why these NTT/NOT SNe deviate and whether this is a sign for SN Ia evolution. As our basic tool we create composite spectra of \emph{deficit} SNe and compare with composite spectra based on normal non-deficit NTT/NOT SNe. 

\begin{figure}[hbt]
  \centering 
  \includegraphics[angle=-90,width=\columnwidth]{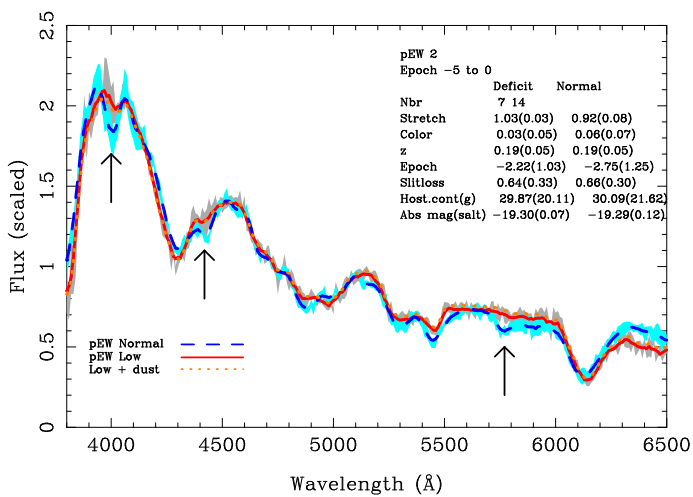}  
  \caption{ 
    \emph{Pre peak, f2 deficit vs normal:} Composite spectra of ``reference normal'' SNe (dashed blue) compared with pEW-deficit SNe (solid red) in the epoch range $-5$ to 0 days past peak brightness. The shaded regions correspond to jackknife dispersion (light blue for normal, grey for deficit). The dotted orange line corresponds to the deficit composite with \citet{cardelli89} dust applied that matches the colour difference. The arrows indicate features discussed in the text. Mean properties of the composites are stated in the Figure.
  }
  \label{fig:f2prepeakcomp}
\end{figure}

\begin{figure}[hbt]
  \centering 
  \includegraphics[angle=-90,width=\columnwidth]{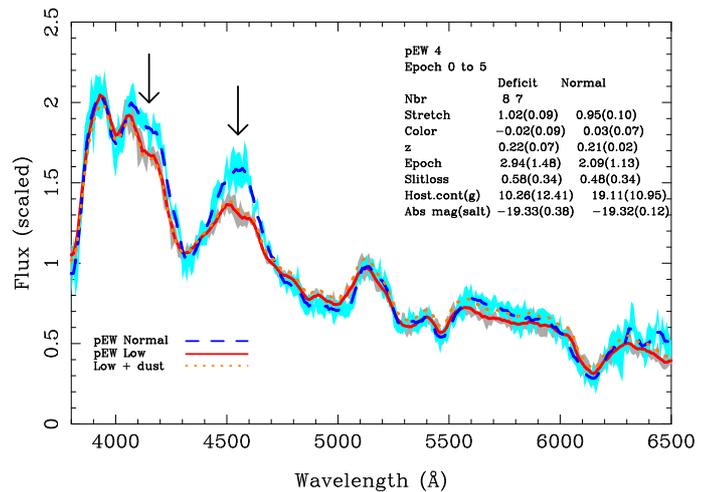}  
  \caption{ 
    \emph{Post peak, f4 deficit vs normal:} Composite spectra of ``reference normal'' SNe (dashed blue) compared with pEW-deficit SNe (solid red) in the epoch range 0 to 5 days past peak brightness. The shaded regions correspond to jackknife dispersion (light blue for normal, grey for deficit). The dotted orange line corresponds to the deficit composite with \citet{cardelli89} dust applied matching the colour difference. The arrows indicate features commented on in the text. Mean properties of the composites are stated in Figure.
  }
  
  \label{fig:f4postpeakcomp}
\end{figure}

As can be seen in Figure~\ref{fig:pew_evo_vsepoch}, most spectra with low pEW f2 can be found \emph{before} lightcurve peak while the deviations in f4 are most clear \emph{after} peak. We therefore make two separate studies: In Figure~\ref{fig:f2prepeakcomp} we compare spectra with deficit pEW f2 in the epoch range $-5$ to 0 with non-\emph{f2deficit} spectra in this epoch range. In Figure~\ref{fig:f4postpeakcomp} we compare spectra with deficit pEW f4 in the epoch range 0 to 5 with non-\emph{f4deficit} spectra.\footnote{\new{Spectra of three SNe (17880, 16215 and 18466) were too deformed (reddening/slit loss) to be used in a composite and were removed from the analysis.}}
%
In each figure we also include the mean and dispersion of a number of properties for each subset. These include both lightcurve properties and estimates for possibly systematic effects like slit loss and host galaxy contamination (See \citet{ostman09} regarding these estimates). We caution that most composites consist of comparably few objects ($\sim10$), and are thus sensitive to random fluctuations.

\paragraph{Description of (physical) differences}

The mean spectra are similar in both cases, except for some limited regions. These are marked with black arrows. Starting with Figure~\ref{fig:f2prepeakcomp} (the pEW f2 deficit before peak brightness) the \emph{deficit} composite has more flux (less absorption) at the silicon lines \si~(f2) and \sifive~(f5), as well as at 4400 {\AA}.

In Figure~\ref{fig:f4postpeakcomp} (pEW f4 deficit after peak brightness) we find more flux (less absorption) in the \emph{normal} composite both at 4150 {\AA} and 4500 {\AA}.

\paragraph{\modifierat{Do the differences agree with} lightcurve correlations?}

We can understand most of the observed differences seen based on the correlations between pseudo-Equivalent widths and lightcurve properties. The pEW f2 deficit SNe in Figure~\ref{fig:f2prepeakcomp} have, on average, higher stretch (1.03) than the normal set (0.92). As was seen above, pEW f2/\si~has a strong negative correlation with stretch. The same is true for \sifive~\citep[see][]{1995ApJ...455L.147N}. Interestingly, the region around 4400 {\AA} show a similar variation as the Si lines. This region was one of the flux ratio ``legs'' used by \citet{2009arXiv0905.0340B} to standardize SNe Ia (as an alternative to lightcurve width). This suggests that the same physical process is responsible for the absorption seen here as at the Si lines.

The f4 composites in Figure~\ref{fig:f4postpeakcomp} have a (small) difference in average stretch, but also in SALT colour: The average colour for the deficit sample is $-0.02$ while it is 0.03 for the normal sample. 
\modifierat{The differences in colour is smaller than the dispersion, but does agree in direction with what we expect form the tentative colour correlation seen in Table~\ref{tab:peakcorr}. As the NTT/NOT SNe have a lower mean colour this could explain why SNe at higher redshifts seem to have smaller pEW f4.}\footnote{We can make a \emph{very speculative} extended comparison with the \citet{2009arXiv0905.0340B} results: If we assume that lightcurve width causes spectral differences as seen in Figure~\ref{fig:f2prepeakcomp} and reddening causes spectral differences as seen in Figure~\ref{fig:f4postpeakcomp} and we want to find one wavelength where flux correlate \emph{both} with lightcurve width and reddening, we are led to the small overlap region at $\sim$4400 {\AA}. \citet{2009arXiv0905.0340B} see most significant correlation at 4430 {\AA}. }

\modifierat{We also find a significant difference at 4130 {\AA}, possibly related to the discussion (also including non-SDSS spectra) in Section~\ref{sec:rest4000} and Figure~\ref{pic:colcomposite}.
More flux is absorbed among low-reddened objects.}

\modifierat{We finally note that some pEW selected subsamples have a smaller absolute magnitude dispersion (\refcom{restframe $B$ band}). Sample sizes are too small to properly evaluate this effect.}

\paragraph{Could the differences be caused by systematic effects?}

Systematic effects such as inaccurate correction of host galaxy contamination, slit loss or dust extinction could significantly affect spectral comparisons. 

To study dust extinction we have applied \citet{cardelli89} type dust matching the difference in SALT colour to the (less reddened) \emph{deficit} composite (dotted orange line in Figures~\ref{fig:f2prepeakcomp} and~\ref{fig:f4postpeakcomp}). In none of the cases does this approach create a better match to the normal set. Standard dust could thus not by itself create the observed differences.

In each figure we have also included mean and dispersion for our (conservatively) \emph{estimated} slit loss and host contamination for each subset. Slit loss is defined as the fractional flux loss at 4000 {\AA} (observed frame), host galaxy contamination as the fraction of galaxy flux in the observed $g$ band \emph{before} host galaxy subtraction.
Slit loss values are consistent among all subsets. Host contamination levels are almost identical for the pEW f2 samples shown in Figure~\ref{fig:f2prepeakcomp}.

For the pEW f4 divided subsets in Figure~\ref{fig:f4postpeakcomp} the mean contamination level is lower for the \emph{deficit} sample ($\sim 10$\%), but it is still comparably low for the reference SNe ($\sim 19$)\%). 
\modifierat{It is very unlikely that this would cause a host contamination systematic difference since (i) the spectral composite differences seen are localized to small wavelength regions and (ii) it is the less contaminated objects that seem to deviate.}

\new{In summary, we find no reasonable combination of systematic effects that
could create the \emph{deficit} subsample.}

\paragraph{Are the pEW deficit SNe ``peculiar''?}

The deficit NTT/NOT SNe were chosen since they do not correspond to \emph{normal} reference SNe Ia. It is natural to ask whether they instead correspond to SNe Ia locally defined as peculiar. None of them seem to be ``very'' odd, in the sense of SN 2000cx or SN 1991T. But they do correspond to several other slightly peculiar \emph{Shallow Silicon} (SS) SNe, like SN 1999aw, SN 1999bp and SN 1999bn (See \citet{2008PASP..120..135B} for a definition of SS). The deficit spectra have, in general, wide lightcurves and small ``shallow'' silicon features.

In total we have 21 SNe belonging to either the f2 or the f4 deficit subgroups.
Among these we find three NTT/NOT SNe that can be classified
as SS SNe (15132, 17497 and 19899). Seven (out of 21) SNe have spectra good enough to be
classified as \emph{not} of this subtype, leaving 14 SNe as possibly peculiar. We thus have between 3 and 14 SNe that would have been identified
as peculiar if observed locally.
In total (deficit \emph{and} normal) we have 41
spectra in this epoch range, which translates into a fraction of SS SNe
between 7\% (3 out of 41) and 32\% (14 out of 41). \modifierat{\citet{2010arXiv1006.4612L} found that 18\% of all SNe in a luminosity limited sample was of the SN1991T subtype but that it gets significantly harder to identify these without early spectra}.

\paragraph{Summary: Evolution}

The ``deviations'' between the NTT/NOT data set and the normal reference
set can be explained through a combination of two (connected) effects:
\begin{itemize}
\item A fraction of ``borderline'' peculiar SNe, mainly similar to Shallow
Silicon SNe, that would have been identified as such if obsereved locally
and thus do not exist in the reference sample of normal SNe.
\item SN features change with lightcurve parameters. Sets with different
lightcurve parameters will thus have different spectral properties.
\end{itemize}

\subsubsection{Sensitivity to evolution models} 

While we cannot exactly determine our sensitivity to (an unknown) spectral evolution, we have simulated how well we would have detected some models of evolution. This process is further described in Appendix~\ref{app:comphost}. These simulations used all property distributions in the NTT/NOT sample and determined how well we would detect \emph{changing SNe subtype distributions}, that is if the fraction of evolved SN increased with redshift. 
\new{We conclude that most of the models studied should have been discovered
at least at low significance ($\sim 2 \sigma$), using one or more indicators
and possibly removing high bias events. These models are, however, not
realistic. 
It is possible that an increased fraction of "deficit" SNe is a sign of
evolving subtype distribution, but if so this does not limit the use of
SNe Ia as standard candles (since these SNe seem to follow the same
luminosity-width relation as other SNe Ia).}

\subsection{Host galaxy properties}
\label{sec:hostdisc}

Different galaxy types give rise to different Type Ia SN
populations. First, the distribution of lightcurve
parameters differ\fixme{REFS?}. Second, as indicated by
\citet{2010ApJ...715..743K,2010arXiv1003.5119S} and \citet{2010arXiv1005.4687L}, Hubble
diagram residuals seem to correlate with host types. However, it is
still not clear how strong these effects are or the causes.

The comparisons between {\si} and host galaxy properties could help clarify these questions. As is discussed in \refcom{Section~\ref{sec:hostprop}} we see indications of relations between spectral indicators and host galaxy properties.

For the epochs after lightcurve peak, we see signs of correlation with
both host mass and specific star formation rate. The measurements could also be interpreted as belonging to different subtypes of SNe Ia \modifierat{(e.g. a subset of SNe produced in low mass hosts with high sSFR whose spectra after peak show very small {\si} values).}

Correlations with host mass could be present
also for pre-max spectral epochs. 
 This is best seen by separating low and high lightcurve stretch SNe.

We have also compared the host galaxy types with the subset of SNe
defined as (possible) shallow silicon (SS) SNe. We find that all clear
SS SNe originate in actively star forming galaxies (Type 2) as well as
three out of four likely SS SNe. These six SNe constitute almost half
of all the 13 SNe with host galaxies of Type 2 (with spectra in the $-5$
to 5 epoch range). Considering the SNe too noisy to be identified as
belonging to a subtype this implies that \emph{half or more of all
SNe in our sample from actively star forming galaxies are
similar to Shallow Silicon SNe.} We also note that these SNe
tend to have blue SALT lightcurve colours and originate in lower mass
host galaxies.


\subsection{Systematic effects}
\label{sec:sysdisc}

\begin{figure*}
  \centering 
  \includegraphics[angle=0,width=\halfcol\columnwidth]{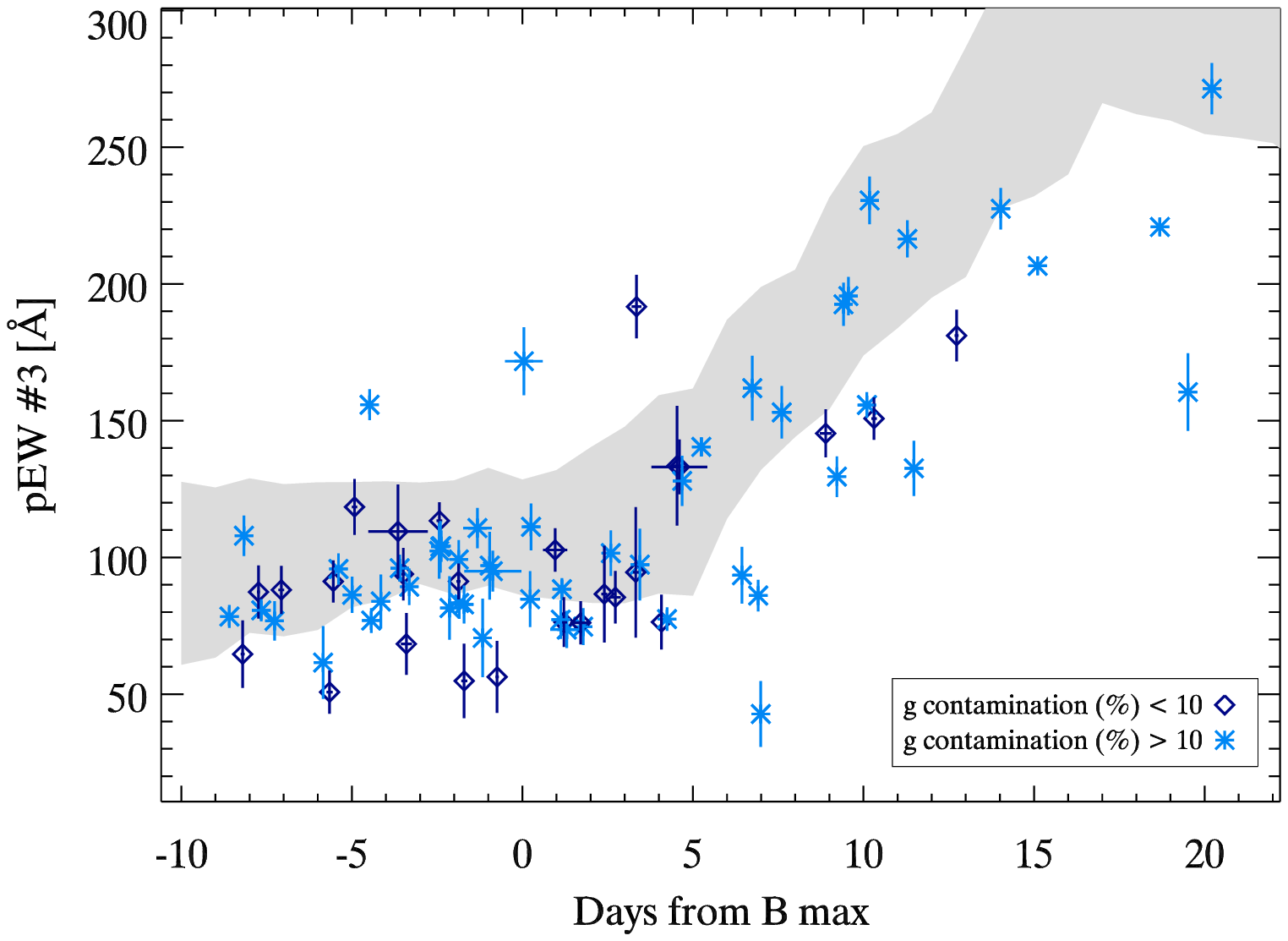}  
  \includegraphics[angle=0,width=\halfcol\columnwidth]{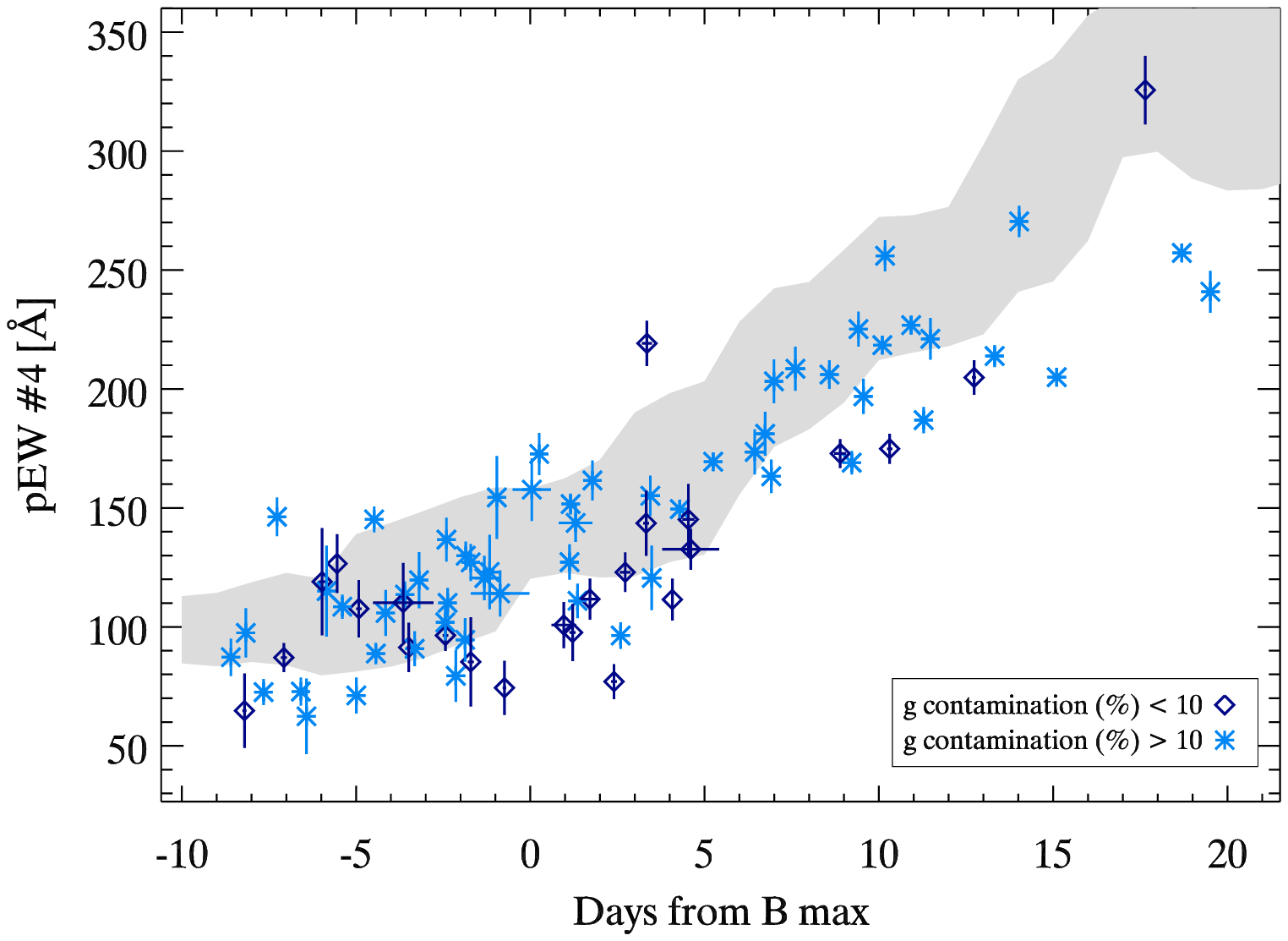} 
  \includegraphics[angle=0,width=\halfcol\columnwidth]{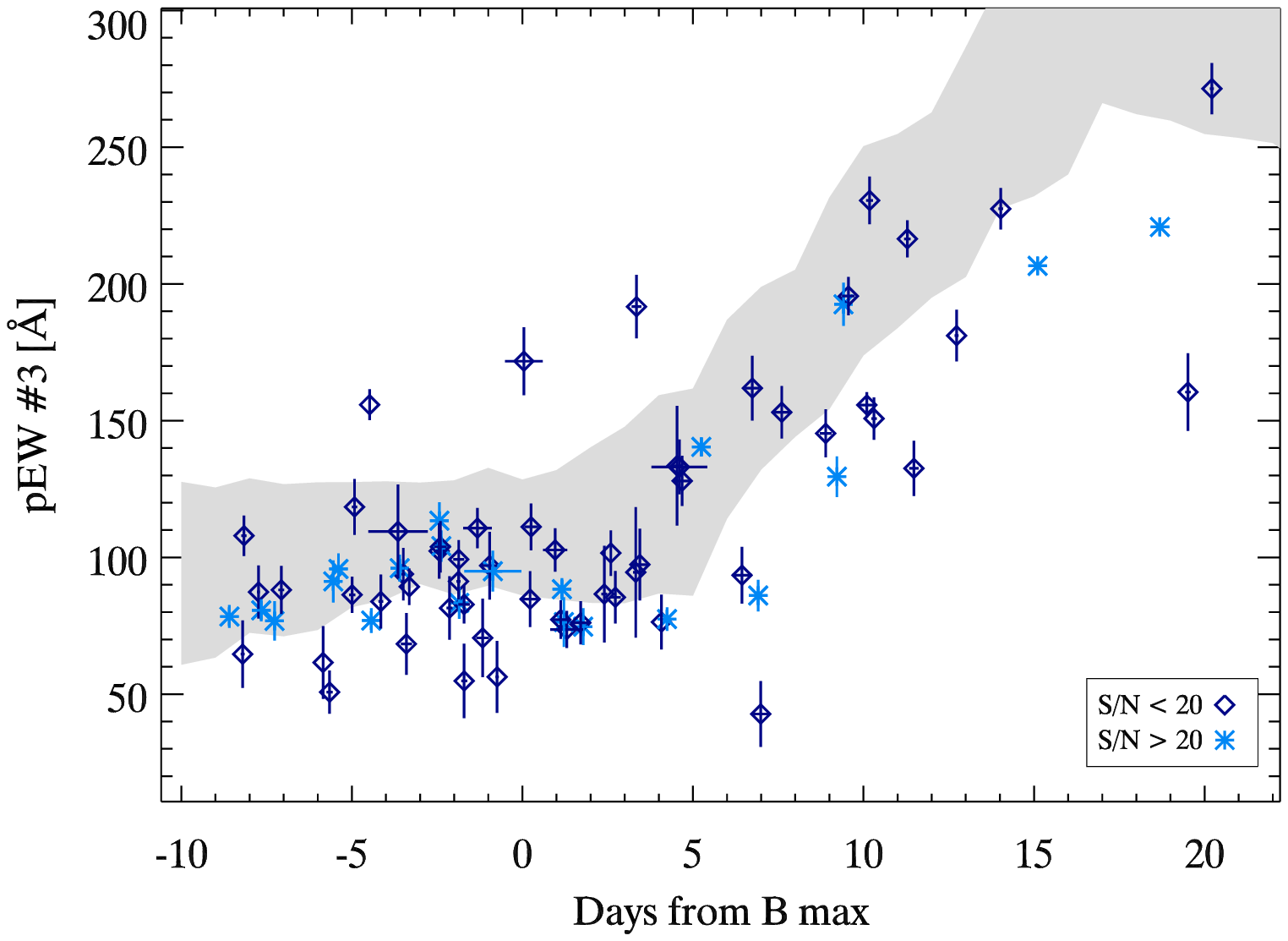}
  \includegraphics[angle=0,width=\halfcol\columnwidth]{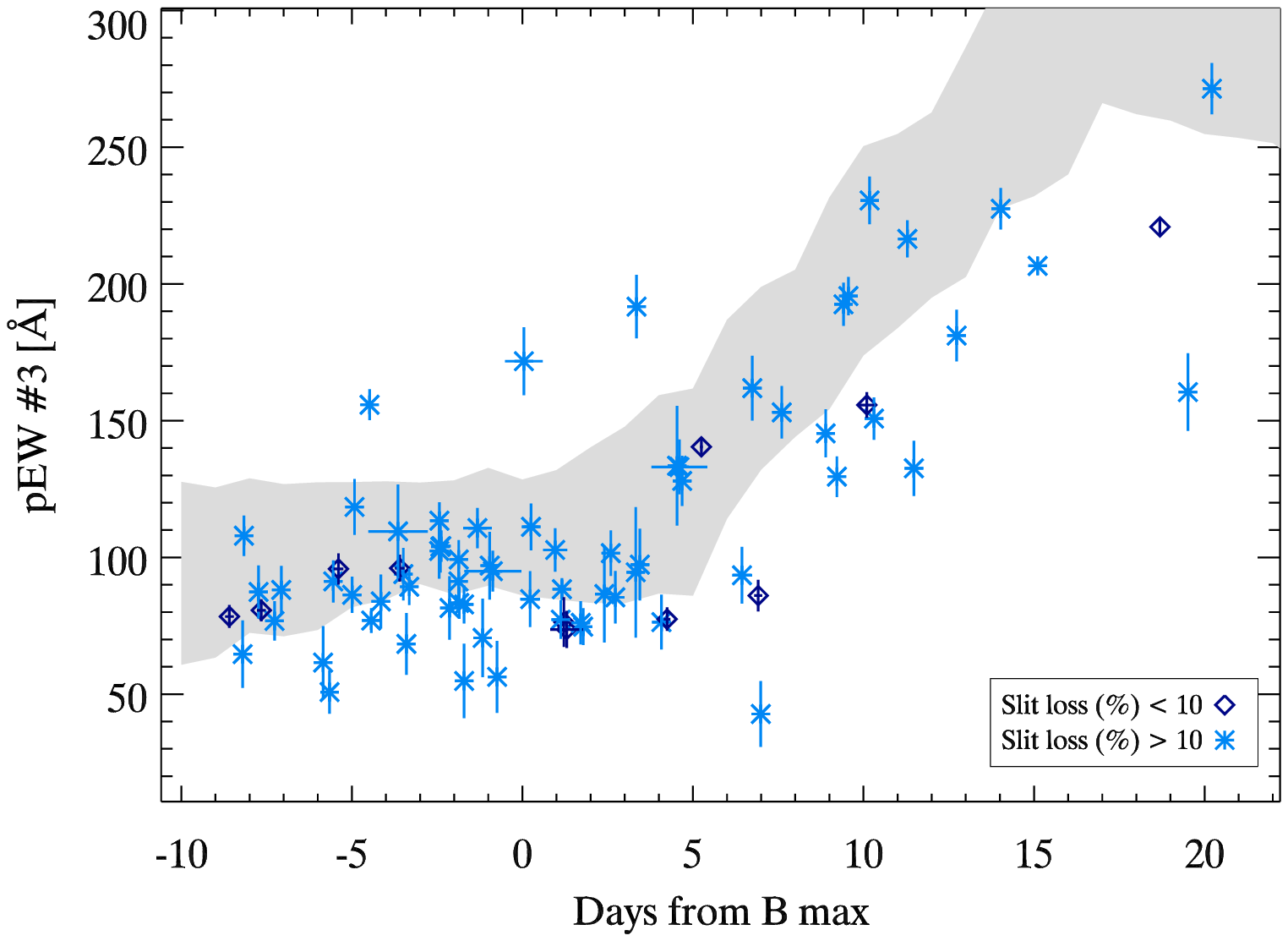}
  \caption{Study of systematic effects: Each panel highlight pEW measurements possibly affected by systematic effects. \emph{Top left:} Feature 3 highlighting low contamination events. \emph{Top right:} Feature 4 highlighting low contamination events. \emph{Bottom left:} Feature 3 highlighting high S/N SNe. \emph{Bottom right:} Feature 3 highlighting low slit loss SNe. No major deviations in subsets with possible systematic errors are detected.
}
  \label{fig:subvarious}
\end{figure*}

We have studied several possible systematic effects. In Figure~\ref{fig:subvarious} we show some sample indicator measurements \modifierat{with NTT/NOT SNe divided into subsamples according to possibly systematic effects}. For example, if host galaxy contamination would yield a systematic bias we would expect low contaminated SNe differ from the high contamination set. We did not detect any such differences or signs of systematic differences.

\paragraph{Use of local templates}

We make use of our knowledge of local SNe and SN templates in a number of different ways. This includes both the host-galaxy subtraction method used here and identification of SNe using templates (e.g. using software like SNID).

This approach is clearly not ideal and creates tension when probing for evolution with redshift. A few items to note regarding this:
\begin{itemize}
\item For moderate evolution, local templates should provide a fair match, and we should be able to accurately type SNe and extract clean spectra. It would be a striking coincidence if the evolution could be completely masked through varying the three galaxy eigenspectra used to model the galaxy. The coefficients of the fit are strongly over-determined by the number of wavelength bins fitted.
\item It is significantly harder to identify completely new subtypes among low S/N distant SNe. While this would be very interesting it is not within the scope of this study.
\item Most cosmological surveys use an identification
mechanism relying on known templates to select the SNe to
include in samples for cosmological fits.
\item The synthetic spectra discussed in Appendix~\ref{app:comphost} are
constructed from SN spectra which differ from the templates
used in the host subtraction. Simulations show that we can still get
correct indicator measurements after subtractions.
\end{itemize}

We thus conclude that the use of SN templates is not a fundamental objection when searching for evolution among SNe Ia used for cosmological studies. 
Our general conclusions would not have changed if only objects with small contamination were included, see Figure~\ref{fig:subvarious}, although they would have had smaller statistical significance.

\paragraph{Noise and slit loss}

Two further possible systematic effects are noise and slit loss. The possible bias effects when comparing high and low S/N data should not be underestimated.To probe these effects, the NTT/NOT sample was split into high/low S/N and high/low slit loss samples and the results compared. In Figure~\ref{fig:subvarious} we show these for feature 3. No significant bias was detected. Simulation results show that we can adjust indicator errors according to noise levels.

However, we note that noise can still limit our ability to determine whether spectral differences are ``real''. In \refcom{Section~\ref{sec:rest4000}} we discussed whether higher noise levels in more reddened SNe could hide small spectral features seen among less reddened SNe.


\paragraph{Selection effects}

The NTT/NOT sample was obtained as part of the SDSS-II Supernova Survey, a rolling search. Very faint targets were usually scheduled for typing at larger aperture telescopes. Even though the NTT/NOT sample has a normal distribution of lightcurve parameters, it is thus possible that these SNe are not representable for the SN Ia population as a whole (e.g. missing some faint objects). We have not attempted a full study of the completeness of the NTT/NOT sample.

The reference sample is very inhomogeneous. These SNe were not observed as part of a rolling survey, and more or less peculiar objects are over represented. Also, since multiple spectra exist of many objects, these objects will carry larger weights.

Finally, the distribution of epoch and colour differ between the
samples. We can thus not expect full agreement between the samples.

As we conclude above the differences between the reference and NTT/NOT sample can be explained by a fraction of semi-peculiar Shallow Silicon SNe. To properly identify these objects, multiple high S/N spectra at early epochs are needed; it is thus not surprising that these are identified in low-z data but not in the NTT/NOT sample.


\section{Conclusions}
\label{sec:conclusions}

We have measured both pseudo-Equivalent widths and line velocities of individual optical spectra observed at the NTT and NOT as part of the SDSS-II Supernova Survey. These spectra cover the redshift range $0.05-0.3$.
Our spectra were compared with a low-redshift sample to probe a
possible evolution between local SNe and SNe at
cosmological redshifts. The samples were then combined and all SNe
were used to investigate possible 
correlations with lightcurve properties.

\modifierat{The differences between reference and moderate redshift SNe can be well described by a fraction ($\sim20$\%) of slightly peculiar, possibly Shallow Silicon, SNe Ia.}

The linear correlation between {\si} pseudo Equivalent-width and lightcurve shape is very significant, both when using SALT stretch and MLCS $\Delta$ parameterisation.
We also found correlations between this feature and SALT lightcurve colour (particularly if highly reddened events are excluded) in spectra observed roughly during the first week after lightcurve peak. This could be an effect from intrinsic colour dependence or a sign of different noise levels. In this epoch range, {\si} correlates with MLCS $\Delta$, but we also see a faint correlation with MLCS $A_V$ for NTT/NOT SNe.
Further studies have to conclude whether an intrinsic physical correlation with reddening exist.

We do not see any significant correlation between spectral properties and absolute magnitude, \modifierat{but we do find a smaller magnitude dispersion among SNe subsamples defined through pEW.}

We also found connections between host galaxy properties and spectral indicators. As for the correlations with lightcurve parameters, these seem strongest for {\si}. Future studies are needed to confirm whether these are real, and whether they, in turn, derive from lightcurve parameters. The correlations could be explained by a subset of SNe with weak {\si} appearing in hosts with low mass and high star formation. 
%


\begin{acknowledgements}

\modifierat{We thank the anonymous referee for valuable comments. We also thank Ryan J. Foley for helpful discussions.}

Funding for the Sloan Digital Sky Survey (SDSS) has been provided by
the Alfred P. Sloan Foundation, the Participating Institutions, the
National Aeronautics and Space Administration, the National Science
Foundation, the U.S. Department of Energy, the Japanese
Monbukagakusho, and the Max Planck Society. The SDSS Web site is
http://www.sdss.org/. The SDSS is managed by the Astrophysical
Research Consortium (ARC) for the Participating Institutions. The
Participating Institutions are The University of Chicago, Fermilab,
the Institute for Advanced Study, the Japan Participation Group, The
Johns Hopkins University, Los Alamos National Laboratory, the
Max-Planck-Institute for Astronomy (MPIA), the Max-Planck-Institute
for Astrophysics (MPA), New Mexico State University, University of
Pittsburgh, Princeton University, the United States Naval Observatory,
and the University of Washington.

The paper is partly based on observations made with the Nordic Optical
Telescope, operated on the island of La Palma jointly by Denmark,
Finland, Iceland, Norway, and Sweden, in the Spanish Observatorio del
Roque de los Muchachos of the Instituto de Astrofisica de Canarias.
The data have been taken using ALFOSC, which is owned by the Instituto
de Astrofisica de Andalucia (IAA) and operated at the Nordic Optical
Telescope under agreement between IAA and the NBI.

\refcom{This paper is partly based on observations collected at  the New Technology Telescope (NTT), operated by  the European Organisation for Astronomical Research in the Southern Hemisphere, Chile. }

The Oskar Klein Centre is funded by the Swedish Research Council.
The Dark Cosmology Centre is funded by the Danish National Research
Foundation. We thank the Swedish Foundation for International Cooperation in Research and Higher Education (STINT) for financial support.

\end{acknowledgements}


\bibliographystyle{aa}
\bibliography{15705}

\appendix

\section{Host-galaxy subtraction uncertainties and evolution detection limits}
\label{app:comphost}

Estimation of the SED of contaminating host galaxy light is an essential step if spectral indicators in contaminated and uncontaminated spectra are to be compared. This will, in turn, be unavoidable when comparing nearby (usually with the SN clearly separated from the host galaxy core) and distant SNe (where SN and galaxy light are degenerate). In \citet{ostman09} we present the host-galaxy subtraction pipeline applied to the NTT/NOT SNe. In short this method consists in matching a SN template with a number of galaxy eigencomponent spectra, including a slit loss/reddening correction. Even if the observed SN SED deviates slightly from the SN templates used in the fit, the very large number of wavelength bins compared with the few fit parameters (five) will allow this SN deviations to remain after the subtraction.

However, the host subtraction produces an increased indicator measurement uncertainty and possibly a bias. It is important that this uncertainty or bias is estimated. In this Appendix we describe the extensive simulations that were run to study the effectiveness of the host subtraction. These simulations were used to calculate a systematic bias and uncertainty for every measurement, depending on the shape of the indicator and contamination level.

As a second step of these simulations we used suggested (metallicity) evolution models to study under which  circumstances these would be detected assuming the properties of the NTT/NOT \dataset.

\subsection{The subtraction pipeline}
\label{sec:pipeline}

The subtraction pipeline is described in detail in \citet{ostman09}. The input parameters are flux density and (optionally) error, observer frame wavelength, redshift and an epoch estimate. This pipeline thus operates identically for real and simulated spectra. A range of internal fit parameters can be changed, including which templates and host galaxy eigencomponent spectra are used as well as the nature of slit loss/extinction approximation. The fit parameters were optimised and fixed during a series of test runs. 

\subsection{Synthetic spectrum simulations}
\label{sec:fake}

To test the reliability of the estimated host galaxy spectra and the
impact on spectral indicators, a large number of simulated
contaminated spectra were created. Besides contamination, these simulations included realistic slit loss and noise levels.  The synthetic spectra are created from
\begin{itemize}

\item {\bf{a supernova spectrum}} \\
The SN spectra used as templates all have high S/N and low contamination. Their epochs are similar to the ones of the NTT/NOT spectra.\footnote{These templates are \modifierat{omitted from} the subtraction pipeline since this would make the fit trivial.} Eleven different SN spectra are used: five of SN 2003du (epochs -6, -2, 4, 9, 10, 17) \citep{2007AA...469..645S}, one of SN 1998aq \citep{2003AJ....126.1489B} at peak brightness, two of the subluminous SN 1999by (epochs -5 and 3) \citep{2004ApJ...613.1120G} and two of the peculiar and luminous SN 1999aa (epochs -5 and 0) \citep{2004AJ....128..387G}. 

\item {\bf{reddening is added to the SN spectrum}}\\
The reddening is \modifierat{added using} the \citet{cardelli89} extinction law using a total-to-selective extinction ratio $R_V$ of 2.1 and a colour excess $E(B-V)$ drawn from the distribution of $E(B-V)$ obtained from the NTT/NOT lightcurve fits.

\item {\bf{a galaxy spectrum}} \\
Four galaxy templates of varying type (elliptical, S0, Sa and Sb) from \citet{1996ApJ...467...38K} are used together with three real galaxy spectra observed at NTT at the same time as the SN spectra analysed here (host galaxy spectra for SDSS SN7527, SN13840 and SN15381). 
The contamination level is randomly chosen between 0 and 70\% for the $g$ band.
These simulations were later extended in a second series where 50 randomly chosen SDSS galaxy spectra were used. Figures displayed here are based on the first run series, but results are similar when including the second set of galaxy spectra.

\item {\bf{redshift}} \\
The object redshift is randomly drawn from the NTT/NOT redshift distribution.

\item {\bf{slit loss is added to the SN spectrum}}\\
The differential slit loss functions are taken from \citet{ostman09} and correspond to typical NTT/NOT situations and range from insignificant to severe.

\item {\bf{noise addition}}\\
A S/N value is randomly chosen from the NTT/NOT spectral S/N distribution. Poisson noise is added to the spectra until the target S/N is achieved. The shape of the noise is determined as a linear combination of the input spectrum and a randomly chosen NTT/NOT sky spectrum. The linear combination is regulated such that the highest S/N value in the NTT/NOT sample corresponds to no contribution from sky noise, the lowest S/N corresponds to complete dominance by sky noise and intermediate values to a combination of the two error sources.

\end{itemize}

All of the created synthetic spectra were then processed through the host
subtraction pipeline and spectral indicators were measured. 
The measured spectral indicators could then be compared with the ones
obtained from the original SN spectrum. The subtractions were thus evaluated only with respect to how well correct indicators were measured.

\subsection{Simulation results}

The simulation results can be analysed in a number of ways: Looking at specific SN spectra, specific galaxy types, spectra with more or less
slit loss or contamination or any combination of these. For each of these subgroups errors in all
equivalent widths and velocities can be calculated. 

In general simulations are stable with the following characteristics: A small bias for very low contamination levels that decrease with added contamination and a random dispersion that increases with contamination. The size of these effects vary slightly from feature to feature.
The small bias for low level contamination means that the subtraction pipeline finds ``something'' to subtract even when no contamination was added. This is fully consistent with having a small amount of host light \emph{already present} in the template spectra. But we cannot rule out that that a part of this bias is caused by the subtraction methods. In practise we do not perform host subtraction on spectra with very low contamination levels. In all cases the full bias as estimated in the simulations is retained, thus generally overestimating the bias levels.

\begin{figure*}[hbtp]
  \centering
  \includegraphics[angle=-90,width= 15cm]{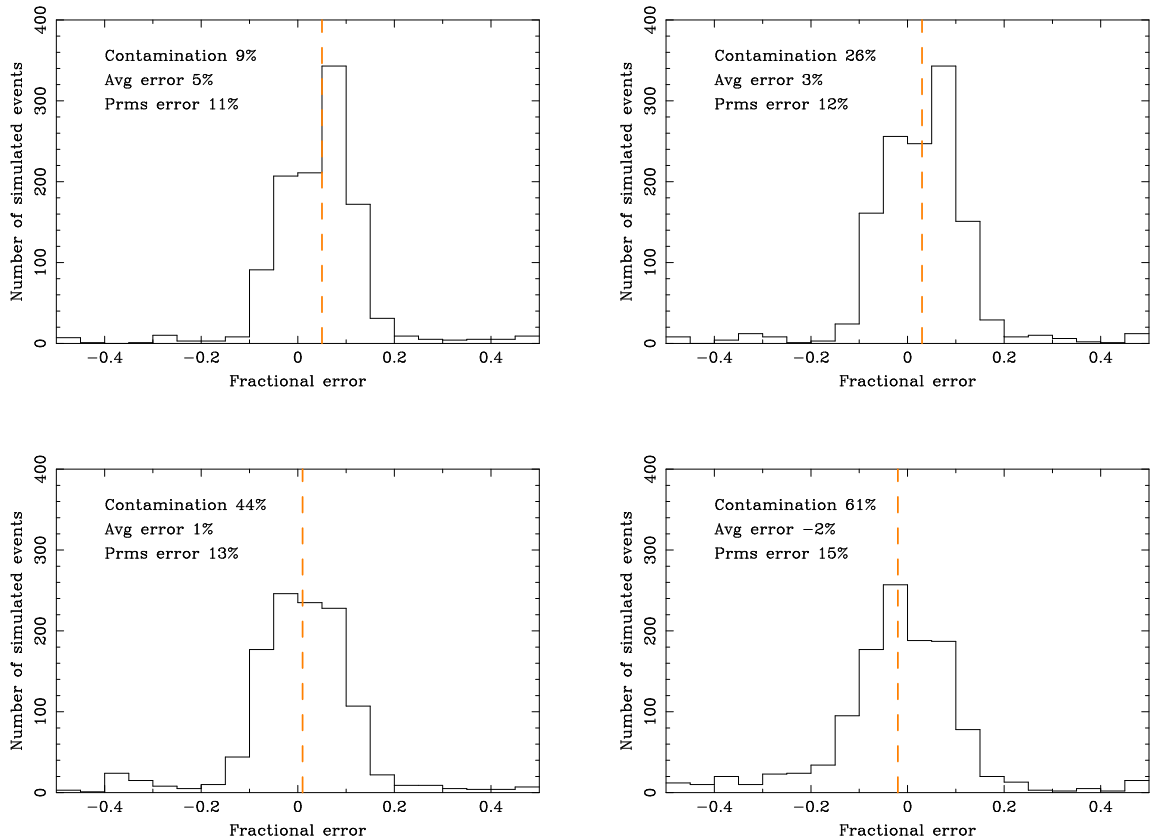}  
  \caption{ Sample host contamination simulations results. For every simulated spectrum the final fractional error is calculated (fractional error is used so that all templates of different epoch and subtypes can be added and analysed as function of contamination). The panels show the distribution of errors, divided into four contamination bins ($0-17.5$,$17.5-35$,$35-52.5$ and $52.5-70$\% in $g$ band). The average contamination, average error and Population RMS (Prms) is printed for each bin. The average error (shown as dashed orange line) indicates a small bias, \emph{decreasing} with contamination. The dispersion indicates a random error from host subtraction, increasing with contamination. These plots are based on pEW f3; other pEWs show similar results.
  }
  \label{fig:constrained}
\end{figure*}

Sample simulation results are presented in Figure~\ref{fig:constrained}.

The simulations were evaluated with and without added noise. Noise was found to increase the error dispersion but not introduce any significant bias. The added dispersion was comparable to uncertainties yielded from the designated noise simulations. We thus separate errors from host contamination and noise. See Appendix~\ref{app:filter} for a further discussion about noise and filtering. For final spectra the systematic uncertainties will be the sum in quadrature of the respective subtraction and noise systematic uncertainties.

\subsection{Alternative subtraction methods}

A number of alternative host subtraction methods were tried. These included two fundamentally different fitting methods: Linear fits using all nearby SN spectra as SN templates and photometry fixed galaxy subtraction where the host galaxy photometry is used to constrain the galaxy shape and proportion. Both methods relax the dependence on the SN template, the first through including a larger variety of such and the second through not using any template at all. However, in general the multiplicative method including the slit loss/reddening correction was found to be superior in most cases and generally more stable.

A number of different implementations of the subtraction pipeline were also tried. These included modifying the number of galaxy eigenspectra, the origin of these eigenspectra and changing constraints on the eigenspectra proportions. The host galaxy subtraction method described above was the final product of these tests.

However, there will be individual objects, for which the host subtraction fails or performs less than ideal. This is a natural consequence of the degeneracy between SN, host galaxy and noise. For some of these objects alternative subtraction methods could have been better suited, but for consistency uniform host subtractions were used.  The simulations were designed to estimate the bias caused by such failed subtractions.

\subsection{Evolution models}

Since it is unknown if evolution exists and how it, if existing, affects the SN Ia SED, it is impossible to predict whether evolution could be detected with the NTT/NOT SNe. But we can still study proposed models to quantify how well these effects would be detected. Two different models were considered here: First ad hoc decrease of the depth of feature 3 and 4, where the \emph{frac} parameter regulates the percent decrease of these depths. This modification was inspired by the indication of changes in these features seen by \citet{2008ApJ...684...68F} and \citet{2009ApJ...693L..76S}. As a second set of models we use the spectral changes caused by one low and one high metallicity model simulated by \citet{2000ApJ...530..966L}. For spectrum templates with epochs less than $-2.5$ the 15 days after explosion model was used, otherwise the day +20 model. 

\begin{figure}
  \centering
  \includegraphics[angle=-90,width=\columnwidth]{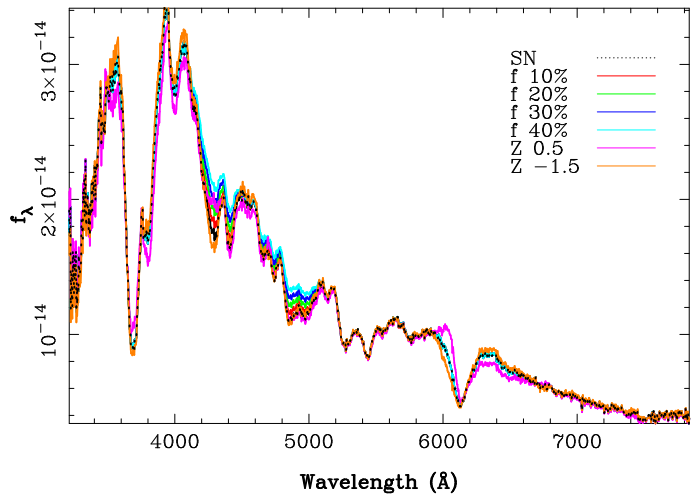}  
  \caption{ Models of evolution/metallicity changes applied to SN2003du. f (frac) models consist of a decrease in the depth of the f3 and f4 features, Z 0.5 corresponds to the \citet{2000ApJ...530..966L} model of increased metallicity, Z -1.5 corresponds to the \citet{2000ApJ...530..966L} model of decreased metallicity. 
  }
  \label{fig:evosample}
\end{figure}

All base SN templates used in the above simulations were modified according to the evolution models, and processed through the subtraction and measurement pipelines again. The modifications as applied to the SN spectrum of SN2003du observed at April 30 2003 is displayed in Figure~\ref{fig:evosample}. 

These models should not be considered realistic evolutionary models to be tested, but rather tests as to what level of evolution can be detected assuming host subtraction uncertainties. They are however, examples of evolution that would not be detected by visual inspection of noisy data but could still effect SN Ia cosmology.

\subsection{Evolution detection limits}

\begin{figure*}
  \centering
  \includegraphics[angle=-90,width=16cm]{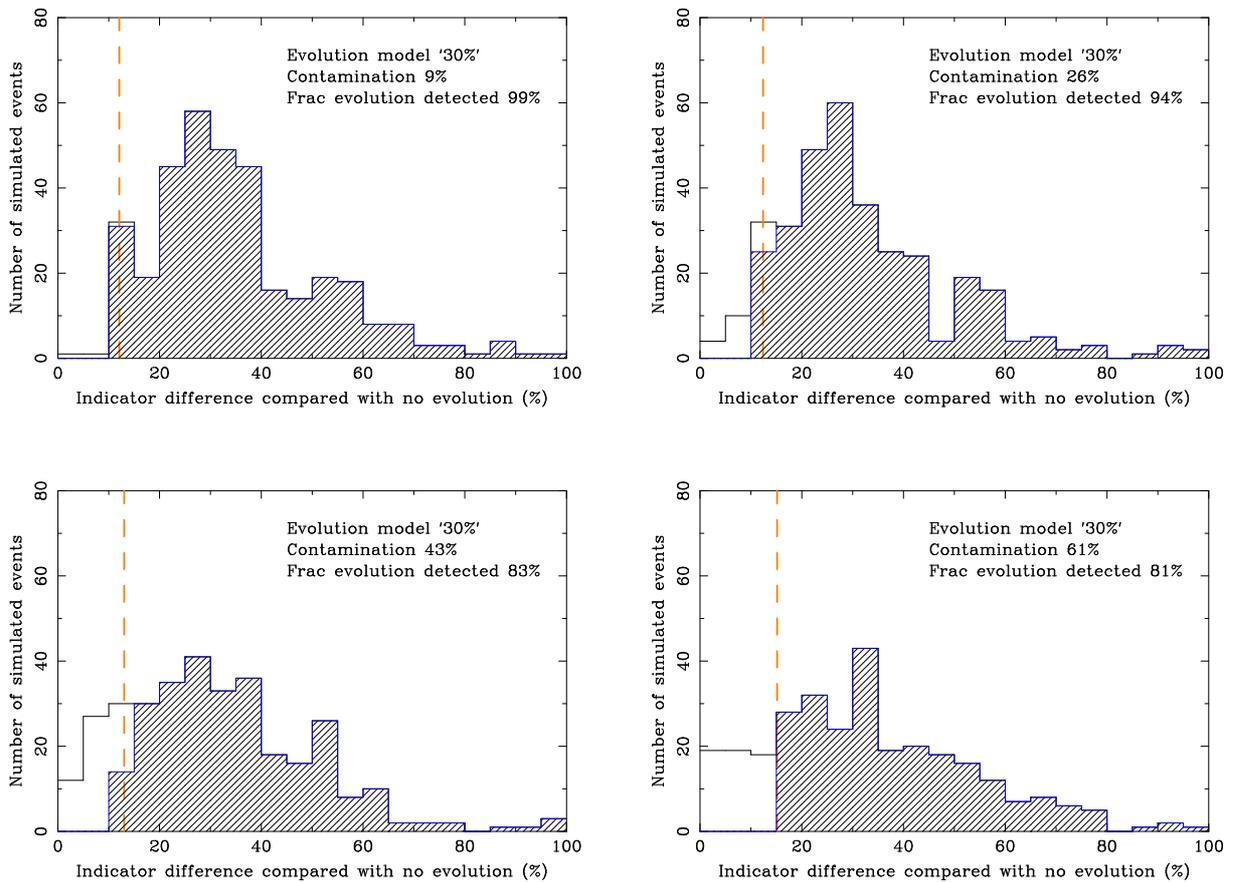}
  \caption{ Sample study of how well evolution is detected in simulated spectra. The ``30 \%'' evolution model was applied to all template spectra and the measured indicators compared with the unevolved measurements. The panels show the distribution of fractional difference, divided into the same contamination bins as in Figure~\ref{fig:constrained}. The total uncertainty in each bin (bias and dispersion) as estimated above is shown as an orange dashed line. \emph{Events where the reported difference is larger than uncertainties would be seen as deviating.} In this sense detectable events are shown as hashed bins. The fraction of detected events is shown in each panel. This fraction decreases with contamination.
  }
  \label{fig:evosec}
\end{figure*}

All measurements on ``evolved'' host galaxy subtracted spectra are collected and compared to the true unevolved reference values. This difference between measurements can then be compared with the estimated statistical and systematic uncertainties and the likelihood of detecting the evolution studied. Sample evolution detection probabilities for evolved SNe is shown in Figure~\ref{fig:evosec}.

These comparisons show that most evolved SNe would be detected. However, the detection limits we are searching for must be realistic: We do not expect all SNe at higher redshift to \modifierat{be} evolved, \modifierat{but} rather the fraction of e.g. low metallicity SNe will change. To study this limit we designed a further simulation based on the NTT/NOT redshift distribution. The probability of each SN to be evolved according to one of the above models, is set to be proportional to redshift and reach 50\% at the average redshift of the NTT/NOT \dataset. For each model we repeat the measurement 5000 times and in each we randomly select which SNe are evolved. The total spectral indicator offset is calculated and compared to the uncertainties, thus obtaining a distribution of the evolution detection limit.

In Table~\ref{table:sigmas} detection limits assuming all NTT/NOT SNe (\emph{including high contamination}) are listed for a number of indicators for the models for evolution/metallicity discussed above. These limits are completely dominated by the systematic bias levels of the high contamination events, since the systematic bias is set to be a systematic floor where the largest bias contained is used. A more realistic and less conservative estimate arises when we remove the highest bias/contamination events; these limits are given in Table~\ref{table:sigmasremoved}. 

These results show that we would be sensitive to all but the very weakest of these evolution models using at least one indicator, albeit at a fairly low significance level.

\begin{table}[ht]
\caption{Probability of detecting models for SN Ia evolution. Each column corresponds to one model (first column is no evolution), see text for further description. Each row corresponds to a search for evolution using the specified spectral indicator, assuming the population changes linearly with redshift. Numbers are the detection level in standard deviations \emph{using max statistical/systematic error as irreducible global error} . The last line is an example where measurements of two indicators are combined to increase sensitivity ('*' = comparison not made).
}
\centering
\begin{tabular}{l c c c c c c c}
\hline\hline
Indicator & 0 & 10 & 20 & 30 & 40 & Low-met & Hi-met \\ [0.5ex]
\hline
pEW f3    & 0 & 0 & 1 & 1 & 1 & 1 & 0 \\
pEW f4    & 0 & 0 & 1 & 1 & 1 & 0 & 0 \\
Vel f3    & 0 & 0 & 0 & 0 & 0 & 0 & 0 \\
Vel f7    & 0 & 0 & 0 & 0 & 0 & 2 & 0 \\
pEW f3+f4 & 0 & 0 & 1 & * & 2 & 1 & 1 \\
\hline
\end{tabular}
\label{table:sigmas}
\end{table}

\begin{table}[ht]
\caption{Probability of detecting models for SN Ia evolution. Each column corresponds to one model (first column is no evolution), see text for further description. Each row corresponds to a search for evolution using the specified spectral indicator, assuming the population changes linearly with redshift. Numbers are the detection level in standard deviations \emph{when removing highest bias events}. The last line is an example where measurements of two indicators are combined to increase sensitivity.
}
\centering
\begin{tabular}{l c c c c c c c}
\hline\hline
Indicator & 0 & 10 & 20 & 30 & 40 & Low-met & Hi-met \\ [0.5ex]
\hline
pEW f3    & 0 & 1 & 2 & 3 & 5 & 3 & 2 \\
pEW f4    & 0 & 1 & 2 & 3 & 4 & 0 & 1 \\
Vel f3    & 0 & 0 & 0 & 1 & 2 & 1 & 2 \\
Vel f7    & 0 & 0 & 0 & 0 & 0 & 4 & 1 \\
pEW f3+f4 & 0 & 1 & 3 & 4 & 5 & 2 & 2 \\
\hline
\end{tabular}
\label{table:sigmasremoved}
\end{table}

\subsection{Velocity host subtraction errors}

Host contamination could affect velocity measurements either through
introducing a false minimum or through modifying the position of the
true minimum. Studies of simulated spectra show that velocity errors
do increase with contamination, but below an $r$-band contamination of
60\%, the errors are small compared to statistical and noise
uncertainties.

Host subtraction methods in general perform similarly. The same
subtractions as for pEWs are used (for consistency). Systematic
uncertainties as estimated from the simulations are added to all
measurements.


\section{Filtering and uncertainties due to noise}
\label{app:filter}

Random noise will degrade data quality, making measurements less accurate. For low S/N SN spectra, the conventional solution is to
apply a filter to remove the high-frequency noise. This technique works well if small
levels of filtering are used (filtering/smoothing are considered
identical processes here), where the true shape is clearly
visible. For noisy data it is no longer obvious what filter to use or how accurate results are.

According to the definition, pseudo-equivalent widths run from one wavelength
extremum point to another. This makes such measurements extremely sensitive
to noise: if any noise peaks remain, the pseudo continuum will be
defined from there. To remove these, and create unbiased data, strong
filtering is needed for low S/N data. We would, however, not want to
filter high S/N spectra (at any redshift) too much since this would
destroy information. We would also like to estimate noise
uncertainties. 

A further complication caused by filtering is that errors in filtered bins are correlated.

\begin{figure*}
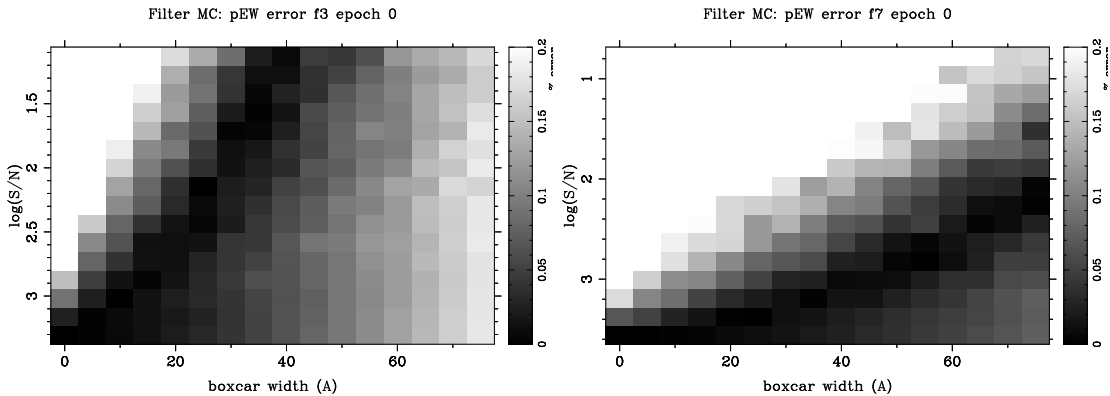

  \centering
  \includegraphics[angle=-90,width=0.8\columnwidth]{15705fig59.ps}  
  \includegraphics[angle=-90,width=0.8\columnwidth]{15705fig60.ps}  
  \caption{Average pEW error for feature 3 (left panel) and feature 7
  (right panel) at the epoch of maximum light. The noise level, expressed through the logarithm of the pseudo-S/N, increases along the y-axis and the filter strength along the x-axis. Darker shades show smaller errors.}
  \label{pic:mcbrusmap}
\end{figure*}

A series of Monte Carlo simulations were run in order to (i) compare
filter methods, (ii) determine filter parameters and (iii) estimate
associated uncertainties (while avoiding having to determine filtered error correlations). These simulations are described below.

\subsection{Filter method comparison}

Three filters easy to implement are (1) the boxcar filter, which is
simple averaging over a wavelength range, (2) the variance-weighted
Gaussian filter where the smoothed value in a pixel is determined from
a surrounding region weighted by a Gaussian determined by the inverse
variance\footnote{See \citet{2006AJ....131.1648B} for a more detailed description.} and (3) the FFT filter, where all frequencies above a
certain maximum frequency are removed from the spectrum.

In order to determine which filter method works best and find optimal
filter parameters, MC simulations were run. Random noise was added to
template SN spectra after which the S/N was determined, the spectra filtered
and indicators measured. For each method the optimal filter parameters
were found through minimisation vs. the true
value. This process was repeated until MC errors were sufficiently
small. It was found that there is no optimal method with a single set
of parameters that worked over the complete range of varying features
and S/N values. All methods \emph{can} yield non-biased values
if correct filter parameters are used. The correct filter parameters
should be determined by the actual noise level and the nature of the
feature studied (broad or sharp).

Since all methods can be made to work but none will work with a single set of parameters, we selected the simplest method, the boxcar filter, as described below.

\subsection{Optimal boxcar filter parameters for pEW measurements}

The above simulations showed that true pseudo-equivalent widths can be measured from noisy spectra after binning, but correct bin widths must be used. A range of MC simulations were run to determine the widths to use and the typical errors caused by noise. This procedure is detailed below.

Noise was generated with a certain amplitude. A gradually
stronger filter was applied, while measuring relevant features at each
stage. Through comparison with the true, noiseless values, the errors
are obtained. For each iteration a ``pseudo-S/N'' is calculated as
follows: A minimal boxcar (spanning three bins) is applied, and a
pseudo-S/N can be calculated by comparing this with the original
spectrum. This value serves as an initial estimation of noise level, and
can later be compared to real spectra (adjusting for bin widths). A
pseudo-S/N is feature relative, and calculated within the maximum
boundaries of the feature in question.

This procedure is repeated 100 times for each noise
amplitude\footnote{Repeated tests were run to verify that results were
not sensitive to the number of iterations.}. For each filter strength and pseudo-S/N we thus have a range of pEW errors, from which we obtain the average and dispersion. Two sample mappings of these errors are displayed in
Figure~\ref{pic:mcbrusmap} (for these maps we have used \emph{absolute} errors). It is seen that for any pseudo-S/N it is
possible to define filter strengths yielding small errors (dark shades in figure), \emph{but}
the optimal filter strength varies with pseudo-S/N.

These maps are used to find the correct filter for a given
feature and pseudo-S/N. Separate maps are created for each feature,
where broad features typically demand stronger filtering. Furthermore,
the \emph{dispersion} of pEW-values in the optimal bin can be used to
approximate the systematic error of doing pEW measurements on noisy
spectra\footnote{This systematic error would only include pure noise
effects and not e.g. effects like host galaxy contamination.}.

Note that it is the shape of the feature that determines correct
filtering, and that this evolves with epoch. To correctly account for
this, the above procedure was repeated for each epoch of the Hsiao
templates \citep{2007ApJ...663.1187H}. The templates were interpolated to 2.5 {\AA} bins in all simulations.

Simulation results are written to a table. These provide, for every
feature and lightcurve epoch, the best filter-width to use to
minimise the risk for noise bias. Since
only the pseudo-S/N is used, we do not require error spectra.

The application to real data can be summarised as:
\begin{enumerate}
\item A minimal boxcar is applied, through which the pseudo-S/N is
determined. 
\item By comparing Monte-Carlo runs for the Hsiao template of the same
epoch and feature, the optimal boxcar width is determined.
\item The average MC error and dispersion around the reference values are
taken as systematic errors from the simulation.
\end{enumerate}

\subsection{Velocity noise errors}

For the well-defined Type Ia SN minima studied here, minimum positions
are stable relative to noise as long as sufficiently wide bins are
used. A constant bin width in velocity space can thus be
used. However, determinations of minima will still be affected by
noise to the degree that on average noisy data will have larger
dispersion. Both these effects, that no bias occurs and the increased
dispersion, were studied using MC simulations of the Hsiao templates
using the same approach as for pEWs. Random noise is added to the
Hsiao templates \citep{2007ApJ...663.1187H} and the velocities
are calculated after binning.

For every template epoch and feature, both bias and dispersion are
obtained as functions of pseudo-S/N. For velocities 2, 3, 5, 6 and 7 (and
reasonable epoch intervals), these results are consistent with no bias and a
gradual increase in dispersion with noise.

For each spectrum studied (in both the reference and NTT/NOT set),
epoch and pseudo-S/N values were used to locate the corresponding MC
dispersion, which was used as systematic velocity error.

For features with more complicated minima (feature 4) or possible additional high velocity absorption features (feature 1), simply determining the minima will not be enough. These features demand either stringent minima criteria or function fitting for optimal study. Automatic minima measurements will show a large scatter.


\section{Data tables}
\label{app:tables}

\include{15705refsn}
\include{15705oursn}

\end{document}

%% file: 15705refsn.tex
\longtab{1}{
\begin{longtable}{p{1.2cm}p{4cm}p{4cm}p{4cm}}
\caption{\label{tab:snspectra} Supernova spectra.} \\
\hline \hline
SN & epochs (days) & Spec source & LC source\\
\hline
\endfirsthead
\caption{continued.}\\
\hline\hline
SN & epochs (days) & Spec source & LC source\\
\hline
\endhead
\hline
\endfoot
\endlastfoot
SN1983g & 6, 7 & \citet{1992AA...259...63C} & - \\
SN1986g & -4, -4, -4, -3, -3, -2, -1, 0, 1, 21 & \citet{2002AJ....124..417H} & \citet{2008AA...492..535A} \\
SN1989b & 0, 6, 11, 21, 22 & \citet{1990AA...237...79B} & \citet{2008AA...492..535A} \\
SN1990n & 2 & \citet{1993AA...269..423M,1998AJ....115.1096G} & \citet{2008AA...492..535A} \\
SN1991bg & 1, 1, 2, 2, 16, 18, 20, 29 & \citet{1996MNRAS.283....1T,1996AJ....112.2094G} & \citet{2008AA...492..535A} \\
SN1991m & 3, 28 & \citet{1998AJ....115.1096G} & \citet{2008AA...492..535A} \\
SN1991s & 14 & \citet{1998AJ....115.1096G} & - \\
SN1991t & -12, -11, -10, -9, -8, -7, -6 & \citet{1995AA...297..509M,1998AJ....115.1096G} & \citet{2008AA...492..535A} \\
SN1992g & 15 & \citet{1998AJ....115.1096G} & - \\
SN1994d & -11, -11, -10, -9, -8, -5, -4, -2, 2, 2, 3, 4, 4, 5, 7, 10, 10, 11, 11, 12, 13, 15, 17, 19, 24, 26 & \citet{1996MNRAS.278..111P,1998AJ....115.1096G} & \citet{2008AA...492..535A} \\
SN1994q & 10 & \citet{1996AJ....112.2094G} & - \\
SN1994s & 22 & \citet{1996AJ....112.2094G} & \citet{2009ApJ...700.1097H} \\
SN1996x & -4, -2, -1, 0, 1, 3, 7, 12, 22, 24 & \citet{2001MNRAS.321..254S} & \citet{2008ApJ...686..749K} \\
SN1997br & -9, -8, -7, -6, -4, 8, 24 & \citet{1999AJ....117.2709L} & \citet{2008AA...492..535A} \\
SN1997cn & 3, 28 & \citet{1998AJ....116.2431T} & \citet{2008AA...492..535A} \\
SN1997do & -12, -11, -8, -7, 8, 10, 11, 12, 14, 15, 20, 21 & \citet{2008AJ....135.1598M} & \citet{2008ApJ...686..749K} \\
SN1997dt & -11, -10, -9, -8, -5, 0, 2 & \citet{2008AJ....135.1598M} & \citet{2008ApJ...686..749K} \\
SN1998ab & -8, 17, 18, 19, 20, 21, 22 & \citet{2008AJ....135.1598M} & \citet{2009ApJ...700.1097H} \\
SN1998aq & -9, -8, -3, 0, 0, 1, 1, 2, 2, 3, 3, 4, 4, 5, 5, 6, 6, 7, 18, 19, 20, 21, 23, 24, 30 & \citet{2008AJ....135.1598M,2003AJ....126.1489B} & \citet{2008AA...492..535A} \\
SN1998bu & -4, -4, -3, -3, -2, -2, 0, 8, 8, 9, 9, 10, 10, 10, 11, 11, 12, 12, 13, 13, 27, 27, 28, 28, 29, 29, 30 & \citet{2008AJ....135.1598M,1999IAUC.7149....2J,2001ApJ...549L.215C,2004AA...426..547S} & \citet{2008AA...492..535A} \\
SN1998dh & -9, -9, -8, -6, -4, -1 & \citet{2008AJ....135.1598M} & \citet{2008ApJ...686..749K} \\
SN1998dm & 4, 5, 7, 10, 12, 15, 17, 24 & \citet{2008AJ....135.1598M} &  \\
SN1998eg & -1, 4, 5, 17, 19, 23 & \citet{2008AJ....135.1598M} & \citet{2009ApJ...700.1097H} \\
SN1998v & 0, 1, 2, 11, 12, 14 & \citet{2008AJ....135.1598M} & \citet{2009ApJ...700.1097H} \\
SN1999aa & -11, -11, -10, -9, -8, -7, -7, -7, -6, -5, -4, -3, -3, -3, -2, -1, -1, 0, 5, 5, 6, 6, 14, 14, 14, 15, 16, 17, 19, 19, 25, 25, 27, 28, 28, 28, 29, 30 & \citet{2008AJ....135.1598M,2004AJ....128..387G,2007AA...470..411G} & \citet{2009ApJ...700.1097H} \\
SN1999ac & -15, -15, -11, -9, -9, -3, 0, 0, 2, 2, 2, 7, 8, 8, 11, 11, 16, 16, 24, 28, 28 & \citet{2008AJ....135.1598M,2005AJ....130.2278G,2006AJ....131.2615P,2007AA...470..411G} & \citet{2008ApJ...686..749K} \\
SN1999af & -5, 1, 15, 17, 17, 25 & \citet{2007AA...470..411G} & - \\
SN1999ao & 5, 7, 9, 12, 17 & \citet{2007AA...470..411G} & \citet{2009ApJ...700.1097H} \\
SN1999ar & 5 & \citet{2007AA...470..411G} & \citet{2009ApJ...700.1097H} \\
SN1999au & 11, 15, 18, 21 & \citet{2007AA...470..411G} & - \\
SN1999av & 2, 5, 9, 30 & \citet{2007AA...470..411G} & - \\
SN1999aw & 3, 5, 9, 12, 15, 23, 30 & \citet{2007AA...470..411G} & \citet{2009ApJ...700.1097H} \\
SN1999be & 14, 19, 26 & \citet{2007AA...470..411G} & - \\
SN1999bi & 5, 11, 12, 26 & \citet{2007AA...470..411G} & \citet{2009ApJ...700.1097H} \\
SN1999bk & 4, 6, 8 & \citet{2007AA...470..411G} & - \\
SN1999bm & 3, 5, 24 & \citet{2007AA...470..411G} & \citet{2009ApJ...700.1097H} \\
SN1999bn & 2, 12, 19, 24 & \citet{2007AA...470..411G} & \citet{2009ApJ...700.1097H} \\
SN1999bp & -2, 0, 1, 6, 16, 21 & \citet{2007AA...470..411G} & \citet{2009ApJ...700.1097H} \\
SN1999bq & 3, 3, 16, 20, 24 & \citet{2007AA...470..411G} & - \\
SN1999by & -5, -5, -4, -4, -3, -3, -2, -2, 1, 2, 3, 3, 3, 4, 4, 5, 5, 6, 6, 6, 7, 7, 8, 9, 10, 11, 16, 24, 25, 28, 29 & \citet{2008AJ....135.1598M,2004ApJ...613.1120G,2007AA...470..411G} & \citet{2008AA...492..535A} \\
SN1999cc & -4, -2, -1, 1, 18, 23, 25 & \citet{2008AJ....135.1598M} & \citet{2009ApJ...700.1097H} \\
SN1999ee & -11, -9, -4, -2, 0, 5, 7, 9, 14, 17, 20, 25, 30 & \citet{2002AJ....124..417H} & \citet{2008AA...492..535A} \\
SN1999ej & -1, 2, 4, 8, 11 & \citet{2008AJ....135.1598M} & \citet{2008ApJ...686..749K} \\
SN1999gd & 2, 9, 27 & \citet{2008AJ....135.1598M} & \citet{2009ApJ...700.1097H} \\
SN1999gp & -5, -2, 0, 3, 5, 7, 22 & \citet{2008AJ....135.1598M} & \citet{2009ApJ...700.1097H} \\
SN2000cf & 3, 4, 14, 16, 24, 25 & \citet{2008AJ....135.1598M} & \citet{2009ApJ...700.1097H} \\
SN2000cn & -10, -9, -8, 8, 10, 12, 21, 26, 27 & \citet{2008AJ....135.1598M} & \citet{2009ApJ...700.1097H} \\
SN2000cx & -4, -3, -2, -1, 0, 1, 5, 6, 7, 9, 11, 14, 19, 22, 24, 26, 28, 30 & \citet{2008AJ....135.1598M,2001PASP..113.1178L} & \citet{2008AA...492..535A} \\
SN2000dk & -5, -4, 1, 4, 9 & \citet{2008AJ....135.1598M} & \citet{2009ApJ...700.1097H} \\
SN2000e & -9, -6, -5, 5 & \citet{2003ApJ...595..779V} & \citet{2008AA...492..535A} \\
SN2000fa & -11, -11, 1, 2, 4, 9, 11, 14, 16, 18, 20 & \citet{2008AJ....135.1598M} & \citet{2009ApJ...700.1097H} \\
SN2001el & 9, 14, 22 & \citet{2003ApJ...591.1110W} & \citet{2003AJ....125..166K} \\
SN2001v & -14, -13, -12, -11, -10, -8, -7, -6, -4, 9, 10, 11, 12, 13, 18, 19, 20, 20, 21, 21, 22, 23, 24, 27, 28 & \citet{2008AJ....135.1598M,2008AJ....135.1598M} & \citet{2003AA...397..115V} \\
SN2002bo & -14, -13, -11, -6, -5, -5, -4, -3, -3, -2, -1, 4, 28 & \citet{2004MNRAS.348..261B} & \citet{2008AA...492..535A} \\
SN2002dj & -11, -10, -9, -8, -6, -4, -3, 9, 10, 13, 17, 22 & \citet{2008MNRAS.388..971P} & \citet{2009ApJ...700.1097H} \\
SN2002er & -11, -9, -8, -7, -6, -5, -4, -3, -2, -1, 0, 2, 4, 5, 6, 10, 12, 13, 16, 17, 20, 25 & \citet{2005AA...436.1021K} & \citet{2008AA...492..535A} \\
SN2003cg & -9, -8, -7, -5, -2, -2, -1, 1, 4, 7, 10, 11, 12, 16, 19, 23, 23, 26, 28 & \citet{2006MNRAS.369.1880E} & \citet{2008AA...492..535A} \\
SN2003du & -13, -11, -11, -11, -8, -7, -6, -5, -4, -3, -2, -1, 0, 0, 1, 2, 2, 3, 4, 6, 6, 7, 8, 9, 9, 10, 13, 15, 17, 18, 19, 21, 24, 26 & \citet{2007AA...469..645S,2005AA...429..667A,2005ASPC..342..250G} & \citet{2008AA...492..535A} \\
SN2004dt & -10, -9, -9, -7, -7, -6, -6, -4, -4, -3, -2, -1, -1, 2, 3, 4, 14, 17, 21 & \citet{2007AA...475..585A} & - \\
SN2004eo & -11, -6, -3, 2, 7, 11, 13, 14, 21, 22, 24, 30 & \citet{2007MNRAS.377.1531P} & \citet{2008AA...492..535A} \\
SN2004s & 1, 7, 12, 13, 13, 18 & \citet{2007AJ....133...58K} & \citet{2008AA...492..535A} \\
SN2005bl & -6, -5, -3, -3, 4, 12, 19, 21 & \citet{2008MNRAS.385...75T} & - \\
SN2005cf & -12, -12, -11, -10, -10, -9, -7, -7, -6, -4, -3, -1, 0, 4, 4, 5, 6, 7, 9, 12, 12, 14, 16, 25, 29 & \citet{2007AA...471..527G,2007AIPC..937..311L} & \citet{2008AA...492..535A} \\
SN2005cg & -10, -9, -4, 0, 5, 7 & \citet{2006ApJ...636..400Q} & - \\
SN2005hj & -6, 0, 2, 5 & \citet{2007ApJ...666.1083Q} & \citet{2009ApJ...700.1097H} \\
SN2005hk & -8, -7, -6, -6, -5, -4, -4, -3, 4, 13, 15, 24, 27 & \citet{2007PASP..119..360P} & \citet{2008AA...492..535A} \\
SN2006gz & -14, -14, -13, -13, -12, -10, -9, -5, -2, 5, 6, 7, 8, 8, 9, 10, 11 & \citet{2007ApJ...669L..17H} & \citet{2008AA...492..535A} \\
SN2006x & -10, -7, 0 & \citet{2009PASJ...61..713Y} & - \\
\end{longtable}
}

%% file: 15705oursn.tex
\longtab{2}{
\begin{longtable}{p{1.0cm}p{1.0cm}p{3cm}p{3cm}}
\caption{\label{tab:sdsspec} NTT/NOT spectra.} \\
\hline \hline
ID & IAU & SPID & Epochs (days) \\
\hline
\endfirsthead
\caption{continued.}\\
\hline\hline
ID & IAU SPID & Epochs (days) \\
\hline
\endhead
\hline
\endfoot
\endlastfoot
12781&2006er&680&10.9\\
12843&2006fa&727&10.2\\
12853&2006ey&685&10.3\\
12856&2006fl&695&-3.2\\
12860&2006fc&688&-1.9\\
12898&2006fw&712&-6.6\\
12930&2006ex&687&10.1\\
12950&2006fy&700&-4.4\\
13025&2006fx&761&3.4\\
13044&2006fm&724, 1062&-8.2, 20.2\\
13070&2006fu&736&6.9\\
13072&2006fi&723&0.0\\
13135&2006fz&739, 998&-7.7, 17.6\\
13796&2006hl&1058, 1058&12.7, 12.7\\
13894&2006jh&1039&9.2\\
14157&2006kj&1040&9.4\\
14437&2006hy&1061&14.0\\
14846&2006jn&1014&-1.7\\
14871&2006jq&1008&-4.2\\
14979&2006jr&1009&-2.1\\
14984&2006js&1027&-1.2\\
15129&2006kq&1015&1.8\\
15132&2006jt&1012&-2.4\\
15161&2006jw&1010&-1.0\\
15171&2006kb&1045, 1045&-5.7, -5.7\\
15203&2006jy&1026&-2.4\\
15222&2006jz&1004&-5.8\\
15259&2006kc&1051&-1.9\\
16021&2006nc&1355&11.3\\
16069&2006nd&1467&11.5\\
16165&2006nw&1326&2.6\\
16215&2006ne&1456&4.3\\
16287&2006np&1449, 1449, 1569, 1569, 1650&2.4, 2.4, 3.3, 3.3, 19.5\\
16352&2006pk&1478&4.1\\
16473&2006pl&1520&1.3\\
16637&&1514&-0.9\\
17332&2007jk&1899&3.3\\
17366&2007hz&1782&8.9\\
17389&2007ih&1811&7.0\\
17435&2007ka&1902, 1902&2.7, 2.7\\
17497&2007jt&1837&-2.4\\
17552&2007jl&1789&3.5\\
17745&2007ju&2161&15.1\\
17784&2007jg&1842&-5.5\\
17790&2007jx&1887&1.0\\
17811&2007ix&1816, 1816&4.6, 4.6\\
17825&2007je&1819&-4.9\\
17875&2007jz&1817&0.3\\
17880&2007jd&1843, 1957&-1.9, 1.2\\
17886&2007jh&1844&-4.5\\
18325&2007mv&2277&8.6\\
18466&2007lm&2270&4.5\\
18768&2007lh&2135&6.7\\
18787&2007mf&2150&0.2\\
18804&2007me&2148&-5.0\\
19023&2007ls&2236&-1.7\\
19101&2007ml&2268, 2268&-6.0, -6.0\\
19149&2007ni&2275, 2275&-7.1, -7.1\\
19155&2007mn&2607&18.7\\
19282&2007mk&2280&-8.2\\
19341&2007nf&2298&-2.4\\
19353&2007nj&2281&-7.3\\
19381&2007nk&2283, 2283&-3.5, -3.5\\
19899&2007pu&2550&1.2\\
19913&2007qf&2585&9.6\\
19953&2007pf&2602&4.2\\
19968&2007ol&2549&5.2\\
20039&2007qh&2584&7.6\\
20040&2007rf&2612&6.4\\
20142&2007qg&2586&4.7\\
20144&2007ql&2541&1.1\\
20345&2007qp&2567, 2567&-0.7, -0.7\\
20364&2007qo&2581&-1.3\\
20430&2007qj&2543&1.4\\
20625&2007px&2551, 2604&-5.4, -3.6\\
21006&2007qs&2566&1.7\\
21033&2007qy&2565&-3.3\\
21034&2007qa&2719&13.3\\
21042&2007qz&2564&-6.4\\
21422&2007rq&2599&-3.6\\
21502&2007ra&2574, 2575&-8.6, -7.7\\
\end{longtable}
}